\documentclass[aps,prx,showpacs,preprintnumbers,twocolumn,superscriptaddress]{revtex4-1}
\usepackage{amsmath,amssymb}
\usepackage{bm}
\usepackage{tipa}
\usepackage{upgreek}
\usepackage{comment}
\usepackage{mathrsfs}
\usepackage{graphicx}
\usepackage{braket}
\usepackage{enumitem}
\usepackage{mathbbol}
\usepackage{gensymb}
\usepackage[normalem]{ulem}
\usepackage{color}
\usepackage[colorlinks,bookmarks=true,citecolor=blue,linkcolor=red,urlcolor=blue]{hyperref}
\usepackage{hyperref}
\usepackage{dsfont}

\begin{document}

\title{Riemann meets Goldstone: magnon scattering off quantum Hall skyrmion crystals probes interplay of  symmetry breaking and topology}

\author{Nilotpal Chakraborty}
\thanks{Corresponding author}
\email{nilotpal@pks.mpg.de}
\affiliation{Max-Planck-Institut f\"{u}r Physik komplexer Systeme, N\"{o}thnitzer Stra\ss e 38, Dresden 01187, Germany}

\author{Roderich Moessner}
\affiliation{Max-Planck-Institut f\"{u}r Physik komplexer Systeme, N\"{o}thnitzer Stra\ss e 38, Dresden 01187, Germany}

\author{Benoit Doucot}
\affiliation{LPTHE, UMR 7589, CNRS and Sorbonne Universit\'e, 75252 Paris Cedex 05, France}

\begin{abstract}
We introduce a model to study magnon scattering in skyrmion crystals, sandwiched between ferromagnets which act as the source of magnons. Thanks to recent experimental advances, such a set-up can be realised in quantum Hall heterojunctions, and it is interesting as skyrmions are topological objects while the skyrmion crystals break internal and translational symmetries, thus allowing to study the interplay of topological and symmetry breaking physics.
Starting from a basis of holomorphic theta functions, we construct an appropriate analytical ansatz for such a junction with finite spatially modulating topological charge density in the central region and vanishing in the leads. We then construct a suitably defined energy functional for the junction in terms of these  spinors and derive the resulting equations of motion, which take the form of a Bogoliubov-de Gennes-like equation. Using a combination of analytical techniques, field theory, heuristic models and fully microscopic recursive transfer-matrix numerics, we calculate the spectra and magnon transmission properties of the skyrmion crystal. We find that magnon transmission can be understood via  a combination of low-energy Goldstone modes and effective emergent Landau levels at higher energies. The presence of the former manifests in discrete low-energy peaks in the transmission spectrum and we show how the these features reflect the nature of the Goldstone modes arising from symmetry breaking. In turn, 
the effective Landau levels, which reflect the topology of the Skyrmion crystal, lead to band-like transmission features, from the structure of which further details of the excitation spectrum of the skyrmion crystal can be inferred.  Such characteristic transmission features are not present in  competing phases of either the quantum Hall phase diagram or in metallic magnets, and hence provide direct and unique signatures of skyrmion crystal phases and their properties.
We discuss experimental considerations regarding the realisation of our model, which most directly apply to heterojunctions in monolayer graphene with the central region doped slightly away from unit filling and the two ends exactly at unit filling, a $\nu = 1:1 \pm \delta \nu : 1$ junction. Such physics is also relevant to junctions formed by metallic magnets which host skyrmion crystal phases, or partly in junctions with artificially realized and periodically modulated gauge fields. 
\end{abstract}

\maketitle

\section{Introduction}
Two central paradigms of condensed matter physics are symmetry breaking and, more recently, topology \cite{anderson2018basic,moessner2021topological}. The concepts involved, down to the language describing them, are quite distinct,  and it is interesting to ask what happens in `mixed situations' where emergent topology and symmetry breaking are both present. One of these is provided by the physics of skyrmions, which are topological objects which also carry a notion of symmetry breaking -- most immediately regarding  internal spin degree of freedom. Natural questions thus arise regarding the demands of the respective paradigms. For instance, Goldstone's theorem demands the existence of stable quasiparticles at low energies, while topological phases tend to come with gapped spectra and low-lying excitations living only at edges and interfaces. Moreover, the natural excitations of topological systems can have quantum numbers which are quite distinct from those of the underlying electronic degrees of freedom. 

A case in point is the SU(2)-invariant quantum Hall effect at $\nu=1$, where a quantized transport plateau coexists with skyrmionic elementary excitations arising from the ferromagnetic ground state.  Quantum Hall skyrmions are special in that they possess quantized electrical charge \cite{sondhiskyr}.  Tuning slightly away from this filling is believed to lead to the formation of a ground state configuration of skyrmion crystals \cite{Breyskyr}. Skyrmion crystals are like Wigner crystals  of composite objects, each of which comprises a group of textured spins and acts as a topological defect \cite{rajaraman1982solitons}. Crucially, these crystals exhibit  spatial symmetry breaking on top of the internal symmetry breaking. Skyrmion crystals have also been heavily studied in metallic magnets, where they arise due to the  Dzyaloshinskii–Moriya interaction, and their detection in such settings was first reported in landmark neutron scattering \cite{muhlbauer2009skyrmion} and electron microscopy experiments \cite{yu2010real}. While there has been some indirect evidence for the existence of a quantum Hall skyrmion crystal, via NMR \cite{gervais2005evidence,desrat2002resistively}, heat capacity \cite{Bayotprl}, Raman \cite{Gallaisraman} and microwave spectroscopy \cite{Hanzhuprl} experiments  direct evidence, such as that in an electron microscopy experiment  imaging the degree of crystalline order is still missing.

Experimental techniques to detect crystalline ordering and to unveil the excitation spectra of ordered structures have a long history in solid state physics. From Bragg scattering of x-rays to detect crystalline structure of solids, to neutron scattering and ARPES experiments to probe the excitation spectra, with the advent of a new experimental technique, new theoretical explorations are called for. 

Recent magnon transport experiments in junctions of quantum Hall states present one such exciting technique \cite{wei2018electrical,zhou2021magnon,zhou2022strong,pierce2022thermodynamics,assouline2021excitonic}. While traditional ARPES and neutron scattering experiments are extremely challenging for thing nanomaterials such as graphene, electron transport experiments  provide a promising route to probe quantum Hall physics \cite{zhang2005experimental,novoselov2007room}, especially due to the ability to tune carrier density with voltage in graphene. However, electron transport experiments are largely insensitive to the underlying spin structure of the ground states. 

Pioneered in \cite{wei2018electrical},  magnon transport techniques involve a coherent source of magnons, usually a quantum Hall ferromagnet, which are injected into an insulating bulk sandwiched between the leads. These techniques allow us to probe the spin structure of the bulk and have been used to study  various ground states expected at different fillings of the zeroth Landau level in monolayer graphene. 

Applying these experimental techniques to questions involving the topology-symmetry dichotomy has to face a number of technically and conceptually unavoidable issues. Concretely, constructing an interface between a skyrmion crystal and a non-topological magnetic state cannot simply be achieved by pasting the two subsystems together along a junction in the way one would, e.g., join a superconductor with a normal metal to observe Andreev reflection. The reason is that the skyrmion is a non-uniform and extended object. An interface will thus minimally need a lateral extent set by the size of the skrymion itself. Moreover, such a problem is theoretically interesting because skyrmions and their crystals are objects that lie in complex projective spaces ($\mathbb{C}\mathrm{P}^{d-1}$ for $SU(d)$ systems), hence the interface problem becomes a non-linear problem as opposed to the conventional  bulk-boundary correspondence in quantum Hall and topological insulators \cite{girvin2002quantum,KaneHasanTI}.

\begin{figure*}[t]
    \centering
    \includegraphics[width = 18cm, height = 10cm]{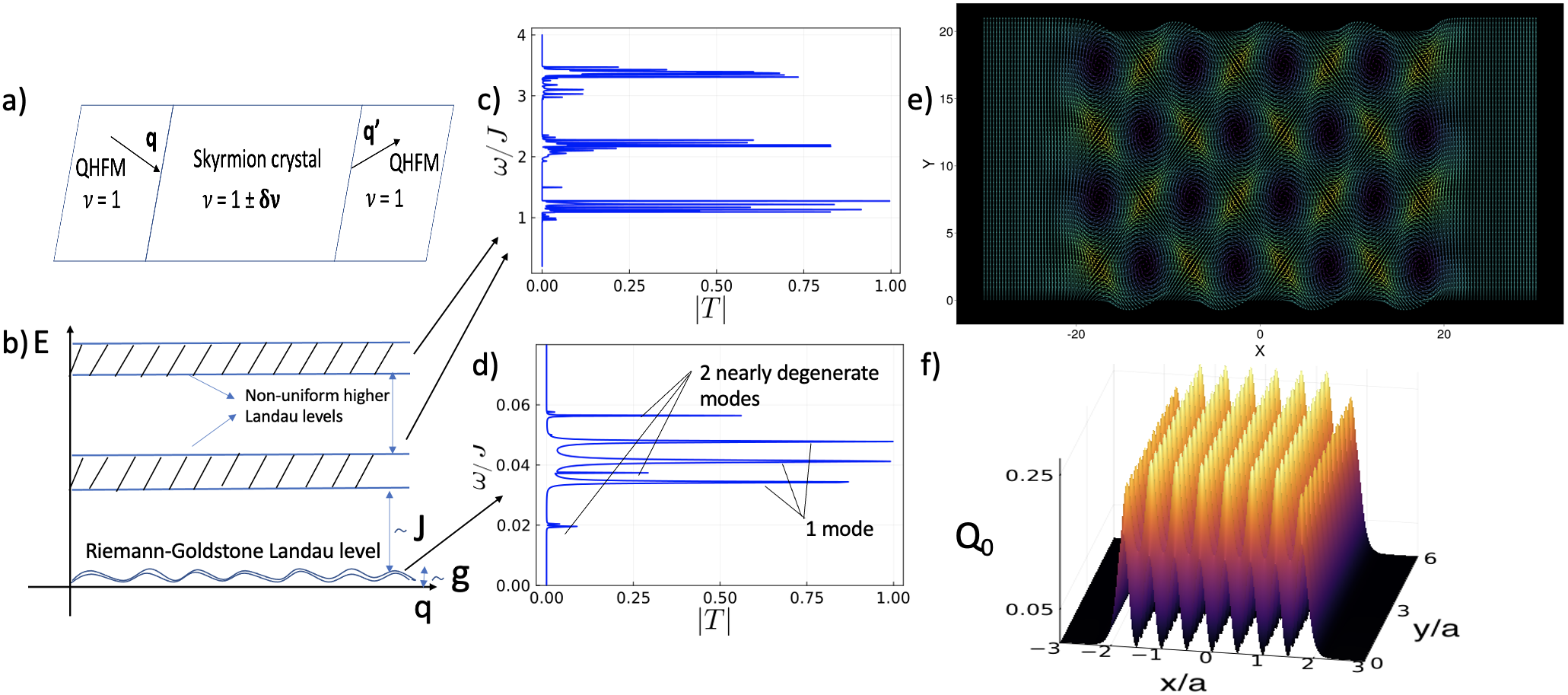}
    \caption{Results for the quantum Hall ferromagnet (QHFM)-skyrmion crystal junction. a) Schematic picture of the scattering problem and the experimental setup - sharp interface drawn only for illustrative purposes. b) Cartoon description of the spectrum of the skyrmion crystal comprising higher non-uniform Landau levels and the lowest Riemann-Goldstone Landau level. c) High-energy band-like transmission features due to effective non-uniform Landau levels (generated due to spatially modulated topological charge density profile (with finite non-zero mean) in (f) and not uniform applied external field). d) Low-energy transmission features due to presence of Goldstone modes in the Riemann-Goldstone Landau level. Two sets of linearly separated peaks indicating linear dispersion and two distinct velocities of the modes. Small split in peaks with larger spacing (velocity), implies two modes are almost degenerate (see Fig. \ref{geff}b for spectrum). e) Spin profile of the junction, from Eq. (\ref{nfrompsi}), generated by the truncated holomorphic theta function ansatz in Eq. (\ref{thetafunc}). The tail of every arrow is a lattice site, its direction is the projection on the z-x plane, and the color is the y-component. f) Topological charge density profile for the junction, obtained from Eq. (\ref{topceq}) and Eq. (\ref{nfrompsi}).  Parameters used for c)-f) are $g= 0.8$, $J = 1$, $N = 140$ and $N' = 1$.}
    \label{Fig1}
\end{figure*}

Here, we devise and study a scattering problem which is motivated by, and amenable to, the above mentioned experimental methods. The set-up consists of a quantum Hall heterojunction (Fig. \ref{Fig1}a) of a Skyrmion crystal sandwiched between two simple quantum Hall ferromagnets, as might be obtained by setting the filling of the outer regions to $\nu = 1$ and doping (or, rather, gating) the central region slightly away from such filling $\nu = 1 \pm \delta \nu$. Such a setup has already been realized in one of the quantum Hall junction experiments \cite{zhou2020solids}.

Our central result is that the energy dependence of the magnon transmission amplitude  reflects the topology-symmetry dichotomy in exquisite detail, establishing such magnon scattering experiments as an excellent platform to probe this dichotomy. The topology of the Skyrmion crystal bequeathes an emergent Landau-level structure to the response; while its lowest Landau level -- which we christen \textit{Riemann-Goldstone Landau level} -- contains the physics of the symmetry breaking itself. Remarkably, from the magnon transmission, one can directly infer the nature of the Goldstone modes which is characteristic of the Skyrmion crystal as well as the effective Landau level structure of the higher levels. 
 
A significant fraction of the following account details important technical advancements that we made to fully solve this problem. We focus on advances which are transferable and useful to other contexts and fields in the main text, and discuss some more specific ones in the appendix. First, we introduce our completely analytical model of a ferrromagnet-skyrmion crystal-ferromagnet interface formed from a basis of holomorphic theta functions, as well as a suitably defined energy functional from which we derive our equations of motion. Second, we introduce a novel method to discretize the topological charge density contributions to the energy functional. Third, we explain our microscopic recursive transfer matrix approach to calculate the full transmission and reflection matrix of the skyrmion crystal scattering problem, even in the presence of evanescent contributions. Fourth, we introduce a recipe to construct sigma models for the coupling between the Goldstone modes of such junctions of regions with different order parameter manifolds. All these advances can find applications in transport problems between interfaces of such topologically trivial and non-trivial structures and possibly also in transport through regions of spatially varying magnetic field.

For the less technically inclined readers, we supply some simple heuristic models to account for the physical phenomena that we have uncovered. While these do not capture the full complexity of the topology-symmetry dichotomy, they do provide a clear rationale for why the proposed set-up is so well-suited for studying this problem, and they yield a transparent and intuitive framework for the interpretation of the full results of our analysis. These heuristic models already provide some predictions which can be tested in future skyrmion crystal junction experiments.

The remainder of our account is structured as follows. Section II provides a short-hand self-contained and largely non-technical summary of our results. Section III contains some general considerations of the dichotomic structure of the problem which leads to an intuitive picture for the formalism developed in later chapters, and for the interpretation of the results thus obtained: a  model of a particle in a heterostructure comprising a modulated magnetic field sandwiched between two zero-field regions provides a simple route to capturing the topological features which are independent of the local symmetry breaking. For the Goldstone sector, we consider a simplified interface between a ferro- and an antiferromagnet on a lattice, which allows us to study the simplest case of a dispersion mismatch problem. We note that it has come as quite a surprise to us that this dichotomy should be so neatly resolvable by this pair of heuristic models. 

Section IV is the most technical section of this paper. In subsection A, we introduce our holomorphic theta functions ansatz and our energy functional. In subsection B, we introduce the method to discretize the topological charge density contributions. 
In subsection C,  we explain our transfer matrix procedure for the magnon scattering problem. We end this section with subsection D, where we introduce the recipe to construct the sigma models. In section V, we present additional results for the quantum Hall ferromagnet-skyrmion crystal-ferromagnet problem which are not discussed in section II, and we highlight similarities with the heuristic models presented in section III. Finally, in section VI, we end by discussing the experimental relevance of our model (subsection A) and further implications of our work (subsection B).

\section{Overview of results}
Our main result is a calculation of the magnon transmission spectrum across the skyrmion crystal.
We find the following characteristic set of  signatures in the transmission spectrum which reflect the topology-symmetry dichotomy:

i) The high-energy transmission spectra comprises sharp peaks in discrete bands of energies, with uniform gaps between the bands as in Fig. \ref{Fig1}c. The sharp peaks within each band arise as a consequence of Fabry-Perot like resonances when the incoming magnon energy coincides with bound states of the scattering problem. Remarkably, these bound states have an emergent Landau-level structure. These levels are emergent because they reflect the spatially modulated topological charge density (which has a finite non-zero mean) of the Skyrmion crystal (instead of the applied external constant magnetic field): magnons experience the Berry flux of the spin texture in the central region as an effective magnetic field.

ii) The lowest emergent Landau level, which we call the \textit{Riemann-Goldstone Landau level} due to it arising from holomorphic constraints \cite{debarre2005complex} (see section IV A for details), hosts the Goldstone modes associated with the symmetry breaking. The transmission spectrum in this low-energy window exhibits discrete sets of uniformly spaced peaks in a small energy window as in Fig. \ref{Fig1}d.  

Indeed, the effects due to topology and symmetry breaking are delicately intertwined. At high energies, the width of and gaps between these discrete bands of transmission corresponds to the same of the emergent Landau levels which reflect the topology of the skyrmion crystal (see Fig. \ref{geff} and section V). Hence magnon transmission allows one to infer the nature of the high energy modes of the Skyrmion crystal (i.e modes just above the Goldstone spectrum) . On increasing energy the  magnon transmission also exhibits  a characteristic angular dependence, with certain preferred angles of transmission and a non-monotonic dependence of transmission on channel number. This dependence reflects the spatial symmetry breaking, i.e.\ it is a consequence of the crystalline order of the Skyrmions. Moreover, the modes in the Riemann-Goldstone Landau level are  associated with the $\rm{SU}(2)$ group manifold
acting on the $\mathbb{C}\rm{P}^1$ local order parameter manifold of the skyrmion crystal. Notably, the uniform spacing within each set of low-energy peaks, indicates the linear dispersion of these Goldstone modes at small momenta. The number of such sets of peaks also directly allows us to infer the presence of three such modes. 
Hence, not only do these results unambiguously indicate the presence of a skyrmion crystal, they also unveil the nature of its excitation spectrum, both at low and at high energies. Figs \ref{Fig1} and \ref{geff} highlight these points clearly.

Section III A presents a simplified heuristic model which accounts for the topological -- but not the symmetry -- aspects of these results.
We prepend this discussion, section III A, to the much more technical analysis by which it was motivated (section IV and Appendix A), where we found that the effective description of the scattering problem bears some resemblance to the problem of a particle scattering off of a region with spatially varying magnetic field: we are led to study a (single) particle scattering  off  a  region with spatially varying magnetic field. 

\begin{figure*}
    \centering
    \includegraphics[width = 18cm,height = 8cm]{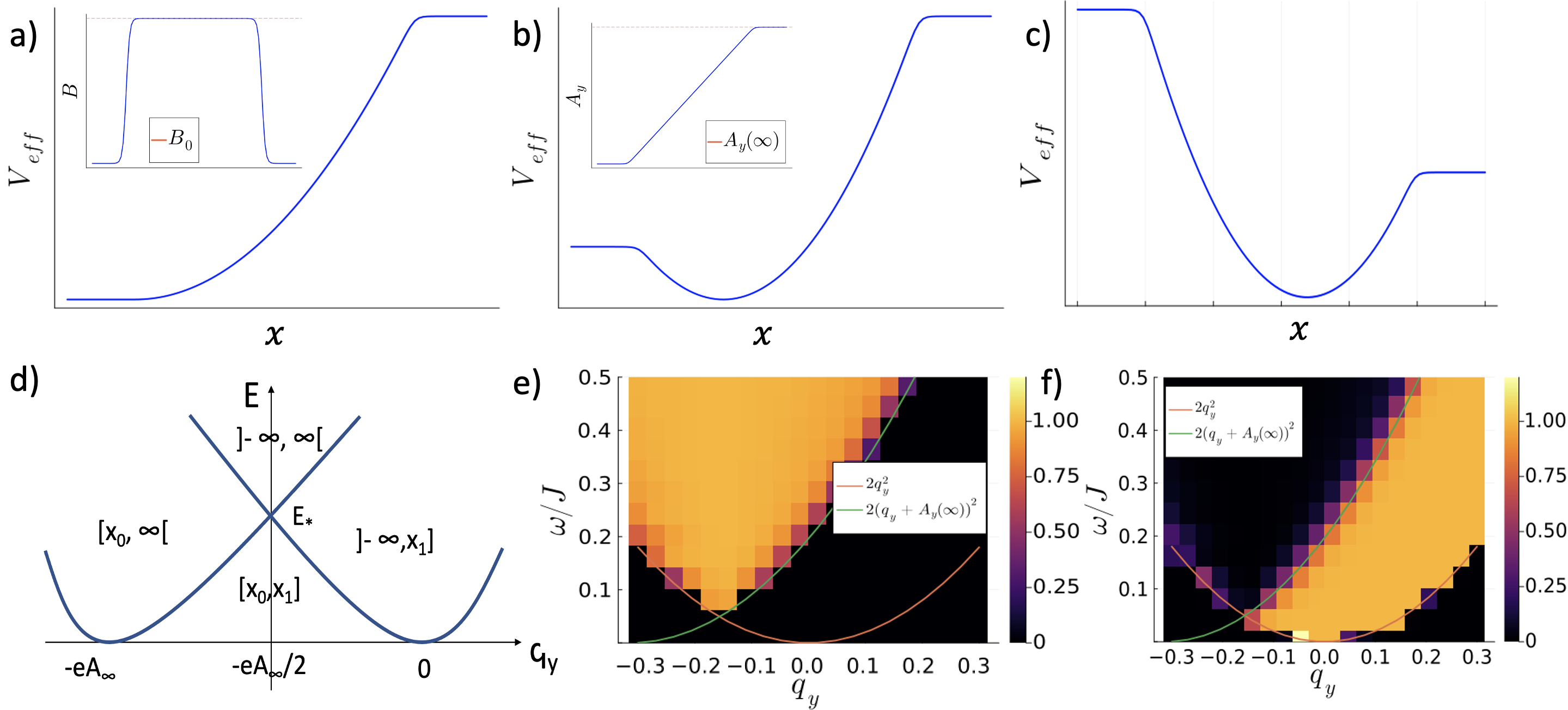}
    \caption{Semi-classical analysis for heuristic model with constant magnetic field in the central region. All arguments and results in this panel are for the Landau gauge ($A_x = 0$). (a-c) Qualitatively different effective potentials for different values of transverse momentum, $q_y \geq 0$ for (a), $-eA_y(\infty)/2< q_y < 0$ for (b) and $-eA_y(\infty)<q_y < -eA_y(\infty)/2$ for (c), the $q_y \leq -eA_y(\infty)$ case is a reflection of (a) . Insets of (a) and (b) show the constant magnetic field profile with smooth decay away from central region and the corresponding vector potential in the Landau gauge respectively. d) Different qualitative regions of scattering in energy-transverse momenta parameter space with the labels in each region indicating the x-support of the corresponding semiclassical trajectories - we get two regions of full reflection, $x \in ]-\infty,x_1]$ or $x \in [x_0, \infty[$, one region with bound trajectories ($x \in [x_0,x_1]$) and one region with full transmission ($x \in ]-\infty, \infty[$) . (g-h) Numerically obtained transmission and reflection coefficients respectively for the quantum problem with Hamiltonian in Eq. \ref{toyH} showing great qualitative agreement with the semiclasical picture in (d). We only consider the case of a particle incident from the left which is why the reflection coefficient in (f) is not fully symmetric as in (d) which considers both left and right incident processes.}
    \label{constBheur}
\end{figure*}

To single out the effect of the variation of the magnetic field, we first consider the problem of a constant magnetic field and map out the transmission and reflection coefficients in energy-momentum space as in Fig. \ref{constBheur}(f-h). We find that such a problem is characterized by a critical energy scale $E_*$ below which there is no transmission, and regions of either full transmission or full reflection with a smooth crossover from one to the other. We also note that there are bound states below $E_*$ which play an important role on introducing spatial modulations of the magnetic field. Also any non-zero transmission is accompanied by an angular deviation which we calculate as a function of incoming energy and present in Fig. \ref{ang}(a). These features can be accounted for in a picture of semi-classical cyclotron orbits.

The physics becomes even richer on introducing modulations along the transverse direction, as is present, in a skyrmion crystal. On doing so, we find that there are sharp transmission peaks in the semi-classically forbidden region, i.e below $E_*$. Moreover, we find that these peaks have a special structure, they occur in certain energy windows, and these windows have uniform gaps between them as in Fig. \ref{heuristicmagymod}(a). Such gaps correspond to the Landau level gaps in the spectrum and the energy windows occur due to the dispersion in the Landau levels introduced by the spatial variation. We also find that the heights of most of the transmission peaks are suppressed. These phenomena are in turn accounted for in terms of resonances from bound states, and interference between more than one propagating mode, respectively.

This brings us to our analysis of the full ferromagnet-skyrmion crystal-ferromagnet junction problem, the solution to which requires several technical advancements. We include four such advances in the main text (less important ones are relegated to appendices) in section IV, which are also applicable to other problems in transport calculations of such junctions between topologically trivial and non-trivial structures. These advancements draw from a wide variety of fields and reflect the richness of this problem.

First, in section IV A, we address the difficult non-linear problem of constructing an interface between such topologically trivial and non-trivial structures as mentioned in the introduction. We use an analytical ansatz constructed from truncated holomorphic theta functions to model our ferromagnetic-skyrmion crystal-ferromagnetic junction. Using the theta functions in eq \ref{thetafunc} as basis functions, we can generate a textured skyrmion crystal with two skyrmions per unit cell in the central region with similarly aligned ferromagnets on the two sides as shown in Fig. \ref{Fig1}(e). On calculating the topological charge density from these truncated theta functions, using eq \ref{topceq}, as in Fig. \ref{Fig1}(f), we see that we get periodic modulations of the topological charge density in the central region and a smooth decay to zero away from it. The region across which we get a smooth decay defines the interface. Similar holomorphic constructions can be used for other topological structures and can also be extended for fractional or entanglement skyrmion crystals \cite{Doucotent}.

Using these theta functions, we reverse engineer an energy functional with short range interactions and with the spin configuration in Fig. \ref{Fig1}(e) as the minima in the continuum limit. 
We calculate the equations of motion for the excitations of this using techniques from linear spin-wave theory in appendix. It is the form of the variation in the energy functional, given in Eq. \ref{efuncfin}, which resembles  the free particle problem discussed in the previous paragraphs.

Any microscopic numerical calculation of the transport properties of such junctions requires real space discretization of the continuum energy functional in Eq. \ref{efuncfin}. Discretizing the exchange term is a standard exercise in finite difference methods, however, discretizing the change in topological charge density is highly non-trivial task. In section IV B we introduce a novel geometrical method to do this discretization in a completely analytical way. Our approach relies on the short-range nature of our interaction which allows us to express the topological charge density in terms of the solid angle subtended by the four geodesics connecting the four spin vectors of a real space plaquette on the Bloch sphere. Our final result in Eq. \ref{varsoltight} expresses the discretized form of the 2nd term in eq \ref{efuncfin} as a tight binding model with upto second nearest neighbor hopping.

To calculate the full magnon transmission matrix, we use a recursive transfer matrix approach explained in section IV C. First, we discretize the energy functional in real space using standard finite difference methods for the exchange term and the topological charge discretization procedure given in IV B. The usual recursive column-wise procedure involves multiplying the transfer matrix at every column and forming a product matrix which relates the left and right ends of the problem. However, such an approach runs into problems in the presence of evanescent contributions which cause a numerical instability in obtaining the final product matrix. We resolve this instability by adapting a method introduced by Pendry for similar problems in optics\cite{Pendryphot}. This method allows us to obtain the full transmission matrix for the magnon and hence gives us access to channel resolved transmission coefficients.

We end this section by focusing on the Goldstone mode sector in section IV D.  In this section we provide a recipe to construct a non-linear sigma model for a long wavelength description for the coupling between the Goldstone modes of the ferromagnet and the skyrmion crystal. To do so, we describe the structure of most general SU(2) invariant coupling terms at the interface, between these regions with different characteristics of their ground state manifold (see Table \ref{tab1}). We find that there are two very general kinds of coupling between the Goldstone modes of the ferromagnet and the Skyrmion crystal as in Eq. \ref{coupH1} and \ref{coupH2}. Such a construction allows one to develop an analytical approach for such scattering problems through coarse grained models. And, our construction lays the foundation for further constructions of such long-wavelength models of coupling between qualitatively different Goldstone modes. 

Section V provides additional numerical results for the full problem of the skyrmion crystal junction, as summarised at the very beginning of this section, and using the technical advancements made in sections IV A-C. We also obtain the spectra of the high energy modes and show how the transmission energy windows correspond exactly to the energies of these emergent Landau levels. Further we comment on how the heuristic model of a particle scattering in a region of spatially varying magnetic field qualitatively captures the behaviour in this energy regime. However, the transmission features at low energies cannot be understood from that heuristic framework, since that framework has no Goldstone modes. We obtain the dispersion of these Goldstone modes, as in Fig. \ref{geff}(b) and show how the transmission varies on varying their dispersion. On increasing dispersion, the discrete low energy peaks in transmission shift in position and their intensity increases as in Fig. \ref{geff}a. The transmission spectra in this regime has qualitative similarities with the heuristic model for the antiferromagnet sandwich which we introduce in section III D and hence one can borrow our intuitive understanding from that analysis. 

As a significant motivation of our work were the experimental advances described in the introduction, in section VI A we present arguments for how our theoretical predictions can be experimentally tested in a $\nu = 1: 1 \pm \delta \nu: 1$ quantum Hall junction on monolayer graphene. There are elements of our model that might not be completely realistic such as absence of anisotropies, delta function interaction potential and a smooth interface. In Section VI A, we comment on how the presence of realistic anistropies might change some low energy signatures by gapping out a subset of the Goldstone modes but some signatures of the remaining gapless modes shall remain. We comment on how to realize short range interaction using metallic gates and finally we comment on situations where the interface is sharper than in our model. We finally close with an outlook in Section VI B.

\section{Heuristic magnon scattering}
In this section we provide details of two heuristic models as mentioned in the earlier sections. In the first three subsections we introduce and study a particle scattering off a region with (i)  a constant and then (ii) a spatially modulated magnetic field. The spatial profile of the magnetic field mimics that of the topological charge density in the skyrmion crystal junction. This heuristic model turns out to be useful since (as shown in the next section and in Appendix A) it turns out to qualitatively describe (primarily the) topological aspects of the skyrmion crystal problem. In the last subsection we introduce a simple model to discuss the coupling between qualitatively different kinds of Goldstone modes, namely  a ferromagnet-antiferromagnet-ferromagnet junction. 
Both these models allow us a simpler and intuitive understanding of complementary parts of the difficult and technically involved problem fleshed out in sections IV and V.

\subsection{Particle scattering off a region of constant magnetic field}
To isolate the effect of spatial modulations in the magnetic field we  first consider a constant magnetic field profile in the central region which exponentially decays to zero across the interface as shown in the inset of Fig. \ref{constBheur}a. Such a system, in the Landau gauge ($A_x = 0$, $B(x) = \partial_x A_y(x)$), has the Hamiltonian
\begin{equation}
    H = \frac{1}{2m}[q_x^2 + (q_y + eA_y(x))^2]
    \label{toyH}
\end{equation}
We consider the following magnetic field profile
$B_c = B_0/2({\tanh}(x - L/2) - {\tanh}(x+L/2))$ as in Fig. \ref{constBheur}(a), where $B_0$ is the value of the magnetic field in the central region of length $L$. Throughout this discussion, we use a gauge in which $A_y(-\infty)=0$ and
$A_y(x)$ is a positive and increasing function of $x$ with a saturation value $A_y(\infty)$ as shown in the inset of Fig. \ref{constBheur}b.

Since we have translational invariance along $y$, we have two degrees of freedom and two conserved quantitities, the total energy $E$ and the transverse momentum  $q_y$. Hence, we have an integrable system. 

Let us understand the semi-classical trajectories for such a system. Since $q_y$ is conserved we get a collection of one dimensional models with an effective potential $V_\mathrm{eff} (x,q_y) = (q_y + eA_y (x))^2/(2m)$.  

As a function of $x$, qualitatively, we have three different types of effective potential depending on $q_y$. i) If $q_y \geq 0$, $V_{\rm{eff}}$ is monotonically increasing with $V_{\rm{min}} = q_y^2/(2m)$ and $V_{\rm{max}} = (q_y + aA_y(\infty))^2/(2m) $ as in Fig. \ref{constBheur}a. For $-eA_y(\infty) <q_y<0$, by contrast,  $q_y + eA_y(\infty)$ changes sign at $x_{*}$ and we get two types of potential curves,  ii) for $-eA_y(\infty)/2 <q_y<0$,  as in Fig. \ref{constBheur}b and  iii) for $-eA_y(\infty) <q_y < -eA_y(\infty)/2$, $V_{\rm{eff}}$ is as in Fig. \ref{constBheur}c. Finally for $q_y \leq -eA_y(\infty$), $V_{\rm{eff}}$ is monotonically decreasing and looks like the reflection of Fig. \ref{constBheur}a. 

For each region, the support in $x$ of the corresponding trajectory depends on the incoming particle energy $E$. If $E < V_{\rm{min}}$, no scattering states exist, if $V_{\rm{min}}<E<V_{\rm{max}}$, the classical trajectories are purely reflected, i.e the radius of the cyclotron orbits is less than the length of the central region. 

Already at this simplistic level, we can see that if the particle is transmitted, i.e the radius of the cyclotron orbit is larger than the length of central region, then particle will exit the central region with a velocity different from its incoming velocity and its direction will be deflected. One can calculate the angle of deflection easily: say the incoming velocity is $\bm{q}$, the outgoing velocity on the right end will be $\bm{q} + \bm{A}_{\infty}$. In the Landau gauge, $A_x = 0$, therefore, the outgoing velocity will be $\bm{q} + A_y(\infty) \hat{\bm{y}}$. Hence, the angle of deviation is given by $\cos^{-1} [\bm{q}\cdot(\bm{q} + A_y(\infty) \hat{\bm{y}})/(|\bm{q}||\bm{q} + A_y(\infty) \hat{\bm{y}}|)]$. This effect resembles that of the magnon Hall effect, studied in magnon scattering off of single skyrmions in metallic magnets \cite{Iwasakimag,Schuttemag}. \\
\begin{figure*}[t]
    \centering
    \includegraphics[scale = 0.55]{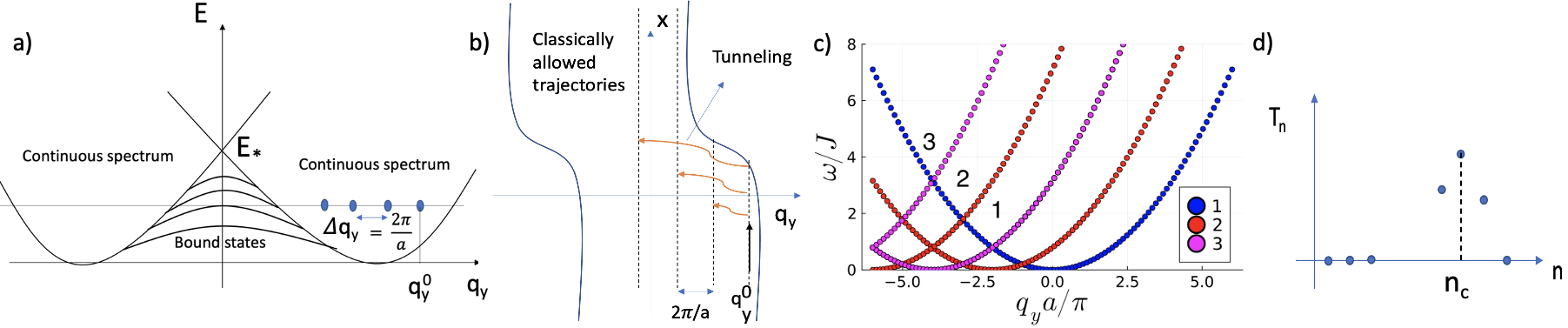}
    \caption{Qualitative arguments for heuristic model with spatially modulated magnetic field along y-axis. a) Continuous spectrum for scattering and discrete spectrum for bound states. Modulation along transverse directions breaks $q_y$ conservation and periodic modulation implies $q_y$ is conserved modulo $2\pi/a$. Transmission below critical energy $E_{*}$ is possible if the energy of one of the channels coincides with the bound state. b) Pictorial description of the possibility of tunneling into other propagating channels due to crystalline order. Each channel has its effective potential profile and regions of allowed transmission as in Fig. \ref{constBheur}(c-f). The presence of multiple propagating modes allows transmission for an incoming magnon due to off-diagonal scattering. c) Number of propagating modes in the $\omega-q_y$ plane in the unfolded zone scheme (for visual reasons). One can transfer this to the first Brillouin zone by standard folding techniques. Each color is for the two curves $2q_{yi}^2$ and $2(q_{yi} + A(\infty))^2$, such that in the region lying above both curves one gets an outgoing propagating mode for the $i$th channel, where $q_{yi} = q_y^{(0)} + 2\pi (i-1)/a$. For this figure we use $B_0 = 2\pi/a^2$ and $L = 20$. d) Pictorial description of non-monotonicity of channel resolved transmission from qualitative arguments presented in section III B and Fig. (b) in this panel (not real data, see Fig \ref{ang}b).}
    \label{heuristicmagymod}
\end{figure*}
The most interesting region is the low energy regime $0<E<E_{*}$, where $E_{*} = (eA_y(\infty)/2)^2/(2m)$. In this case, the semi-classical solutions (and also the corresponding eigenstates for the quantum version) depend on $q_y$, but there is no extended state going from $x = -\infty$ to $x = \infty$: the transmission coefficient across the central region exactly vanishes for $E<E_{*}$. However, in a window in the $E-q_y$ parameter space, there exist classically bound trajectories (closed cyclotron orbits in the central region) for $E<E_{*}$. In the quantum problem, the bound states of the classical picture correspond to Landau levels, which will play an important role once the magnetic field in the central region is modulated, as is the case in the actual skyrmion crystal junction. \\
To confirm this above picture, we calculate the reflection and transmission probabilities starting from our Hamiltonian in eq \ref{toyH}. We see that our numerical results in Fig. \ref{constBheur}(e-f) agree very well with the semi-classical analysis in presented above and summarized in Fig. \ref{constBheur}d. There exists a minimum energy $E_{*}$ below which there is no transmission, the threshold energy for transmission depends on the transverse momentum and as expected for the quantum problem there is a smooth evolution from full reflection to full transmission on increasing energy at fixed $q_y$. Also, there is an angular deviation in the region of full transmission as in Fig. \ref{ang}a.
\subsection{Effect of periodic modulation along $y$-axis}
Our analysis of semi-classical trajectories showed that a constant magnetic field in the central region implies that there is no transmission below a certain threshold $E_*$, yielding distinct regions of full transmission and full reflection in $E-q_y$ parameter space. Quantum mechanically, regions where the $x$-support of classical trajectories is infinite have a continuous spectrum whereas the bound state region, which has only a finite support $[x_0,x_1]$, has a discrete Landau level spectrum. The bound states have slightly bent dispersions because the local potential wells around $x_*$ become very shallow as $x_* \rightarrow \pm \infty$, see Fig. \ref{heuristicmagymod}a. In this  and the following subsection we address the effects of periodic modulation of $B$ about its mean $B_0$ in the central region. Again, we examine the heuristic model given by Eq. (\ref{toyH}), but now, first with a periodic modulation of $B$ along the transverse ($y$) direction.

A periodic modulation of period $a$ in the $y$-direction breaks $q_y$ conservation and hence generates matrix elements between states with $q_y$ values differing by integer multiples of $2\pi/a$. This mechanism generates a tunneling amplitude to an order $N_p$ in perturbation theory given approximately by $2 \pi/N_p \approx e A_y(\infty) = B_0 L/\phi_0$, so $N_p \approx B_0aL/(2\pi \phi_0)$, where $B_0$ is the average magnetic field in the central region and $\phi_0$ is the flux quantum. If $E$ lies in the gap of the bound state spectrum, since $N_p \sim L$, the corresponding transmission amplitude will be exponentially small in $L$. However, importantly there will be some resonances for ($q_y^{(0)},E$) values such that E coincides with a bound state (i.e., a state of the  Landau level) with energy at $q_y^{(n)} = q_y^{(0)} + 2n\pi/a$, $n \in \mathbb{Z}$. Such resonances  permit transmission at energies below the threshold $E_*$.

Once $q_y$ conservation is broken, the scattering problem becomes a multi-channel problem, where the number of channels depends on the discretization procedure. The presence of multiple channels makes the problem very rich and we devise a  transfer matrix procedure which calculates the full transmission matrix, which allows us to obtain the channel resolved transmission. Say, we consider an $N$ channel problem based on the discretization of the $a \times a$ unit cell into $a/N \times a/N$ grids. Out of the $N$ possible values of $q_y$, some values will represent propagating channels, Im($q_x = 0$), whereas, for relevant energy scales and $N$ values, most channels will be evanescent, Im($q_x \neq 0$). 

\begin{figure*}[t]
    \centering
    \includegraphics[width = 18cm, height = 4cm]{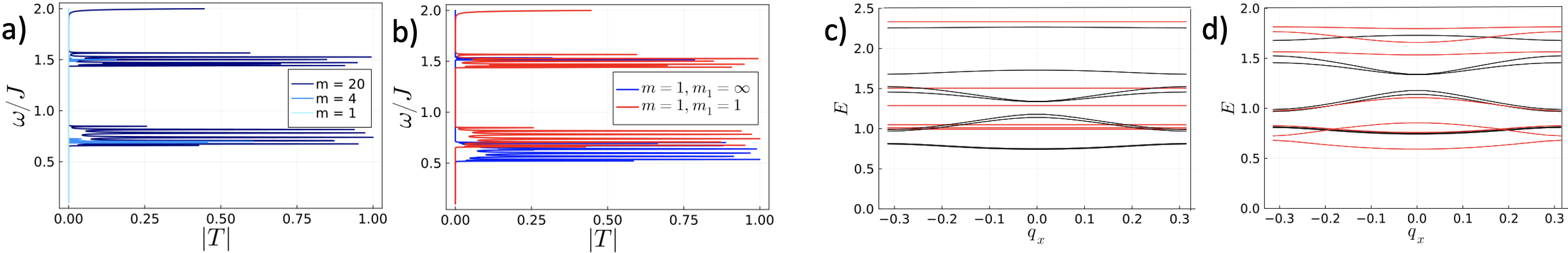}
    \caption{Results for heuristic model with spatially varying magnetic field. a) Transmission spectra for a particle scattering off a region with periodic modulation, see main text, of magnetic field along $y$-axis, where $m$ is the parameter that controls the amplitude of modulation, $B = B_c (\sin(4\pi y/a)/m+1)$, and $B_c$ is the magnetic field used in the last subsection. Small $m$ implies large modulation and vice-versa. The case with $m = 1$ resembles the same amplitude of modulation along $y$-axis as in the topological charge density profile in Fig. \ref{Fig1}f. b) Transmission spectra for the case of a magnetic field modulated along $x$ and $y$-axes, where $m_1$ has the same properties as $m$ described, but for the $x$-axis in (a). The case with $m = 1, m_1 = 1$ resembles the same modulation along both axes as in the topological charge density profile in Fig. \ref{Fig1}f.  c) Effect of modulation along y on Landau levels - adds bandwidth. Red lines are for the spectrum with negligible modulation and black lines are for large modulations along $y$-axis. d) Effect of modulation along x on Landau levels - reduces gap. Red lines are for the spectrum with negligible modulations along $x$ and black lines are for the case with large modulations along $x$ (both cases have the same large modulation along $y$).  The spectra in (c-d) are shown as an illustrative example, for a quantitative comparison of energy and transmission spectra in the actual (not heuristic) model, see Fig. (\ref{geff}).}
    \label{specyv}
\end{figure*}

To extend our heuristic picture to this multi-channel problem, one can examine the effective potentials, as in the constant $B$ case, for each of the $q_y$ channels. For energies $E_* < E < V_{\rm{max}}(q_y^{(0)})$, in the constant magnetic field case there is no transmission and hence full reflection, $|T| = 0, |R| = 1$. However, for the modulated case, there exist channels such that $E_* < V_{\rm{max}}(q_y^{(n)}) < E$, hence there will be transmission through these channels. Moreover, there will be a non-monotonic dependence of the transmission amplitudes on the channel number, with maximal transmission for $n = n_c$, as depicted pictorially in Fig. \ref{heuristicmagymod}d. Further, the channel number of the maximally transmitted channel will increase on increasing energy.

Following from this qualitative picture based on the multi-channel scattering analysis, we proceed to implement the above problem numerically using our transfer matrix approach described in section IV C. In the Landau gauge the right end of the junction had a finite non-zero vector potential $A_y(\infty)$. However, in the actual experiment there is no such vector potential in the ferromagnetic end, hence to enable direct comparison we also implement a gauge fixing procedure using a string of Aharanov-Bohm fluxes, aided by our problem being discretized on a lattice, to ensure that the vector potential vanishes on the right end (see appendix D for details).
% \rim{not sure this following paragraph fits into the flow here}
% To get a sense of quantitative similarity with the full skyrmion crystal calculation presented later, we note that for two skyrmions in a unit cell, we get four flux quanta acting on a spin-1 magnon, hence the analogous average magnetic field would be $B_0 = 8\pi/a^2$ (see appendix A and \cite{douccot2018zero}). 
% For the skyrmion crystal junction, experimentally we have two quantum Hall ferromagnets on either side, each of which has no topological charge, hence, on transferring to our heuristic setting, no magnetic field or vector potential. However, in the Landau gauge as seen in the previous problem, we see that the presence of a magnetic field induces a finite vector potential on the right. To make a direct comparison with the skyrmion crystal problem, we make a gauge transformation aided by our problem being discretized on a lattice. We use a string of Aharonov-Bohm fluxes along the transverse direction at the mid-point of the skyrmion lattice to fix the vector potential to zero on both sides of the skyrmion lattice. Details of this procedure are given in appendix C. 

We choose a magnetic field profile with sinusoidal modulations along the $y$-axis with period $a/2$, $B = B_c (\sin(4\pi y/a)/m+1)$, where $B_c$ is the magnetic field used in the last subsection.
$m$ controls the amplitude of transverse modulation, with large $m$ implying small modulations and vice-versa. 
From Fig. \ref{specyv}a we see that for small or negligible variations along the $y$-axis, there is no transmission for the plotted energy range, since for these energies and for this value of $B_0$, the incoming energy is lower than the critical energy required for transmission. For $B_0 = 8\pi/a^2$, $a = 10$ and $L = 4a$, we get  $E_* = (16\pi/a)^2/2 \gg 2$, in units of $e = m = J = 1$ (see section V A for reasoning for such values of parameters).  However, for large modulations, and more importantly for modulations which mimic the topological charge modulations (in the $y$ direction) of the $SU(2)$ skyrmion crystal (see Fig. \ref{Fig1}(b)), we see a dramatic change in behaviour, characterized by the appearance of resonant peaks of finite transmission. Moreover, these peaks appear in discrete regions of energy centered around energies corresponding to the different Landau levels of the constant magnetic field problem. This confirms the qualitative picture we developed in the last section, in which finite transmission below the threshold energy takes place when the incoming particle energy coincides with the bound state energy for certain channels. The transmission windows also allow one to infer the width of such effective Landau level bound states.
\begin{figure}[t]
    \centering
    \includegraphics[scale = 0.4]{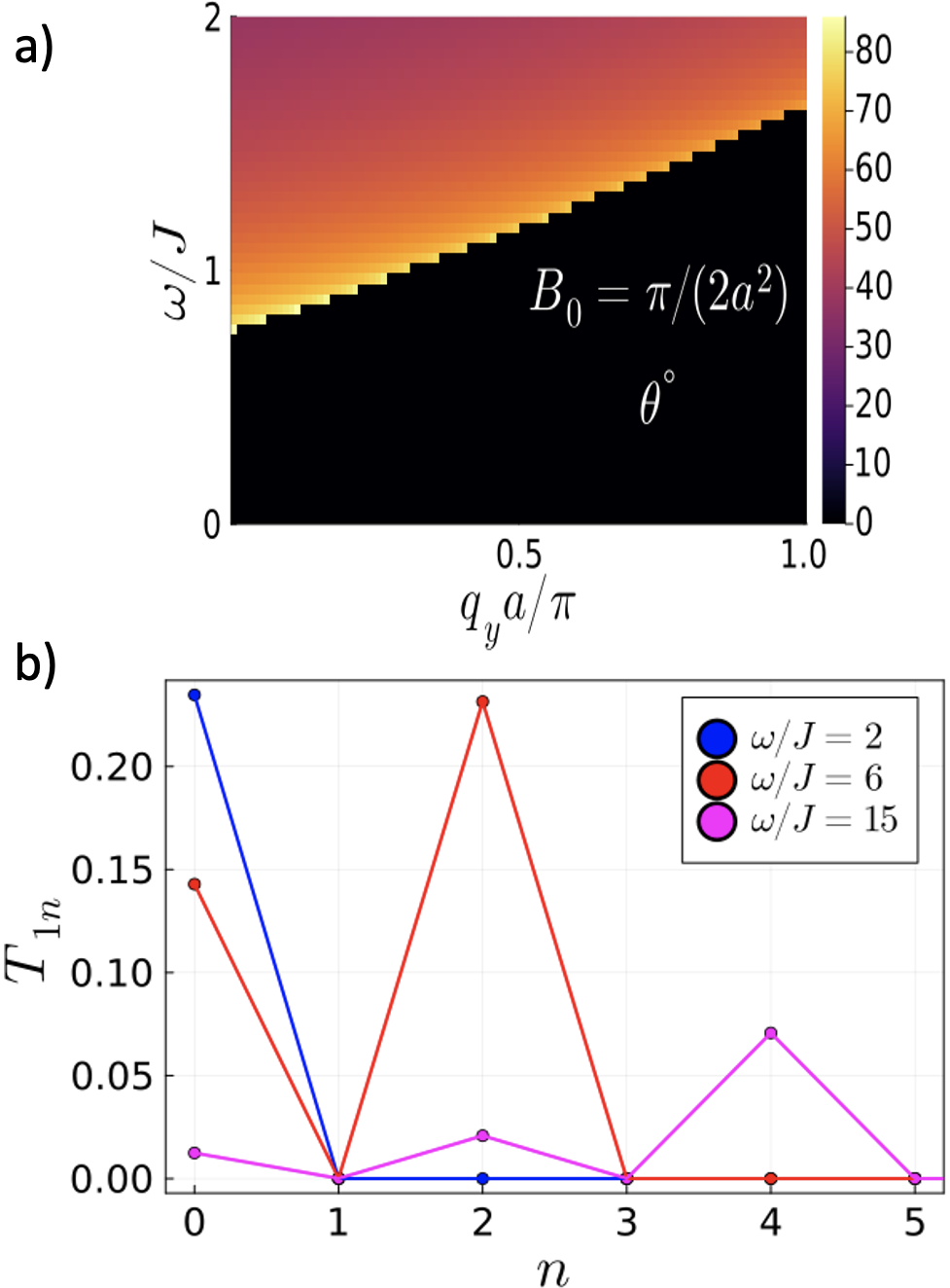}
    \caption{Channel dependence and angular deviation of transmission for the heuristic model in sections III A-C. a) Angular deviation of particle scattering off a constant magnetic field region - similar to magnon scattering off a single skyrmion. b) Channel dependence of transmission coefficient for spatially modulated magnetic field ($B_0 = 8\pi/a^2$, $L = 20$, $m = 1$, $m_1 = 1$, $q_y^{(0)} = 0$). Certain channels dominate in certain energy regions, and there is a non-monotonic dependence of transmission coefficient on outgoing channel number $n$, where $q_y^{(n)} = q_y^{(0)} + 2(n-1)\pi/a$. For low energies, below $E_*$, only one propagating channel is present $q_y^{(0)}$. One increasing energy the number of propagating channels increases as shown in Fig. \ref{heuristicmagymod}c, and the particle can scatter into these channels (the transmission  matrix has non-zero off diagonal elements). This agrees with the qualitative picture of tunneling into other propagating channels, as presented in section III and Fig. \ref{heuristicmagymod}(b)}.
    \label{ang}
\end{figure}

Besides the resonant transmission features below the critical energy, we also verify the non-monotonic channel dependence of transmission for energies above the critical energy by plotting the channel-resolved transmission coefficients in Fig. \ref{ang}(b). Similar non-monotonic transmission and angular dependence will be observed in skyrmion crystal junctions. In Fig. \ref{ang}(b), we see that in different energy windows different channels dominate transmission, and there is a non-monotonic channel number dependence. Moreover, from our intuitive picture of off diagonal scattering one can predict exactly which channel  dominates transmission in the different energy windows. However, the details for which channel dominates depend on the length of the skyrmion crystal region and we leave that analysis for future more experimentally specific work.

One surprising result that is not captured by our earlier  qualitative analysis is the varying height of the transmission peaks corresponding to the different Landau levels. Some peaks in the energy range of the higher Landau levels appear to be suppressed. Such a suppression in peak height can be understood as a consequence of interference between multiple propagating channels in the central region. Appendix E presents a detailed technical discussion of the effect of multi-mode interference on the peak heights.

\subsection{Effect of periodic modulation along $x$-axis}
From the above two subsections, we see that a periodic modulation of the magnetic field along $y$-axis induces resonant peaks of finite transmission at energies corresponding to low-lying Landau level energies of the constant magnetic field problem. Hence, due to the gap between Landau levels, we also see a gap between regions of finite transmission as in Fig. \ref{specyv}(a). We now complete the analogy of our heuristic model with the skyrmion crystal by introducing its final ingredient, the modulation of the magnetic field along the $x$-axis on top of the modulations along $y$. We use a similar sinusoidal variation dependent on parameter $m_1$ (large $m_1$ corresponds to small variation and vice-versa).

Variations of the magnetic field along the $x$-axis  broaden the Landau levels and hence the gap between the regions of finite transmission  decreases. This is shown in Fig. \ref{specyv}(b) and (d), obtained from our numerics, where we see that for a modulation amplitude that mimics the topological charge density modulation along the $x$-axis of the skyrmion crystal (i.e for $m_1 = 1$), the gap is reduced. Hence a modulation of the magnetic field along the $x$-axis increases the energy range of finite transmission due to Landau-level broadening.

\subsection{Ferromagnet-antiferromagnet-ferromagnet junction - effect of dispersion mismatch}
Our heuristic model in the previous three subsections did not involve the physics of Goldstone modes arising from symmetry-breaking of the skyrmion crystal. To highlight the issues involved in the transmission properties of a magnon  through structures with not only different dispersion relations but also a different number of collective modes, we consider a very simple model of an antiferromagnet sandwiched between two ferromagnets. The dispersion relation of a ferromagnetic magnon is $\sim J_F k^2$ whereas that of an antifferomagnetic magnon is $\sim J_{AF} k$. Moreover, the antiferromagnet has two branches of Goldstone modes as opposed to the single one in the ferromagnet. 

\begin{figure}[t]
    \centering
    \includegraphics[width = 8cm, height = 12cm]{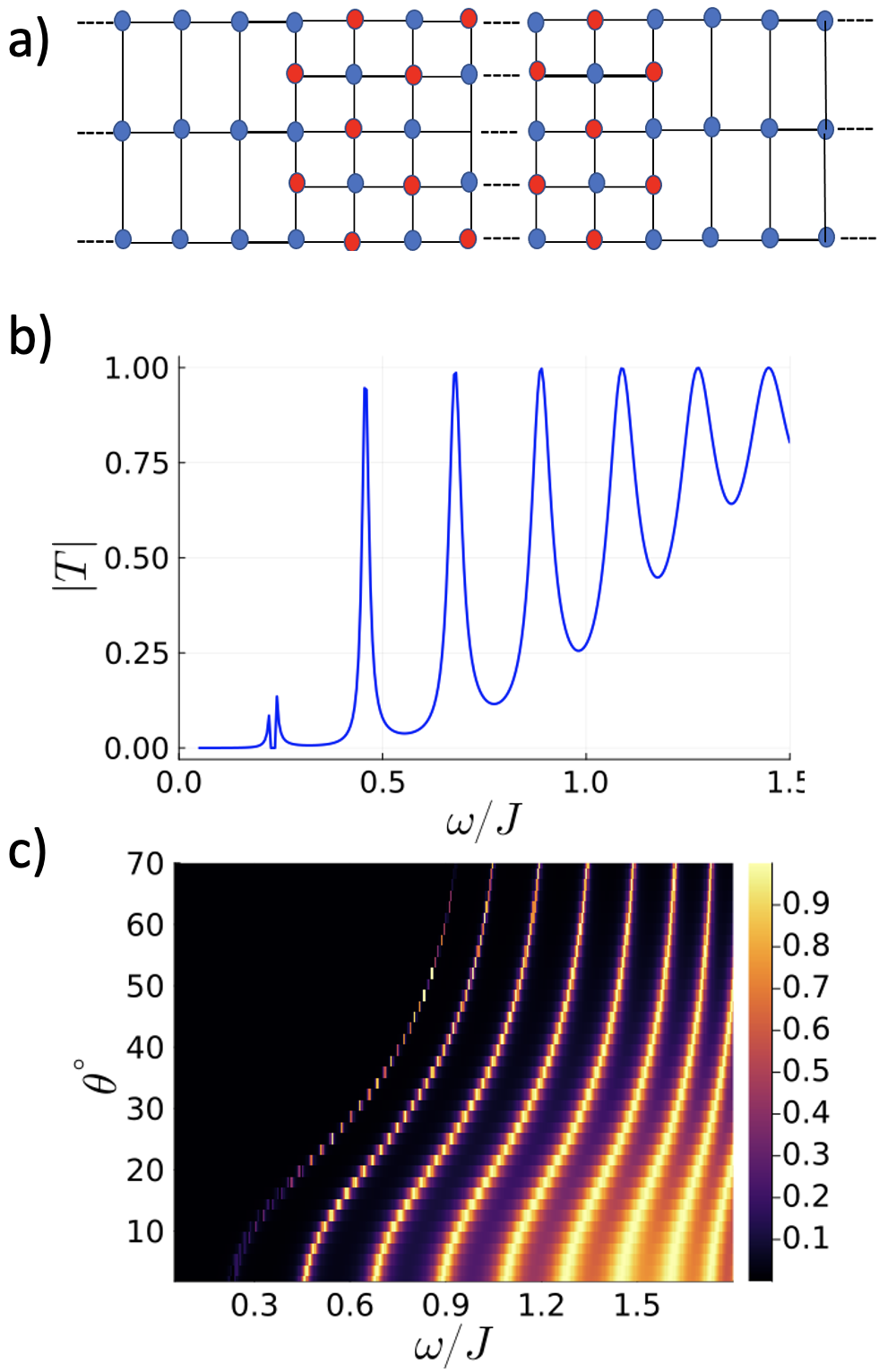}
    \caption{Results for the ferromagnet-antiferromagent-ferromagnet quantum Hall junction heuristic model- parameters used $J_F = 1,J_{AF} = 1, L_I = 18a_x$, where $a_x$ is the lattice spacing along $x$. a) Lattice structure of the heuristic model with the $y$-spacing for the ferromagnet twice that of the $x$-spacing. Such a simplification is made to simplify the sublattice matching across the interface. b) Fabry-Perot resonance peaks at normal incidence ($q_y = 0, \theta = 0$) in the total transmission. c) Transmission as a function of angle of incidence and energy of incident magnon. At fixed angle of incidence, there is a critical energy for transmission following which there is a set of equally spaced peaks in the low-energy regime, reflecting the linear nature of the antiferromagnetic Goldstone modes. The cutoff energy at $\theta = 0$, is a finite size effect, as we increase the length of the middle region, this value will go closer to zero.}
    \label{Affig}
\end{figure}

We construct a very simple sandwich structure which makes our calculations entirely analytically tractable. We consider an antiferromagnet with half the lattice spacing in the $y$ direction of the ferromagnet, so that only the $A$ sublattice sites in the antiferromagnet are connected to the ferromagnet (as shown in Fig. \ref{Affig}a). We then solve the scattering problem for a magnon injected from the ferromagnetic region on the left with the following form of the complex wavefunction in the ferromagnetic regions, $x \geq L_I$, which describes the spin deviation perpendicular to the equilibrium magnetization
\begin{multline}
    \delta n_F(x,y) =
\left\{
	\begin{array}{ll}
		 A e^{ik_x x + ik_yy} + B e^{-ik_x x + ik_yy}; x \leq -L \\
		G e^{ik_x x + ik_yy} \\
	\end{array}
\right.
\end{multline}
whereas in the antiferromagnet, $|x| \leq L_I$, due to the presence of two modes, the same can be written as
\begin{multline}
   \delta n_{AF}(x,y) = C e^{ik_{1x} x + ik_yy} + D e^{-ik_{1x} x + ik_yy}+\\
		E e^{ik_{2x} x + ik_yy} + F e^{-ik_{2x} x + ik_yy} \\
\end{multline}
where $x = -L_I$ and $x= L_I$ are the positions of the interfaces, $k_x$ and $k_y$ are the incoming parallel and transverse momenta of the magnon, $k_{(1/2)x}$ are the parallel momenta of the two modes in the antiferromagnetic region and the capital letters denote the amplitudes of the various left and right-moving waves. The transverse momentum $k_y$ is conserved and is hence a good quantum number for the scattering process. We then match the wavefunction across the two interfaces as in standard scattering problems to get the transmission and reflection amplitudes. The total transmission coefficient for the outgoing magnon in the right ferromagnetic lead is given by $|T| = |G|^2/|A|^2$ and the total reflection coefficient for the reflected magnon in the left ferromagnetic lead is given by $|R| = |B|^2/|A|^2$. 

This very simple model already exhibits various qualitative features which carry over to the case of the skyrmion crystal we are interested in. First, in much of the parameter space in Fig. \ref{Affig}c,  transmission is suppressed. Second, at a fixed angle of incidence of the incoming magnon, there is a cutoff energy due to the dispersion mismatch below which the entire wave is reflected for all angles. For low energies, this cutoff energy can be rephrased as a cutoff angle above which one gets no transmission.

Beyond the cutoff energy we get a series of peaks in the transmission amplitude  which broaden on increasing energy. These peaks are essentially Fabry-Perot interference peaks due to multiple reflections within the sandwiched structure. One can also verify that the width of these peaks depends on the length of the sandwiched structure, as expected for Fabry-Perot peaks. Also, at a fixed transverse momentum, or incident angle, the low energy peaks are equidistant. This reflects the characteristic linear dispersion of the Goldstone modes in the antiferromagnet. As we increase energy, the equidistant nature disappears as the dispersion relation ceases to be linear. 

In closing, we note that despite its simplicity, this model reproduces similar qualitative features (the interference pattern and critical angle curve) as the $\nu = 1:0:1$ quantum Hall junction, where the sandwiched structure hosts a canted antiferromagnetic ground state, which was studied analytically \cite{atteia2022beating} and numerically \cite{Weimagn} using Hartree-Fock methods.

\section{Setup and solution of the scattering problem for the skyrmion crystal junction}
In this section, we present in turn the central technical aspects of our solution of the full scattering problem of magnons off the symmetry-breaking skyrmion crystal. 
\subsection{Basis for smoothly decaying topological charge - truncated theta functions}
A quantum Hall junction with an SU(2) skyrmion crystal sandwiched between two ferromagnets 
appears as the result of an externally imposed spatial variation of the electrostatic potential seen by electrons in the 2D layer. Neglecting all anisotropic couplings in spin space, the total energy of the quantum Hall ferromagnet is given by the following functional \cite{sondhiskyr, MoonPRB}
\begin{multline}
    E({\textbf{n}}) = J \int [(\partial_x \textbf{n})^2 + (\partial_y \textbf{n})^2]d^2\textbf{r} + \\
    \int (Q(\textbf{r}) - Q_0 (\textbf{r})) V(\textbf{r}-\textbf{r'}) (Q(\textbf{r'}) - Q_0 (\textbf{r'}))
    d^2\textbf{r}\:d^2\textbf{r'}
    \label{efuncinitgen}
\end{multline}
where the unit vector $n(\textbf{r})$ denotes the local spin orientation, $V(\textbf{r}-\textbf{r'})$ is the two-body (possibly screened) Coulomb potential, $Q(\textbf{r})$ is the local topological charge density (which is proportional to the local charge density), and $J$ is a local exchange energy also due to Coulomb interactions. In the case of un-screened Coulomb interactions,
$J=e^2/(32 \sqrt{2\pi} \epsilon l_B)$ in Gaussian units, $l_B$ being the magnetic length and $\epsilon$ is the dielectric constant. The presence of the imposed external potential is taken into account through the background charge $Q_0 (\textbf{r})$, which we assume to be significant in an infinite (along $x$ axis) slab of finite width parallel to the $y$ axis. The topological
charge density is given by:
\begin{equation}
 Q(\textbf{r}) = \frac{1}{4\pi} \textbf{n} \cdot (\partial_x \textbf{n} \times \partial_y \textbf{n}) \ .
 \label{topceq}
\end{equation}
Minimizing the above energy functional in the presence of the prescribed background charge $Q_0$ is a  difficult and highly non-linear problem. Casting this in an analytical form is yet more challenging. Furthermore, in a given experimental setting, determining precisely the actual $Q_0 (\textbf{r})$ is also not at all straightforward. 

For these reasons, and because our goal is to investigate magnon dynamics, we   start by constructing a plausible Ansatz for the spin configuration $n(\textbf{r})$ which interpolates between a region of finite and spatially modulating topological charge for the skyrmion crystal in the middle to a zero charge region to the two ferromagnetic ends. 

Skyrmion crystals with periodic boundary conditions were previously studied using a basis of theta functions, which are used to construct holomorphic spinors with values in the complex projective space $\mathbb{C}\mathrm{P}^{d-1}$ \cite{Dimaskyrmion}. Such theta functions were first introduced by Haldane and Rezayi in the quantum Hall setting for constructing Laughlin-Jastrow wavefunctions under periodic boundary conditions \cite{Haldanetheta}. 

In the present work, we  focus on SU$(2)$ spins described by a two-component spinor field $|\psi(\textbf{r})\rangle$. The relation between this local spinor
and the spin orientation vector $n(\textbf{r})$ is given by
\begin{equation}
    \textbf{n}(\textbf{r}) = \dfrac{\langle \psi(\textbf{r})| \boldsymbol{\sigma}| \psi(\textbf{r})\rangle}{\langle \psi(\textbf{r})|\psi(\textbf{r})\rangle}
    \label{nfrompsi}
\end{equation}
where $\sigma^{x}$,$\sigma^y$ and $\sigma^z$ are the Pauli matrices.
Because multiplying the local spinor by an arbitrary phase
factor does not change the physically observable spin orientation, 
$|\psi(\textbf{r})\rangle$ can be considered as an element of the complex projective space
$\mathbb{C}P^{1}$, which is the same manifold as the $S^2$ sphere, which is the $d=2$ case, although the present construction easily
generalizes to arbitrary integer values of $d$.

In our model for the skyrmion crystal junction, we have periodic boundary conditions in the $y$-direction and open boundary conditions in the $x$-direction. To model the finite-$x$ support of the crystal we sharply truncate the theta functions whose sum, instead of taken to infinity as is done for periodic skyrmion crystals, is taken to some integer $N'$. For $d$ skyrmions in a $b \times a$ unit cell, the relevant truncated theta functions are given by 
\begin{equation}
    \theta_p^{(N')}(z) = \sum_{|n| \leq N'} e^{-\pi bd\frac{d}{a}(n+p/d)^2 + 2 \pi \frac{d}{a}(n + p/d)z} 
    \label{thetafunc}
\end{equation}
where $z = x + iy$.
The zeros of the theta function indicate the position of skyrmion cores, and $p$ runs from $0$ to $d-1$ in agreement with the Riemann-Roch theorem \cite{debarre2005complex}. Usual $\theta$ functions (corresponding to $N'$ infinite)
are characterized by the following relations:
\begin{align}
    \theta_p(z+ia) &= \theta_p(z)\\
    \theta_p(z+b) &= e^{(\pi b + 2\pi)\frac{d}{a}z}\theta_p
\end{align}
The different $\theta_p$ functions are related by the following translation operators
\begin{equation}
    \theta_{p+1}(z) = e^{-\frac{\pi b}{d a} + \frac{2 \pi}{a}z} \theta_p(z-b/d)
\end{equation}
At finite $N'$, translational symmetry along $y$ is preserved, but not along the $x$ axis.
Using these truncated $\theta$ functions we construct the holomorphic spinor
defined by $|\psi(\textbf{r})\rangle_0 = (\theta_0^{(N')}(z),\theta_1^{(N')}(z))^T$. From Eq. (\ref{nfrompsi}), this defines the reference spin configuration $\textbf{n}_{0}(\textbf{r})$.\\
The minimal spin configuration and corresponding topological charge density profile for the ferromagnet-skyrmion crystal-ferromagnet junction generated by these theta functions, and using Eq. (\ref{topceq}), is shown in Fig. \ref{Fig1}e and f. We see that the topological charge density is non-zero and spatially modulated along both $x$ and $y$ axes, with period $a/2$ in the central region outside which it decays smoothly to zero. A sharp cutoff in the theta functions thus leads to a smooth decay of the topological charge density. This allows us to define the notion of an interface for the junction as the region across which the topological charge density goes to zero. Furthermore, one can tune the length of the  skyrmion crystal formed the by these truncated theta functions by varying the cutoff $N'$.

In order to investigate magnon dynamics, we need to specify the energy functional,
which is minimized by the reference spin configuration $\textbf{n}_{0}(\textbf{r})$. 
Because holomorphic spinors always generate local minima for the local exchange term,
it is sufficient to set the background charge $Q_0 (\textbf{r})$ in Eq. (\ref{efuncinitgen})
equal to the topological charge density of the reference configuration $\textbf{n}_{0}(\textbf{r})$.
In our calculations, we have replaced the non-local Coulomb interaction in the second term of Eq. (\ref{efuncinitgen})
by a local `Coulomb interaction' (delta function in real space) to simplify the calculations and make our problem partly analytically tractable. Therefore, the corresponding functional becomes
\begin{equation}
    E({\textbf{n}}) = J \int [(\partial_x \textbf{n})^2 + (\partial_y \textbf{n})^2]d^2\textbf{r} + g \int (Q(\textbf{r}) - Q_0 (\textbf{r}))^2 d^2\textbf{r}
    \label{efuncfin}
\end{equation}
 Moreover, such a delta function interaction term can be realized in quantum Hall junction experiments in graphene with metallic gates (see discussion in section VI). 
 
 Without any interaction term, i.e $g = 0$, all holomorphic functions give the same exchange energy. This renders the magnons non-dispersive and localized, which motivated us to call the collection of the corresponding set of states the {\em Riemann-Goldstone Landau level}. This also shows that we need $g>0$ to get a finite dispersion of these Goldstone modes.

 Equation \ref{efuncfin} is the starting point from which we derive equations of motion using  spin-wave  techniques. There are various subtleties in this procedure such as introducing a set of local orthonormal frames, accounting for the holomorphic constraint or ensuring gauge invariance. These are  addressed in detail in appendices A-C. 
 
 The resulting linearized Landau-Lifshitz equations of motion for such a system can be expressed as a time-dependent Schrodinger equation (see Eq. (\ref{schroeq}) in appendix A), which forms the basis of our transfer matrix analysis. The second order variation of the energy functional, obtained from the spin-wave theory analysis,  resembles the form of a particle in a vector potential generated by a magnetic field $Q_0$ discussed above (see Eq. (\ref{finalE}) in appendix A). This justifies the use of the first heuristic model in section III.
 \subsection{Real space discretization of topological charge from geodesics}
\begin{figure}[t]
    \centering
    \includegraphics[scale = 0.25]{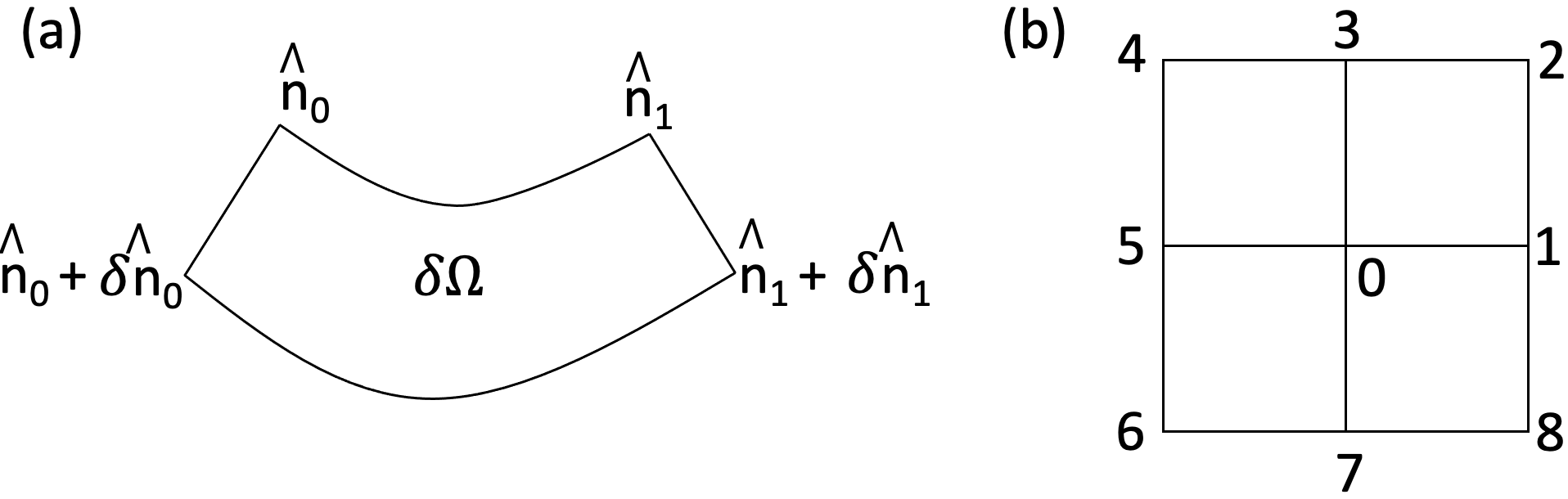}
    \caption{a) Change in solid angle between two geodesics on the sphere. b) Plaquette-wise discretization of topological charge. Tight-binding model includes contributions of nearest and second nearest neighbours (shown here for site 0)}.
    \label{topchargediag}
\end{figure}

To calculate magnon transmission coefficients through the skyrmion crystal we need to discretize the energy functional in eq \ref{efuncfin} on a finite grid, and hence in turn we need to discretize the exchange and the topological charge density terms. Discretizing the exchange term is a standard exercise in finite difference methods (see appendix F), however, discretizing the variation in the topological charge density is a highly non-trivial task.
Here we present an geometrical approach based on solid angles between geodesics. Such a discretization procedure should carry over for similar settings in metallic magnets hosting skyrmion crystals and could be transferred to other topological spin textures. \\
We associate a topological charge to each plaquette with the topological charge density being equal to to the solid angle subtended by the four spin vectors associated with the vertices of the plaquette. Before calculating the variation of such a solid angle, let us first consider the much simpler problem of the variation of the solid angle subtended by two spin vectors on the sphere with spherical coordinates $(\theta,\phi)$ and $ds^2 = (d\theta)^2 + \sin^2 \theta \,(d\phi)^2$. The path between the end points of the two spin vectors $\hat{n}_1$ and $\hat{n}_2$ on the sphere describes a geodesic and the fluctuations in these spin vectors due to the spin waves describe a new geodesic, hence the problem reduces to finding the variation in solid angle between these two geodesics, as shown in Fig. \ref{topchargediag}(a). The standard equations of motion for geodesics are
\begin{align}
\begin{split}
    \ddot{\theta} &=  \frac{1}{2}\sin(2\theta)\,\dot{\phi}^2 \\
    \sin^2 \theta \,\dot{\phi} &= \rm{const.}
    \label{geoeq}
\end{split}
\end{align}
Without loss of generality (due to rotational invariance) we can choose $\bm{n}_0$ and $\bm{n}_1$ along the equator, so the corresponding geodesic becomes
\begin{equation}
    \theta(t) = \pi/2 ; \; \phi(t) = \phi(0) + (\phi_1 - \phi_0)t
\end{equation}
Considering the first order variations in eq \ref{geoeq} we get 
\begin{align}
\begin{split}
    &\delta \ddot{\theta} = \cos(2 \theta)\dot{\phi}^2 \delta \theta + \sin(2\theta)\dot{\phi} \,\delta\dot{\phi} \\
    &\sin(2 \theta) \dot{\phi} \,\delta \theta + \sin^2\theta \,\delta\dot{\phi} = \rm{const}
\end{split}
\end{align}
Focusing on the vicinity of geodesic we get 
\begin{equation}
    \begin{split}
        \delta \theta &= \dfrac{\delta \theta_0 \sin(\alpha (1-t)) + \delta \theta_1 \sin(\alpha t)}{\sin \alpha} \\
        \delta \phi &= \delta \phi_0 (1-t) + \delta \phi_1 t
    \end{split}
\end{equation}
where $\alpha = \phi_1-\phi_0$.

In a radial gauge $A = a(\theta)d\phi$. The solid angle between two parallel circles at $\theta$ and $\theta = d\theta$ is equal to $2 \pi \sin \theta d \phi$ and should be equal to (via Stokes' formula) $2\pi (a(\theta + d\theta)-a(\theta))$, therefore $\frac{da}{d\theta} = \sin\theta$. We require that $a = 0$ as $\theta = 0$, hence $a(\theta) = 1- \cos(\theta)$. The first order variation in $\delta \Omega$ is equal to $\oint A$ along the closed path formed by the parallelogram with vertices as the four spin vectors (2 bare and 2 perturbed) and is given by
\begin{multline}
    \delta \Omega = \int_0^1[(1-\cos(\theta + \delta \theta (t)))\frac{d(\phi + \delta \phi(t))}{dt}  - (1-\cos\theta)\frac{d\phi}{dt}]\\
    +\delta \phi_0 - \delta \phi_1
\end{multline}
which after a bit of algebra can be written as 
\begin{equation}
   \delta \Omega = \dfrac{1-\cos\alpha}{\sin\alpha} (\delta \theta_0 + \delta \theta_1)
\end{equation}
To connect the above formula with the original spin vectors $\bm{n}_0$ and $\bm{n}_1$ we can also write the first order variation as 
\begin{equation}
    \delta \Omega_{01} = -\dfrac{\bm{n}_0 \times \bm{n}_1}{1 + \bm{n}_0 \cdot \bm{n}_1} (\delta \bm{n}_0 + \delta \bm{n}_1)
    \label{varsolid2}
\end{equation}
where we have defined the $z$-axis such that $\bm{n}_0 \times \bm{n}_1 = \sin\alpha \,\hat{z}$, with $\alpha \in [0,\pi]$ and $\cos \alpha =\bm{n}_0 \cdot \bm{n}_1$, which in turn also implies that $\delta \theta_i = -\delta \bm{n}_i \cdot \hat{z}$\\

To tackle the problem of discrete $\int \delta Q^2$ we consider all four plaquettes connected to a particular site $0$. The variations in the solid angle and hence the topological charge density term that appears in the energy functional of Eq. (\ref{finalE}) has the form
\begin{equation}
    \delta E^{(0)}_C = \sum_{0 \in \square} \delta \Omega^2_\square \ .
    \label{discplaq}
\end{equation}
On a plaquette with vertices $i$,$j$,$k$ and $l$ in anticlockwise order  $\delta \Omega_{ijkl} = \delta \Omega_{ij} + \delta \Omega_{jk} + \delta \Omega_{kl} + \delta \Omega_{li}$ with each term being given by Eq. (\ref{varsolid2}). Expanding these above terms and keeping only terms including the vertex $0$ we get (see Fig. \ref{topchargediag}(b) for vertex numbering)
\begin{multline}
    \delta E^{(0)}_C = \delta \Omega^2_{01} + \delta \Omega^2_{03} + \delta \Omega^2_{05} + \delta \Omega^2_{07} \\
    + \delta \Omega_{01}\delta \Omega_{30} + \delta \Omega_{03}\delta \Omega_{50} + \delta \Omega_{05}\delta \Omega_{70} + \delta \Omega_{07}\delta \Omega_{10}\\
    + (\delta \Omega_{01}+\delta \Omega_{30})(\delta \Omega_{12}+\delta \Omega_{23}) + (\delta \Omega_{03}+\delta \Omega_{
    50})(\delta \Omega_{34}+\delta \Omega_{45})\\
    + (\delta \Omega_{05}+\delta \Omega_{70})(\delta \Omega_{56}+\delta \Omega_{67}) + (\delta \Omega_{07}+\delta \Omega_{
    10})(\delta \Omega_{78}+\delta \Omega_{81})
    \label{varsoltight}
\end{multline}
We can view the above complicated expression in a tight-binding formulation which will help us for the transfer matrix formalism. The first two lines in Eq. (\ref{varsoltight}) include the onsite energy terms, all the lines include nearest neighbour hopping terms and the last two lines include second nearest neighbour hopping terms. While on-site and nearest neighbour terms also arise in the exchange part of the functional (see appendix F), second nearest neighbour contributions come only from the variation in the topological charge density. 

\subsection{Recursive transfer matrix approach to calculate magnon transmission}
\begin{figure}[t]
    \centering
    \includegraphics[scale = 0.3]{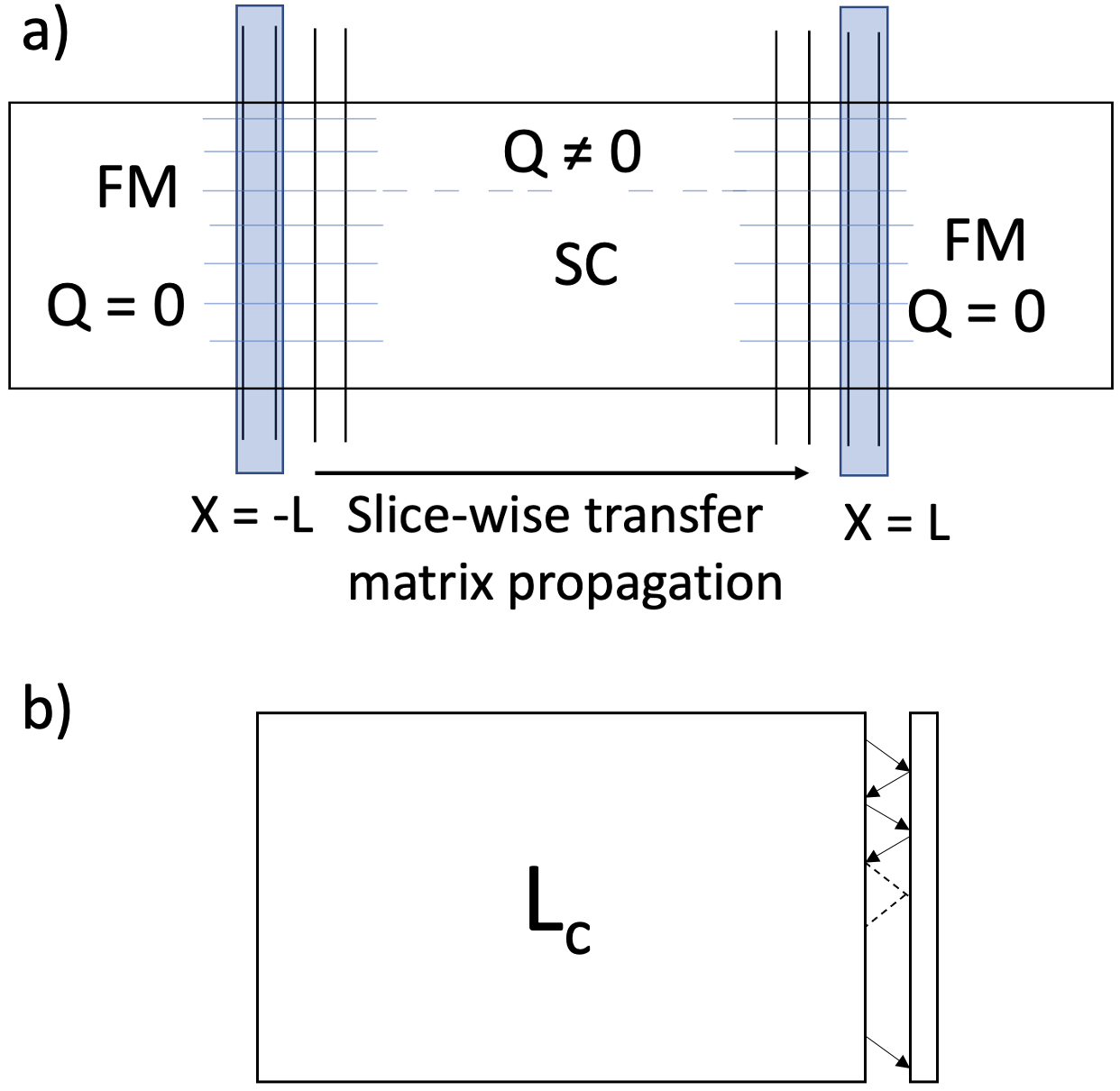}
    \caption{a) Slice-wise recursive transfer matrix procedure for the ferromagnet-skyrmion crystal-ferromagnet junction. The solid lines highlight the real space discretization and their intersection points are the lattice sites. b) Multiple scattering processes for propagating transmission and reflection matrices. Calculate the transfer matrix for the slab of length $L_c$ the usual way and then do successive slice wise rotation to propagate the transmission matrices using the infinite series.}
    \label{TMprop}
\end{figure}
In the previous two sections, we have set up the machinery needed to discretize the ferromagnet-skyrmion crystal-ferromagnet quantum Hall junction problem on a real space grid. Now, to numerically obtain the transmission properties of an incoming magnon from the left ferromagnetic end and outgoing on the right end (as in Fig.  \ref{Fig1}a), one needs to perform either a recursive transfer matrix or recursive Green's function calculation. For transfer matrices, one usually recursively calculates the full transfer matrix of the system by multiplying matrices column by column and then performing a rotation to obtain transmission and reflection matrices \cite{pichard1981finite}. For Green's functions one does the same procedure and then arrives at the conductivity using the Fisher-Lee relation \cite{Mackramer,Fisherlee}. However, both these problems suffer from numerical instabilities due to the presence of (growing) evanescent modes, which cause the product matrix to blow up. This instability is common also in the optics community, where one discretizes Maxwell's equations on a real space lattice. In this section, we adapt a method proposed by Pendry \cite{Pendryphot} for the optics problem, to obtain the full transmission and reflection matrices of the problem despite the instability. All our transmission and reflection matrices, unless explicitly mentioned, are for a wave entering from the left and exiting on the right. 

To calculate the magnon transmission across the junction we discretize the unit cell of size $a \times a$ into $N$ slices in the $x$ and the $y$ direction. We then consider a semi-infinite strip of unit cell width along the $y$-direction and impose periodic boundary conditions along $y$. Now, consider a magnon entering the skyrmion crystal region with incident energy $\omega$, transverse momentum $q_y$ and parallel momentum $q_x$ which are related by the standard ferromagnetic dispersion. Using the discretization procedure for the energy functional in Eq. (\ref{efuncfin}), we can recast the resulting time dependent Schrodinger equation in Eq. (\ref{schroeq}) as a tight binding equation which in turn can be written as a matrix equation of the form
\begin{equation}
    D \Psi_X + A_R \Psi_{X+1} + A_L \Psi_{X-1} = 0
\end{equation}
where $\Psi_X = (\chi(X,1),\bar{\chi}(X,1),...,\chi(X,N),\bar{\chi}(X,N))^T$ is a column vector for all the $\chi,\bar{\chi}$ with $x = X$ (refer to appendix A for definitions of $\chi$ and $\bar{\chi}$), and the $2N \times 2N$ matrices, $D$, $A_R$ and $A_L$ are matrices of coefficients for the wavefunctions at column $X$, $X+1$ and $X-1$ respectively (see appendix A, for expressions for $\chi$ and $\bar{\chi}$ and appendix F for expressions for matrix elements). Such an equation can then be recast as a transfer matrix equation
\begin{equation}
\begin{split}
    \begin{bmatrix}
    \Psi_{X+1} \\
    \Psi_{X}
    \end{bmatrix} &= \hat{T}_X \begin{bmatrix}
    \Psi_{X} \\
    \Psi_{X-1}
    \end{bmatrix} \\
    \hat{T}_X  &= \begin{bmatrix}
    -A_R^{-1}D && -A_R^{-1}A_L \\
    \mathds{1} && 0
    \end{bmatrix}
\end{split}
\end{equation}
where $\hat{T}_X$ is the $4N \times 4N$ transfer matrix which relates the values of $\chi$ in the nearest neighbouring columns. Using the standard properties of transfer matrices we can propagate the wavefunction from the left ferromagnet region at $x = x_L$ through the skyrmion lattice to the right ferromagnet region at $x = x_R$, column by column, resulting in the final equation
\begin{equation}
\begin{split}
    \begin{bmatrix}
    \Psi_{x_R+1}\\
    \Psi_{x_L}
    \end{bmatrix} &= \hat{T} \begin{bmatrix}
    \Psi_{x_L+1} \\
    \Psi_{x_L}
    \end{bmatrix}\\
    \hat{T} &= \prod_{i = x_L}^{x_R} \hat{T}_i
    \end{split}
    \label{Teq1}
\end{equation}
Since in section III we use truncated theta functions to smoothly interpolate between zero topological charge regions on the two ends and a finite periodic topological charge in the middle, we do not have sharp boundaries between the three separate regions. Hence we start our transfer matrix procedure on the left from a column in the region with zero topological charge density and we end on a slice on the right, again deep in the region, with zero topological charge density as shown in Fig. \ref{TMprop}(a).\\
The transfer matrix procedure as illustrated relates the values of the wavefunctions of the two end-point columns and not the amplitudes we need to calculate the transmission and reflection coefficients. To convert such a transfer matrix in the tight binding formulation to the transfer matrix which relates the amplitudes of the waves in the left region to those in the right, we first need to express the wavefunction in terms of these amplitudes using the standard scattering ansatz. For the wavefunction in the starting column at $x = x_L$, we can write 
\begin{align}
\begin{split}
    \chi_{x_L,y} &= \sum_{i =1}^{N} A_{i'} e^{iq_{xi}x_L+iq_{yi}y} + B_{i'} e^{-iq_{xi}x_L+iq_{yi}y}\\
    \bar{\chi}_{x,y} &= \sum_{i =1}^{N} A_{i''} e^{iq_{xi}x_L+iq_{yi}y} + B_{i''} e^{-iq_{xi}x_L+iq_{yi}y}
\label{scatansatz}
\end{split}
\end{align}
where $i' = 2i-1$ and $i'' = 2i$, and $q_{xi},q_{yi}$ are the wavevectors of the $N$ different modes, with $\triangle q_y = 2\pi/a$. Note here that generally, $\chi_{x_L} \neq \bar{\chi}_{x_L}^*$, even though $\chi = \chi_1 + i\chi_2$ and $\bar{\chi} = \chi_1 - i\chi_2$, because $\chi_1,\chi_2 \in \mathbb{C}$ as a result of complex phase factors (further details on this are given in appendix I). Using the form of the scattering ansatz we can define the following rotation relating the $4N$ sized column vectors
\begin{equation}
\begin{split}
    &\begin{bmatrix}
    \Psi_{x_L},
    \Psi_{x_L-1}
    \end{bmatrix}^T = \begin{bmatrix}
    \chi_{x_L,1},...,
    \bar{\chi}_{x_L-1,N}
    \end{bmatrix}^T\\
    &= Q_{x_L} \begin{bmatrix}
    A_1 e^{ik_{x1}},
    A_2 e^{ik_{x1}},...,
    B_{2N-1}e^{-ik_{xN}},
    B_{2N}e^{-ik_{xN}}
    \end{bmatrix}^T\\
    \end{split}
\end{equation}
where $Q_{x_L}$ is a $4N \times 4N$ matrix and $k_{xi}$ and $k_{yi}$ are the different modes obtained from the discretization on the grid, $k_{yi} = k_{y0} + 2\pi i/N$, with $i = 0,..,N-1$ and $k_{xi} = \omega/(2N^2) (2-\cos(k_{yi}))$. These wavevectors on the grid are related to the continuous ones by $q_y a = k_y N$. We can then express the transfer matrix equation in Eq. (\ref{Teq1}) in terms of the scattering amplitudes on both ends as
\begin{equation}
\begin{split}
    \begin{bmatrix}
    C_1 e^{ik_{x1}}\\
    C_2 e^{ik_{x1}}\\
    .\\
    .\\
    D_{2N-1}e^{-ik_{xN}}\\
    D_{2N}e^{-ik_{xN}}
    \end{bmatrix} &= \Tilde{T} \begin{bmatrix}
    A_1 e^{ik_{x1}}\\
    A_2 e^{ik_{x1}}\\
    .\\
    .\\
    B_{2N-1}e^{-ik_{xN}}\\
    B_{2N}e^{-ik_{xN}}
    \end{bmatrix}\\
    \Tilde{T} &= Q^{-1}_{x_R}TQ _{x_L}
    \end{split}
\end{equation}
where $C_i,D_i$ represent the amplitudes for the wavefunction on the right-most slice at $x = x_R$. The transmission and reflection coefficients can be expressed from the elements of the rotated transfer matrix $\Tilde{T}$, see Eq. (\ref{TrRslice}).\\
Such a column-wise multiplied transfer matrix procedure runs into numerical instabilities due to growing evanescent modes which blow up on increasing the length of middle region (largest eigenvalue $> 1$) and are a common cause of instability in such recursive transfer matrix methods. To overcome this we use a method common in optics \cite{Pendryphot}, in which one uses the usual recursive approach up to a certain column and then propagates the transmission and reflection matrices column-wise thereon using multiple scattering, instead of propagating the whole  transfer matrix, as shown in Fig. \ref{TMprop}a. Such an approach does not suffer from numerical instabilities since the reflection and transmission matrices are bounded because of unitarity of the scattering matrix. To obtain the transmission and reflection matrices for a slab of length $L_c$, we perform the rotation shown above to obtain
\begin{equation}
    Q^{-1}_{L_c} T_{L_c} Q_{0}=  \Tilde{T}_{L_c} = \begin{bmatrix}
    \Tilde{T}_{L_c11} && \Tilde{T}_{L_c12}\\
    \Tilde{T}_{L_c21} && \Tilde{T}_{L_c22}\\
    \end{bmatrix} 
    \label{slicerot}
\end{equation}
where $T_{L_c}$ is the transfer matrix obtained relating the columns at the two ends of the slab of length $L_c$ and $Q$ is the rotation matrix as defined earlier. One can obtain the transmission matrices $Tr(L_c)$ and $R(L_c)$ for the slab from the $\Tilde{T}_{Lcij}$. They are given by 
\begin{align}
\begin{split}
    R(L_c) &= -\Tilde{T}_{L_c22}^{-1}\Tilde{T}_{L_c21} \\
    Tr(L_c) &= \Tilde{T}_{L_c11} +   \Tilde{T}_{L_c12}^{-1}R(L_c)
    \label{TrRslice}
\end{split}
\end{align}
\begin{table*}[t]
\begin{tabular}{ |p{4cm}|p{4cm}|p{4cm}|  }
 \hline
 Property& Ferromagnet&Skyrmion crystal\\
 \hline
 Number of 0-energy deformations  & 2    &3\\
 \hline
 Ground state manifold &   Sphere $\bm{n}\cdot \bm{n} = 1$  & SO(3) group parameterized by $x_1,x_2,x_3$ \\
 \hline
 Total angular momentum on ground state manifold & $\bm{L} \sim \bm{n} \neq 0$ & $\bm{L} = 0, (\pi_1,\pi_2,\pi_3 = 0)$\\
 \hline
 Dynamical variables  & $\bm{n}$ with $\{n_i,n_j\} = \epsilon_{ijk}n_k$; $1 \leq n_i,n_j,n_k \leq 3$ & $x_1,x_2,x_3,\pi_1,\pi_2,\pi_3$; $\{x_i,x_j\} = \{\pi_i,\pi_j\} = 0$; $\{x_i,\pi_j\} = \delta_{ij}$\\
 \hline
\end{tabular}
\caption{Different properties of the bulk in the ferromagnet and skyrmion crystal highlighting the different order parameter manifolds. The most general form of coupling between these two that respects the individual properties shown in this table are given in Eq. \ref{coupH1} and \ref{coupH2}. }
\label{tab1}
\end{table*}
where $\Tilde{T}_{L_cij}$, $i,j \in [1,2]$ are the $2N \times 2N$ blocks in Eq. (\ref{slicerot}).
Say we have a situation in which we calculate the transmission and reflection matrices, $Tr(L_c)$ and $R(L_c)$ for a slab of length ($L_c$) within numerical accuracy using the standard recursive protocol. We can then propagate the transmission and reflection matrices by summing up the infinite series from multiple scattering events from the additional slice, as shown in Fig. \ref{TMprop}b, to get:
\begin{equation}
\begin{split}
    Tr(L_c+1) &= Tr(1)(I-T_{1}R(1))^{-1}Tr(L_c)\\
    R(L_c+1) &= R(L_c) + T_{2} R(1)(I-T_{1}R(1))^{-1}Tr(L_c)
\end{split} 
\label{multscateq}
\end{equation}
where $Tr(1)$ and $R(1)$ are the transmission and reflection matrices for a wave incident from the left on the added slice, obtained using the same procedure as in Eqs. \ref{slicerot},\ref{TrRslice} (but now with $T(1)$ instead of $T_{Lc}$). $T_{1}$, $T_{2}$ are the reflection and transmission matrices for a wave incident from the right on the slab of length $L_c$ respectively\cite{Pendryphot}. Such an approach, although more numerically expensive due to more matrix inversions, resolves the numerical instabilities and allows one to calculate the full transmission and reflection matrices.
\subsection{Coupling between Goldtsone modes - recipe for a non-linear sigma model}
    \begin{figure*}[t]
    \centering
    \includegraphics[width = 18cm, height = 4cm]{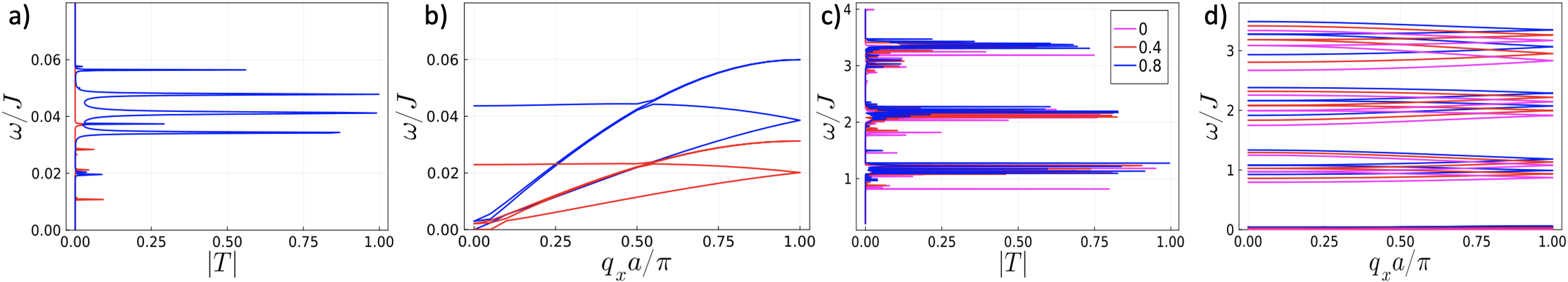}
    \caption{Transmission and energy spectra for the ferromagnet-skyrmion crystal-ferromagnet junction at normal incidence $q_y = 0$ at g/J = 0.4 (blue),0.8 (red) and 0 (magenta, only in (c-d)). The effect of increasing $g$ in the 2nd term of the energy functional in Eq. \ref{efuncfin} on a) low-energy transmission - peaks shift and increase in height b) Goldstone mode dispersion - modes become more dispersive and c) high energy transmission- slight shift in position of peaks but no significant change in height}
    \label{geff}
\end{figure*}
On top of the microscopic numerical calculations of the scattering problem, facilitated by the developments in the last three subsections, it is desirable to have  a long-wavelength description for such junctions. Such a construction provides an analytical coarse-grained framework without delving into the specifics of the microscopic structure. These coarse grained constructions usually take the form of non-linear sigma models in magnetic systems. However, the construction of such a non-linear sigma model for our problem is complicated by the presence of coupling between the ferromagnet and the skyrmion crystal at the interface, since both have different order parameter manifolds (see Table \ref{tab1}). To understand what kind of couplings may arise at the interface of such a junction, one has to first construct a good parametrization of SO$(3)$ to express the total angular momentum $L$ in the $x_i,\pi_j$ variables (see Table \ref{tab1}). Such a parametrization will allow us to construct SO(3) invariant Hamiltonians in the vicinity of the degenerate ground state manifold of the non-collinear skyrmion crystal, and will help us find ways to couple it to $\bm{n}$ in an SO(3) invariant way.

We start by writing rotation matrices in terms of SU(2) matrices. Consider the family of such rotation matrices of the form
\begin{align}
\begin{split}
    U(\bm{x}) &= \begin{bmatrix}
        \sqrt{1-\bm{x}^2}+ix_3 && ix_1+x_2 \\
        ix_1-x_2 && \sqrt{1-\bm{x}^2} -ix_3
    \end{bmatrix} \\
    &= \sqrt{1-\bm{x}^2}\mathds{1} + i\bm{x}\cdot \bm{\sigma}
\end{split}
\end{align}
Since SO(3) is non-Abelian we have two distinct actions of SU(2) on itself, either by left or right multiplication. Note that any left multiplication commutes with any right multiplication, $U(VW) = (UV)W$, however two left or two right multiplications do not commute. We choose left actions to correspond to global SO(3) symmetries, and generators of right actions therefore, to commute with generators of global symmetries - they can be used to construct effective low energy Hamiltonians.\\
Let us first study the left action. Consider an infinitesimal rotation $\exp(-i\frac{\bm{\epsilon}}{2} \cdot \bm{\sigma})$ where $\bm{\epsilon}$ is a small vector in $\mathbb{R}^3$. Now,
\begin{multline}
    (\mathds{1} - i\frac{\bm{\epsilon}}{2} \cdot \bm{\sigma} )U(\bm{x}) = U(\bm{x} + X_{\bm{\epsilon}}(\bm{x})) + O(\epsilon^2)  \\
    X_{\bm{\epsilon}}(\bm{x}) = -\sqrt{1-\bm{x}^2}\frac{\bm{\epsilon}}{2} + \frac{\bm{\epsilon} \times \bm{x}}{2}
    \label{defXepsilon}
\end{multline}
 When $\epsilon \rightarrow 0$, $X_{\bm{\epsilon}}(\bm{x})$  may be seen as a tangent vector to the SO(3) group manifold at point $\bm{x}$, so $x \rightarrow X_{\bm{\epsilon}}(\bm{x})$ is a vector field associated to the infinitesimal rotation.

 Vector fields form a Lie-algebra under the Lie-bracket. We can check that $[X_{\bm{\epsilon}},X_{\bm{\eta}}]= X_{\bm{\eta} \times \bm{\epsilon}}$, which is the Lie-algebra structure of SO(3) in $\mathbb{R}^3$ (see appendix J for details). Now, coming to the right action of the rotation matrix, we find that 
 \begin{multline}
    U(\bm{x}) (\mathds{1} +i\frac{\bm{\epsilon}}{2} \cdot \bm{\sigma} ) = U(\bm{x} + Y_{\bm{\epsilon}}(\bm{x})) + O(\epsilon^2) \\
    Y_{\bm{\epsilon}}(\bm{x}) = \sqrt{1-\bm{x}^2}\frac{\bm{\epsilon}}{2} + \frac{\bm{\epsilon} \times \bm{x}}{2}
\end{multline}
We can recover the same Lie-algebra structure as for the left action here as well.

In Hamiltonian mechanics, angular momentum defined by $\tilde{L} = \epsilon_{ijk} x_j \pi_k$ generates rotations in phase space owing to its commutation relations with the $x$ and $\pi$ variables. From the equations of $X_{\bm{\epsilon}}$ and $Y_{\bm{\epsilon}}$, one can show that the left action is generated by $\bm{\epsilon} \cdot \bm{L}$ where 
\begin{equation}
    L_i = \frac{1}{2} (-\sqrt{1-\bm{x}^2}\pi_i + \tilde{L}_i)
\end{equation}
and the right action is generated by $\bm{\epsilon} \cdot \bm{R}$, where 
\begin{equation}
    R_i = \frac{1}{2} (\sqrt{1-\bm{x}^2}\pi_i + \tilde{L}_i)
\end{equation}
These functions satisfy the commutation relations $\{L_i,L_j\} = \epsilon_{ijk} L_k$, $\{R_i,R_j\} = \epsilon_{ijk} R_k$ and $\{L_i,R_j\} = 0$.

For the uniform ($\bm{q} = 0$) sector, the non-collinear skyrmion crystal Hamiltonian contains only a kinetic term, which we take to be quadratic in the $R_i$ variables, since $R_i = 0$ identically on the degenerate ground state manifold and $R_i$ deviates linearly from zero when $\pi_i$'s are small. We can write this term as
\begin{equation}
    H_{K} = \frac{1}{2} \sum_{a,b} I^{-1}_{ab}R_a R_b
\end{equation}
where $I_{ab}$ is a positive definite symmetric matrix with real entries which may be regarded as a generalized inertia matrix for a kind of top.

The simplest left-invariant coupling between the ferromagnetic magnetization $\bm{n}$ and the skyrmion crystal system is given by
\begin{equation}
    H_{c1} = g_c{\bm{n} \cdot \bm{L}}
    \label{coupH1}
\end{equation} \\
where $g_c$ is some coupling constant.\\
Since the coupling terms occur at the interface one can also consider coupling $\bm{n}$ to a fraction of the spins composing the skyrmion crystal (those belonging to the interface) in an SU(2) invariant way. So we should be able to construct triples of functions over the SU(2) group manifold (with $x$ coordinates) which transform as the three components of the usual vectors under usual SO(3) rotations. To express this in the coordinates obtained for SU(2), we use the Heisenberg picture for observables. We set $\bm{\sigma}(\bm{x}) \equiv U^{\dagger}(\bm{x})\bm{\sigma} U(\bm{x})$. Now, let us change $\bm{x}$ to $\bm{x} + X_{\bm{a}'}$, where $\bm{a}'$ is an infinitesimal vector. By construction, this amounts to sending $U(\bm{x})$ to $\exp(-i \frac{\bm{a}'}{2} \cdot \bm{\sigma}) U(\bm{x})$. Using the above and the relations $[\sigma_i,\sigma_j] = 2i\epsilon_{ijk}\sigma_k$ and $[\bm{a'} \cdot \bm{\sigma},\bm{\sigma}] = -2i \bm{a'} \times \bm{\sigma}$ one gets that
\begin{equation}
    \bm{\sigma} (x+X_{\bm{a}'})= \bm{\sigma}(\bm{x}) + \bm{a'} \times \bm{\sigma} (\bm{x}) + O(\bm{a'}^2)
    \label{sigmaeq}
\end{equation}
We can then choose any density matrix $\rho_0$ (with $\rho = \rho^{\dagger}$, positive eigenvalues and Tr$\rho_0 = 1$) and form a vector valued function
\begin{align}
\begin{split}
    \langle \bm{\sigma} \rangle_{\rho_0} &: SU(2) \rightarrow \mathbb{R}^3 \\
    &U(\bm{x}) \rightarrow \langle \bm{\sigma} \rangle_{\rho_0} \equiv Tr(\bm{\sigma}(\bm{x})\rho_0)
\end{split}
\end{align}
Then, Eq.(\ref{sigmaeq}) implies that
\begin{equation}
  \langle  \bm{\sigma} (x+X_{\bm{a}})\rangle_{\rho_0} = \langle \bm{\sigma} (\bm{x})\rangle_{\rho_0} + \bm{a} \times \langle \bm{\sigma} (\bm{x}) \rangle_{\rho_0}
\end{equation}
Hence, one can write down a second kind of coupling term 
\begin{equation}
    H_{c2} = g'_c \bm{n} \cdot \langle \bm{\sigma} (\bm{x}) \rangle_{\rho_0}
    \label{coupH2}
\end{equation}
where $g'_c$ is some other coupling constant. Therefore the full sigma model Hamiltonian including the standard gradient potential terms for the bulk would be
\begin{equation}
    H_{\sigma} = H_{bulk} + H_{K} + H_{c1} + H_{c2}
    \label{fullHsigma}
\end{equation}
where the coupling terms would be evaluated at the coordinates of the interface. One can use this model to calculate the equations of motion and the corresponding transmission coefficients. Following this, one can fit the results to the values obtained numerically from our transfer matrix calculations to get the values of all the coupling constants in the sigma model. However, we leave a detailed analysis of such sigma models for future work.
\section{Magnon transmission through a Skyrmion Crystal}
Using the technical advancements described in sections V A-C, one can numerically solve the scattering problem of magnon scattering in the ferromagnet-skyrmion crystal-ferromagnet setup. The main results of the problem  provide a unique set of transport signatures for the skyrmion crystal. They are summarized in the section II and in Fig. \ref{Fig1}(c-d). In this section we provide some additional results of the scattering problem that reflect the topology-symmetry dichotomy.

With the detailed analysis of our heuristic model of a particle scattering in a magnetic field in hand, we return to the full problem of the ferromagnet-skyrmion crystal energy functional in Eq. (\ref{efuncfin}). As pointed out earlier, the effect of a non-zero coupling constant in the topological charge density term is to provide some sense of stiffness to the skyrmion crystal, and thereby inducing dispersion in the Goldstone modes. To separate the dispersion effect first we look at the $g = 0$ case in which all Goldstone modes are pinned to zero and the Riemann-Goldstone Landau level is a zero energy flat band. From the full expression of the variation of the energy functional in eq \ref{finalE} in appendix A, one can see that for $g = 0$ the expression resembles that of the heuristic model. Hence, we can use the intuitive understanding developed in section III. 
\subsection{Similarities with heuristic picture - high energy sector}
For $g= 0$, based on our transfer matrix analysis, we see that the transmission spectra in Fig. \ref{geff}c reflects the underlying topology of the skyrmion crystal, since the non-zero transmission occurs in energy regions which reflect the emergent Landau levels of the problem. We also note the remarkable qualitative similarity of the response with that of the heuristic model in Fig. \ref{specyv}(a-b). For two skyrmions in a unit cell, we get four flux quanta acting on a spin-1 magnon, which justifies our use of the analogous average magnetic field $B_0 = 8\pi/a^2$ for the results in Fig.\ref{specyv}(a-b)(see appendix A and \cite{douccot2018zero}). We see that the transmission peaks are suppressed, i.e we do not get full transmission at these resonant energies, and one can understand this using similar multi-channel interference arguments presented in appendix E for the heuristic model. 

Besides calculating transmission coefficients we also obtain the spectra for the skyrmion crystal from the energy functional in Eq. (\ref{efuncfin}). For the $g = 0$ "skyrmion crystal", we find that the high energy modes resemble the dispersive Landau levels, similar to those observed in the heuristic model for a spatially varying magnetic field. The transmission peaks occur in energy regimes of the effective Landau levels and the gaps in non-zero transmission correspond exactly to the gaps in the Landau level dispersion.

The qualitative similarity with the heuristic model also implies that magnon transmission at high energies will be characterized by certain preferred angles of transmission and a non-monotonic dependence of transmission on the channel momenta. Such a non-monotonic dependence on channels and corresponding angular spread is a clear consequence of crystalline order.
\subsection{Effects of Goldstone mode dispersion - low energy sector}
The lowest energy modes  for the $g= 0$ case are pinned to zero energy since one can deform $\bm{n}_0(\bm{r})$ continuously in the space of holomorphic textures while keeping the exchange energy constant (first term in Eq. \ref{efuncfin}). Hence, for $g = 0$ we get localized modes in the Riemann-Goldstone Landau level. On introducing a finite $g$, we see from the Goldstone mode spectra in Fig. \ref{geff}(b), that the Goldstone modes acquire a finite dispersion. We get three low energy Goldstone modes,  as expected for an SU(2) skyrmion crystal. Out of these three, two modes are almost degenerate and have a higher velocity than the third.  All these modes have a linear dispersion at low $q$, as behooves an  antiferromagnet. Note that,  remarkably, one can infer all this information about the Goldstone modes just by looking at the transmission spectra in Fig. \ref{geff}a. We see that there are two sets of peaks, within each set, the peaks are equally spaced and increase in height on increasing energy. 

Such behaviour is qualitatively consistent with our results from the heuristic model of the ferromagnet-antifferomagnet-ferromagnet junction (see Fig. \ref{Affig}b). These two sets of peaks correspond to the two Goldstone mode branches, and their different peak positions imply that the velocity of these two modes are different, as verified by our results of the Goldstone mode spectrum in Fig. \ref{geff}(b). We also see that there is a very small splitting in one set of peaks, indicating the fact that the two higher velocity modes are almost-degenerate. On increasing $g$, we see that the Goldstone modes become more dispersive, as in Fig. \ref{geff}b and one can also infer this by looking a the transmission spectra, the peaks of which shift and become more prominent, as in Fig. \ref{geff}a. 

Hence, the magnon transmission spectra encode the nature of the Goldstone mode spectra in the emergent Riemann-Goldstone Landau level sector of the skyrmion crystal.

We also note from Fig. \ref{geff}d that the peaks in the higher energy effective Landau levels are slightly shifted but the heights are relatively unaffected on increasing $g$, which confirms that the physics of the Riemann-Goldstone Landau level, associated with symmetry breaking, is indeed distinct from that induced by the underlying topology of the spin texture. Such separation of energy scales highlights the topology-symmetry dichotomy of the problem and is very nicely elucidated by the magnon transmission. Moreover, the qualitative similarities of the two heuristic models with the two different energy sectors also presents a simplified and intuitive understanding of this very rich problem.

We note that the Goldstone modes present in Fig. \ref{geff}(b) don't go down to exactly zero energy.  This is a consequence of real space discretization and the holomorphic ansatz being an exact minimum of the exchange terms only in the continuum limit. Because of this, for a finite discretization scheme, the Goldstone mode eigenvalues will actually have a small imaginary part (compared to the real part). We have plotted only the real part of these eigenvalues in Fig. \ref{geff}(b). However, as we approach the continuum limit, the ansatz  exactly minimises the energy and so for larger values of $N$, the complex part becomes numerically insignificant and the real part of the modes will go down to exactly 0 as $q\rightarrow 0$ in Fig. \ref{geff}(b). The phonon mode in our low-energy spectra however, shall remain gapped due to the nature of our energy functional: the spatial modulations of the topological charge density $Q_{0}(\mathbf{r})$ explicitly break translational
symmetry. This gapped phonon branch appears as the upper branch in Fig. \ref{geff}b. We comment more on the implications and feasability of the gapped magnetophonon in the next section.

\section{Discussion}
\subsection{Anisotropies and experimental considerations}
In this work we have used an effective continuous model derived from a holomorphic ansatz motivated from the physics of isotropic skyrmion crystals. This approach carries a long way in terms of physical intuition, analytical control and a full qualitative understanding from such an effective theory. However, there are features beyond the model that could be present in experiment. In this section, we discuss how such features could modify the results we presented. Just as importantly, we also discuss how the results from our model could be realized in ongoing experiments.

The price we pay for using the holomorphic ansatz as a starting point, is the absence of anisotropies. While the energy scale for anisotropies is smaller than that of the Coulomb interaction, they still play a role in the low-energy physics of monolayer graphene in the zeroth Landau level \cite{kharithall}. The leading anisotropy in such systems would be the Zeeman term $g_1 \mu_b \bm{n} \cdot \bm{B}$. The dispersion for realistic models of skyrmion crystals in graphene with such terms were studied in \cite{Cotecoll} using Hartree-Fock methods. The authors showed that the Zeeman term gaps out one of the three Goldstone modes. Hence, we expect that the  transmission signatures we predict for the Goldstone modes can still be observed in experiments on monolayer graphene, with the modification that the spacing of the peaks  would be less linear in the low-energy sector of the transmission spectra. The higher energy signatures from the effective Landau levels should also be robust to the presence of any relevant anisotropies such as the Zeeman, or even the lattice scale, terms. We also note, as briefly mentioned in the section above, due to our energy functional being constructed to have our truncated theta-function ansatz as the minima, the magnetophonon mode obtained from our spectra in Fig. \ref{geff}(b) is also gapped. In isotropic and fully periodic skyrmion crystals such a mode is expected to be gapless and have the characteristic $\sim q^2$ dispersion for short range and $q^{3/2}$ dispersion for Coulomb interactions in two dimensions \cite{Dimaskyr}. However, one nonetheless expects the phonon mode to be gapped in the presence of the junction between regions of  different filling. Moreover, anisotropies also gap out the phonon mode \cite{CoteCP3}, hence we do not expect its presence to alter our results much.

Another source of potential mismatch between experiment and our theory would be the range of our interaction term. The interaction term in our effective energy functional is a delta function (in real space) terms, whereas the Coulomb interaction is long-range. However, such an interaction can be engineered in graphene using metallic gates which screen the Coulomb interaction. A typical magnon transport experiment on graphene involves a sheet of monolayer graphene sandwiched in between hBN substrates and additional metallic gates on top and bottom. The  gate-screened potential in momentum space is given by 
\begin{equation}
    V(q) = \dfrac{4\pi e^2}{\sqrt{\epsilon_x \epsilon_z}} \dfrac{\sinh(qd\sqrt{\frac{\epsilon_x}{\epsilon_z}})\sinh(qd'\sqrt{\frac{\epsilon_x}{\epsilon_z}})}{q\sinh(q(d+d')\sqrt{\frac{\epsilon_x}{\epsilon_z}})}
\end{equation}
where $d$ and $d'$ are the distance from the top and bottom gates to the graphene sample and $\epsilon_x$,$\epsilon_z$ are the static in and out-of plane permittivites of the hBN \cite{kim2020control}. On tuning the parameters $d$ and $d'$, it is plausible to expect that one can realize a potential structure that is fairly flat in momentum space, leading to a localized delta function in real space. Moreover, by tuning the strength of the interaction, one can also tune the coupling constant $g$, which controls the dispersion of the Goldstone modes and hence one can observe the variations in the transmission spectra as mentioned in the previous section.

For current experiments on graphene, the external field $B_{\rm{ext}} \sim 10 T$, hence $l_B \sim 10 \rm{nm}$. In our model, we assume that all spatial variations are on scales larger than the magnetic length. In particular, the two important scales are the skyrmion crystal lattice constant and the interface width, governed by electrostatics. In our ansatz these two length scales are comparable. Hence, to exactly derive results from our model for experiment, these two length scales should be at least an order of magnitude larger than $l_B$. If, in experiments the skyrmion period is of the order of $l_B$, or if the ferromagnet-skyrmion crystal interface is much sharper then one might need  to resort to more
microscopic time-dependent Hartree-Fock treatments which will likely change some quantitative details, but should retain the structure of transmission from Goldstone modes in the Riemann-Goldstone Landau level and higher energy effective-Landau levels presented here. Such qualitative similarity is fair to expect for low momentum physics given the early work on skyrmions which compared Hartree-Fock and effective continuous theory treatments \cite{sondhiskyr,Breyskyr,Fertigskyrmions,MoonPRB}, and is also bolstered by the qualitative similarities between our heuristic  model for the ferromagnet-antiferromagnet junction and a full Hartree-Fock calculation for the junction with $\nu = 0$ sandwiched in the middle \cite{Weimagn}.

In our model, for theoretical purposes, the effective Landau level gap is  $\sim J$, the exchange coupling constant. However, we can estimate what this effective gap, $\hbar \omega_G$, will be in experiment. To do so we neglect the $g$ terms since as we have seen in the last section they only have a qualitative effect   on the Goldstone modes. Now, we can use the similarity with the heuristic model in section III, to consider a magnetic field $B = 4 \pi Q_0$ with $Q_0 = \delta \nu/(2\pi l^2_B)$ where $\delta \nu$ is the deviation from unit filling in the central region. The spectral gap for the simplified energy functional is $2JB$. Using Eq. (\ref{schroeq}), we get $\alpha \omega_G = 4JB$, where $\alpha $ is defined in appendix A. Using the values of $\alpha$ and $B$ we can write $\hbar \omega_G = 32\pi J \delta \nu$. Now using the standard value of $J = e^2/(32\sqrt{2\pi}\epsilon l_B)$ we obtain a spectral gap $\hbar \omega_G = \sqrt{\frac{\pi}{2}}\frac{e^2}{\epsilon l_B} \delta \nu$. The gap is linear in $\delta \nu$. This is important as $\delta \nu$ is easily tunable in experiment.

The experiment in \cite{zhou2020solids}, which was part of our motivation for this project, prepared a junction similar to the one suggested in our paper and reported the observation of a possible skyrmion crystal due to suppression of transmission on doping slightly away from $\nu = 1$ in the central region. At such a filling of the central region, theoretically one would expect the formation of a skyrmion crystal \cite{Breyskyr}, which has a qualitatively different Goldstone mode dispersion compared to the ferromagnet, and the observed suppression would agree with the picture of magnon decay into some of these. This experiment raised the important question of the non-trivial interaction between ferromagnetic magnons and excitations with qualitatively different dispersions. \\
While consistent with the hypothesis of the formation of the skyrmion crystal, the reported suppression does not tell us much about the nature of its Goldstone/high-energy modes. Moreover, such suppression can also arise within the context of elastic scattering, due to any other spin structure which hosts a qualitatively different dispersion as compared to the incoming magnon, for example similar suppression is seen for the case of the ferromagnet-antiferromagnet junction (section III and \cite{Weimagn,atteia2022beating}). Our results provide concrete signatures in non-local response which are unique to the skyrmion crystal and as far as we can see do not appear in any other  phase  in the quantum Hall phase diagram. The combination of Landau-level like transmission and equally spaced low energy peaks due to the linear nature of the Goldstone modes would elucidate both the degree of crystalline order and the nature of the skyrmion crystal. Further experiments in which the non-local response is studied as a function of the incoming magnon energy should be able to detect such signatures.
\subsection{Outlook}
We have shown how magnon transport through skyrmion crystals probes the interplay of topology and symmetry breaking. We have shown that the magnon transmission spectra allows one to probe the topology arising from the high-energy effective Landau level structure which comes from the texture of skyrmion crystal. Moreover, and perhaps more interestingly, low-energy transmission spectra can also probe the nature of the Goldstone modes in the Riemann-Goldstone Landau level, which arises from a complex interplay of the topology as well as SU(2) symmetry breaking. Therefore, not only does our work provide a rich example of the salient features of the confluence of topology and symmetry breaking, it also presents a set of results which allow one to probe crystalline order and map out the excitation spectrum of a quantum Hall skyrmion crystal -- direct experimental evidence of which has not been established conclusively -- in current ongoing experiments.

We have also provided a simpler tool set comprising two heuristic models which allow us to intuitively understand parts of the complex problem. Moreover, to solve the complex problem, we have made several technical advances which are easily transferable to analogous problems elsewhere. Firstly we have provided an analytical framework to study junctions of topologically trivial and non-trivial structures. Secondly, we have provided a novel method for the discretization of topological charge in real space, of possible use in various fields, including metallic magnets. Thirdly, we have provided an example of the construction of a novel type of non-linear sigma model for such a junction-like structure between two different ground state manifolds. Such a construction and its extensions can be used in metallic magnets as well, where two magnetic materials with different collective excitations are separated by domain walls.

Besides monolayer graphene, where quantum Hall skyrmion crystals are expected to form near unit filling of the zeroth Landau level, there are various other platforms which host skyrmion crystals. Metallic magnets in two and three dimensions, for example, as mentioned in the main text have been a rich source of skyrmion crystal physics. Besides these usual suspects, with the advent of twistronics, spurred by the experiments on twisted bilayer graphene \cite{cao2018correlated,cao2018unconventional}, there have been several proposals for the realization of skyrmion crystal phases in such  settings. For example, in twisted bilayer graphene, skyrmions have been proposed as the lowest energy charged excitations of the insulating phase, and possibilities of such skyrmions forming a crystal have also been put forward \cite{khalaf2021charged,khalaf2022baby,kwan2022skyrmions,chatterjee2020symmetry}. Also, a recent experiment in twisted bilayer graphene has used SQUID measurements to map out the inhomogeneous spatially varying Berry curvature-induced magnetism at zero external field \cite{grover2022chern} near the magic angle. While we have considered periodically varying effective magnetic fields in our problem, one could extend this to incorporate disordered profiles. Such profiles should have a distinct signature in the magnon response. Hence, our work motivates the possibility of the exploring the zero-field Chern mosaic in twisted bilayer graphene using magnon transport.
Besides graphene, skyrmion crystal phases have also been proposed in other twisted van der Waals magnets \cite{xu2021emergence,akram2021moire}. Our work presents a route to detect skyrmion crystals in all these systems using magnon scattering.
 
 One can also use our analytical ansatz to formulate the scattering problem for other topologically non-trivial structures such as meron or bimeron crystals \cite{Breymeron}. Meron crystals have a different collective mode dispersion \cite{CoteCP3}, hence it would be interesting to see how their transport signatures differ for magnon scattering. Most of the theoretical work thus far has focused on integer fillings in the central region. One could also ask the  question of what response ground states of fractional fillings have in such magnon scattering. Our analytical ansatz of the theta functions, as mentioned in the main text, is closely related to the analytic part of the Laughlin-Jastrow wavefunctions under periodic boundary conditions \cite{Haldanetheta}. Hence using such truncated versions of similar holomorphic functions could be a good starting point for such a theoretical analysis. Moreover, fractionally charged skyrmions have also been predicted near certain fractional fillings \cite{kamilla1996skyrmions,balram2015fractionally,doretto2005spin,wojs2002spin}, and similar suppression of the non-local response as for integer charged skyrmion crystals was also observed \cite{zhou2020solids}. Hence, studying their response and comparing with our results would be an interesting direction to pursue. 
 
Another avenue of theoretical research would be to explore the scattering problem for crystals of entanglement skyrmions. Entanglement skyrmions are textured of entangled spin-valley degrees of freedom \cite{Doucotent}. Recent work has shown that such skyrmions could be realized in monolayer graphene under realistic values of anisotropies \cite{Lianent}. It would be interesting to explore if the injection of spin-waves could detect the degree of entanglement between spin and valley degrees of freedom. The non-linear sigma model construction shown in this paper would also be a much richer theoretical problem for the entangled skyrmion case due to the entanglement skyrmions living in $\mathbb{C}\rm{P}^3$ space.

Moreover, as also mentioned in the main text, the presence of effective Landau levels for the magnons presents such skyrmion crystals as a fertile platform for topological magnonics, a point appreciated also in a recent work of a skyrmion crystal in a three dimensional metallic magnet \cite{weber2022emergent}. Such connections allow one to transfer the physics of Chern bands, edge states and bulk-boundary correspondence from topological band theory to magnons.
Besides quantum Hall junctions, such junction like structures have also been considered for domain walls in two dimensional magnets \cite{Yantorque,Kimpropulsion}. Two dimensional antiferromagnets host stable skyrmions \cite{vsmejkal2018topological} and recently, skyrmion domain walls between a ferromagnet and antiferromagnet have also been considered \cite{lee2022magnon}. Hence, it would be interesting to study how the signatures of an antiferromagnetic skyrmionic crystal would differ from our results of a ferromagnetic one.

Given the angular dependence of the transmission predicted in our work for magnon scattering off skyrmion crystals, several interesting experimental possibilities also emerge. One could create geometrically optimized  junctions to maximize magnon transmission, and perhaps also place a series of such junctions to create a narrow beam of magnons with very little angular spread. 

Overall, the new experimental capacities are remarkably well-suited to study phenomena arising from the combination of symmetry-breaking and topology in two-dimensional systems, and we hope this work will motivate further studies of this complex of questions in both theory and experiment.

\section{Acknowledgements}
The authors thank Dmitry Kovrizhin and Mark Goerbig for valuable discussions and especially Preden Roulleau for insights into connections with experimental implementations. This work was in part supported by the Deutsche Forschungsgemeinschaft under grants SFB 1143 (project-id 247310070) and the cluster of excellence ct.qmat (EXC 2147, project-id 390858490). B. D. thanks the PkS Max Planck institute for its generous hospitality during several extended visits, which were crucial for the realization of this project. NC thanks the physics department of Sorbonne for their generosity and hospitality during his visit.

\appendix
\section{Equations of motion and mapping to Schr\"odinger equation}
In this appendix we provide the details of the spin-wave theory calculations starting from Eq. (\ref{efuncfin}) in the main text. Due to our construction of the energy functional, we have seen that the holomorphic texture $|\psi(\textbf{r})\rangle_0$ and hence $\textbf{n}_0 (\textbf{r})$ forms a locally stable minimum,
so we have a well defined collective mode (magnon) spectrum for fluctuations around $\textbf{n}_0 (\textbf{r})$.\\

We introduce small deviations such that 
$\textbf{n}(\textbf{r},t) = \textbf{n}_0(\textbf{r}) + \delta \textbf{n}(\textbf{r},t)$. 
First, we need to construct local coordinates $\chi_1(\textbf{r})$ and $\chi_2(\textbf{r})$ around $\textbf{n}_0(\textbf{r})$ on the sphere. 
To do so, we introduce local orthonormal frames ($\bm{n}_0$,$\bm{e}_1$,$\bm{e}_2$) such that 
$\bm{n}_0 (\bm{r}) = \bm{e}_1 (\bm{r}) \times \bm{e}_2 (\bm{r})$, using which we can write 
\begin{equation}
    \delta \textbf{n}^{(1)}(\textbf{r}) = \chi_1(\textbf{r}) \bm{e}_1(\textbf{r}) + \chi_2 \bm{e}_2(\textbf{r})
    \label{1rstorder}
\end{equation}
Since $\big| \textbf{n}(\textbf{r},t)\big|^2 = 1$, we get $\textbf{n}_0(\textbf{r}) \cdot \delta \textbf{n}^{(1)}(\textbf{r},t) = 0$.
To study collective modes, we need to expand the total energy to second order in $\chi_1(\textbf{r})$ and $\chi_2(\textbf{r})$. Normalizing $\textbf{n}(\textbf{r})$ and then expanding up to second order we get 
$\delta \textbf{n}=\delta \textbf{n}^{(1)} + \delta \textbf{n}^{(2)}$, with
\begin{equation}
    \delta \textbf{n}^{(2)}(\textbf{r}) =  - \frac{1}{2} [\chi_1(\textbf{r})^2 + \chi_2(\textbf{r})^2]\textbf{n}_0(\textbf{r})
    \label{2ndorder}
\end{equation}
Now, by expanding $\partial_i \textbf{n} \cdot \partial_i \textbf{n}$ to 2nd order and using the fact that $\textbf{n}_0(\textbf{r})$ is a local minimum of the energy functional we get the following expression for the energy functional
\begin{multline}
    E = E_0 + g \int \delta Q(\textbf{r})^2 + J \int [\partial_x (\delta \bm{n}^{(1)})]^2 + [\partial_y (\delta \bm{n}^{(1)})]^2 - \\
    [(\partial_x \textbf{n}_0)^2 + (\partial_y \textbf{n}_0)^2](\chi_1(\textbf{r})^2 + \chi_2(\textbf{r})^2)
    \label{efunc1}
\end{multline}
where $\delta Q$ is the 1st order variation of the topological charge density. Using eq \ref{2ndorder}, \ref{efunc1} and the holomorphic constraint arising from minimizing the exchange energy at fixed topological charge , we can write the change in energy as
\begin{multline}
    \delta E^{(2)} = g \int \delta Q(\textbf{r})^2 + J \int \big[ |i \partial_x \chi + A_x \chi |^2 + |i \partial_y \chi + A_y \chi|^2 \\
    -(c_{1x}^2 + c_{2x}^2)|\chi|^2 \big] dxdy
    \label{finalE}
\end{multline}
where $A_{x/y} = \hat{\bm{e}}_1 \cdot \partial_{x/y} \hat{\bm{e}}_2$ , $c_{1x/y} = \hat{\bm{e}}_{1} \cdot \partial_{x/y} \bm{n}_0$ and $c_{2x/y} = \hat{\bm{e}}_{2} \cdot \partial_{x/y} \bm{n}_0$. We find that the second order variation of the exchange term can be interpreted as the energy of a quantum particle described by a wave-function 
$\chi(\textbf{r})=\chi_{1}(\textbf{r})+i\chi_{2}(\textbf{r})$, $\bar{\chi(\textbf{r})}=\chi_{1}(\textbf{r})-i\chi_{2}(\textbf{r})$ 
and subject to a vector potential $\textbf{A}$, an effective magnetic field $B = 4 \pi Q_0$ and a scalar potential $c_{1x}^2 + c_{2x}^2$ (see appendix B and C for more details on the effect of the holomorphic constraint and gauge invariance of the energy functional). The physical origin of this effective magnetic field, as mentioned earlier, comes from the Berry phase picked up by the magnon when traversing through the skyrmion crystal.

In order to get linear equations of motion we now expand the standard Landau-Lifshitz equations to first order in $\delta n$. Since $\delta E/\delta n = 0$ for the configuration $\textbf{n}_0 (\textbf{r})$, the linearized version of the standard Landau-Lifshitz equations gives us 
\begin{equation}
    \alpha\dfrac{\partial}{\partial t} \delta n^a = 
    \epsilon^{abc} n^b \dfrac{\delta E^{(2)}}{\delta \delta n^c}
\end{equation}
where $\alpha=\hbar/(4\pi l_B^2)$, \cite{sondhiskyr, MoonPRB} assuming that
the Landau level filling factor $\nu$ remains everywhere close to 1.
Now, on using the equations derived in this section we can express the linearized equation in matrix form as
\begin{equation}
\begin{split}
    \alpha \dfrac{\partial}{\partial t} \begin{bmatrix}
    \chi_1 \\
    \chi_2 
    \end{bmatrix} &= \begin{bmatrix} 0 && -1 \\ 1 && 0 \end{bmatrix} \begin{bmatrix}\delta E^{(2)}/\delta \chi_1 \\ \delta E^{(2)}/\delta \chi_2 \end{bmatrix}\\
    \alpha \dfrac{\partial \chi}{\partial t} &= 2i \dfrac{\delta E}{\delta \bar{\chi}}\\
    \alpha \dfrac{\partial \bar{\chi}}{\partial t} &= -2i \dfrac{\delta E}{\delta \chi}\\
\end{split}
\label{schroeq}
\end{equation}
which is a time-dependent Schr\"odinger equation for the Bogoliubov-de Gennes like energy functional $E$.

\section{Effect of holomorphic constraint}
The holomorphic constraint results from minimizng the exchange energy at fixed total topological charge. Let's see how this arises. We denote $\bm{v}_x = \partial_x \bm{n}_0 $ and $\bm{v}_y = \partial_y \bm{n}_0$, both of these quantities belong to the plane perpendicular to $\bm{n}_0$, so we can regard them as 2-component vectors.\\
The exchange energy density is $\bm{v}_x^2 + \bm{v}_y^2$ and the local topological energy density is $1/(4\pi) \bm{v}_x \times \bm{v}_y = 1/(4\pi) J'(\bm{v}_x) \cdot \bm{v}_y = -1/(4\pi) \textbf{v}_x \cdot J'(\textbf{v}_y) $, where $J' = (0,-1; 1,0)$. Let us minimize $\bm{v}_x^2 + \bm{v}_y^2$ at fixed $\bm{v}_x \times \bm{v}_y$, i.e we extremize the function $(\bm{v}_x,\bm{v}_y) \rightarrow (\bm{v}_x^2 + \bm{v}_y^2)/2 - \lambda \bm{v}_x \times \bm{v}_y $, where $\lambda$ is a Lagrange multiplier. We get
\begin{align}
\begin{split}
    \bm{v}_x + \lambda J'(\bm{v}_y) &= 0 \\
    \bm{v}_y - \lambda J'(\bm{v}_x) &= 0
\end{split}
\end{align}
which implies that $\bm{v}_x^2 + \lambda^2J'^2(\bm{v}_x) = 0$ and $(1-\lambda^2)\bm{v}_x = 0$, so $\lambda = \pm 1$. Since $\bm{v}_y = \lambda J'(\bm{v}_x)$ and $\bm{v}_x \times \bm{v}_y = \lambda J'(\bm{v}_x) \cdot J'(\bm{v}_x)$, $\lambda = 1(-1)$ implies a positive (negative) local topological charge density. In our case $Q_0$ is positive, therefore the holomorphic constraint corresponds to $\lambda = 1$. This implies, $\bm{v}_y = J'(\bm{v}_x)$, so
\begin{equation}
    \begin{pmatrix}
        c_{1y}\\
        c_{2y}
    \end{pmatrix} = \begin{pmatrix}
        -c_{2x}\\
        c_{1x}
    \end{pmatrix}
    \label{holconstraint}
\end{equation}
where $c_{(1/2)(x/y)} = \bm{e}_{(1/2)} \cdot \partial_{(x/y)}\bm{n}_0$ as in the main text. Therefore 
\begin{align}
\begin{split}
    Q_0(\bm{r}) &= \frac{1}{4\pi} \bm{n}_0 \cdot (\partial_x \bm{n}_0 \times \partial_y \bm{n}_0)\\
    &= \frac{1}{4\pi}(c_{1x}c_{2y} - c_{1y}c_{2x})\\
    &= \frac{1}{4\pi} (c_{1x}^2 + c_{2x}^2)
    \label{holQ0}
\end{split}
\end{align}
\section{Checks for gauge-invariance}
In our choice of local frames in the spin-wave theory calculations, we have gauge freedom. Instead of choosing $\bm{e}_1, \bm{e}_2$ we could also choose $\bm{e}'_1 = \cos(\lambda (\bm{r}))\bm{e}_1 + \sin(\lambda (\bm{r}))\bm{e}_2$ and $\bm{e}'_2 = -\sin(\lambda (r))\bm{e}_1 + \cos(\lambda (\bm{r}))\bm{e}_2$. Then we would have
\begin{align}
\begin{split}
    \begin{pmatrix}
        \chi_1(\bm{r})\\
        \chi_2(\bm{r})
    \end{pmatrix} &= \begin{pmatrix}
       \cos(\lambda (\bm{r})) && -\sin(\lambda (\bm{r})) \\
       \sin(\lambda (\bm{r})) && \cos(\lambda (\bm{r}))
    \end{pmatrix} \begin{pmatrix}
        \chi'_1(\bm{r})\\
        \chi'_2(\bm{r})
    \end{pmatrix}\\
    \chi(\bm{r}) &= e^{i\lambda(\bm{r})}\chi'(\bm{r})
    \label{gauge1}
\end{split}
\end{align}
where $\chi = \chi_1 + i\chi_2$ as in the main text. Under the gauge transformation above, $(c_{1x},c_{2x})^T$ and $(c_{1y},c_{2y})^T$ transform as $(\chi_1,\chi_2)^T$. The relations in Eq. \ref{holconstraint} and \ref{holQ0} expressing the holomorphic nature of $\bm{n}_0(\bm{r})$ are preserved under gauge transformations, since $J'$ commutes with $R(\lambda)$ (the rotation matrix in the top line of the above equation). Now we look at the influence of a gauge transformation on the terms in the energy functional in eq \ref{finalE}. 
\begin{align}
\begin{split}
    A'_x &= \bm{e}'_1 \cdot \partial_x \bm{e}'_2 = \bm{e}'_1 \cdot (-\bm{e}'_1 \partial_x \lambda - \sin (\lambda) \partial_x \bm{e}_1 + \cos (\lambda) \partial_x \bm{e}_2)\\
    & = -\partial_x \lambda + A_x
\end{split}
\end{align}
and the same result holds for $A_y$. Together with Eq. \ref{gauge1} this implies that 
\begin{equation}
    \bm{\nabla}\chi - i\bm{A}\chi = e^{i\lambda}(\bm{\nabla}\chi' - i\bm{A}'\chi')
    \label{vecgaugeinv}
\end{equation}
which ensures that the all the exchange terms in the energy functional are gauge invariant. To show the gauge invariance of the $\delta Q$ terms let us first expand $\delta Q$ to first order in $\delta \bm{n}^{(1)}$ (see Eq. (\ref{1rstorder}) for expression). We can write
\begin{multline}
    4 \pi \delta Q = \delta \bm{n}^{(1)} \cdot (\partial_x \bm{n}_0 \times \partial_y \bm{n}_0) + \bm{n}_0 \cdot (\partial_x \delta \bm{n}^{(1)} \times \partial_y \bm{n}_0) + \\
    \bm{n}_0 \cdot (\partial_x \bm{n}_0 \times \partial_y \delta \bm{n}^{(1)})
\end{multline}
The first term vanishes, since $\partial_x \bm{n}_0$ and $\partial_y \bm{n}_0$ are both orthogonal to $\bm{n}_0$ as well as to $\delta \bm{n}^{(1)}$. To evaluate the last two terms we need to project $\partial_x \delta \bm{n}^{(1)}$ on the plane orthogonal to $\bm{n}_0$ which is equal to $(\partial_x \chi_1 + A_x \chi_2)e_1 + (\partial_x \chi_2 - A_x \chi_1)e_2$. Using this and Eq. \ref{holconstraint} we can write the 2nd term in the above equation as
\begin{align}
\begin{split}
    \bm{n}_0 \cdot &(\partial_x \delta \bm{n}^{(1)} \times \partial_y \bm{n}_0) = \begin{vmatrix}\partial_x \chi_1 + A_x \chi_2 && -c_{2x} \\
    \partial_x \chi_2 - A_x \chi_1 && c_{1x}
    \end{vmatrix}\\
    &= c_{1x}(\partial_x \chi_1 + A_x \chi_2) + c_{2x}(\partial_x \chi_2 - A_x \chi_1)
    \label{2term}
\end{split}
\end{align}
Similarly, once can also write the 3rd term as,
\begin{align}
\begin{split}
    \bm{n}_0 \cdot &(\partial_x \bm{n}_0 \times \partial_y \delta n^{(1)}) = \begin{vmatrix}c_{1x} && \partial_y \chi_1 + A_y \chi_2\\
    c_{2x} && \partial_y \chi_2 - A_y \chi_1
    \end{vmatrix}\\
    &= c_{1x}(\partial_y \chi_2 + A_y \chi_1) - c_{2x}(\partial_y \chi_1 + A_y \chi_2)
    \label{3term}
\end{split}
\end{align}
Now, we can use Eq. \ref{vecgaugeinv} to show that $(\partial_x \chi_1 + A_x \chi_2 , \partial_x \chi_2 - A_x \chi_1)^T$ and  $(\partial_y \chi_1 + A_y \chi_2 , \partial_y \chi_2 - A_y \chi_1)^T$ transform like $(\chi_1 , \chi_2)^T$. This ensures the gauge invariance of eqs \ref{2term} and \ref{3term}, since the determinant between two column vectors is invariant under rotations. Hence, this also ensures the gauge invariance of the $\delta Q$ term in the energy functional
\section{Gauge-fixing procedure}
\begin{figure}[t]
    \centering
    \includegraphics[scale = 0.35]{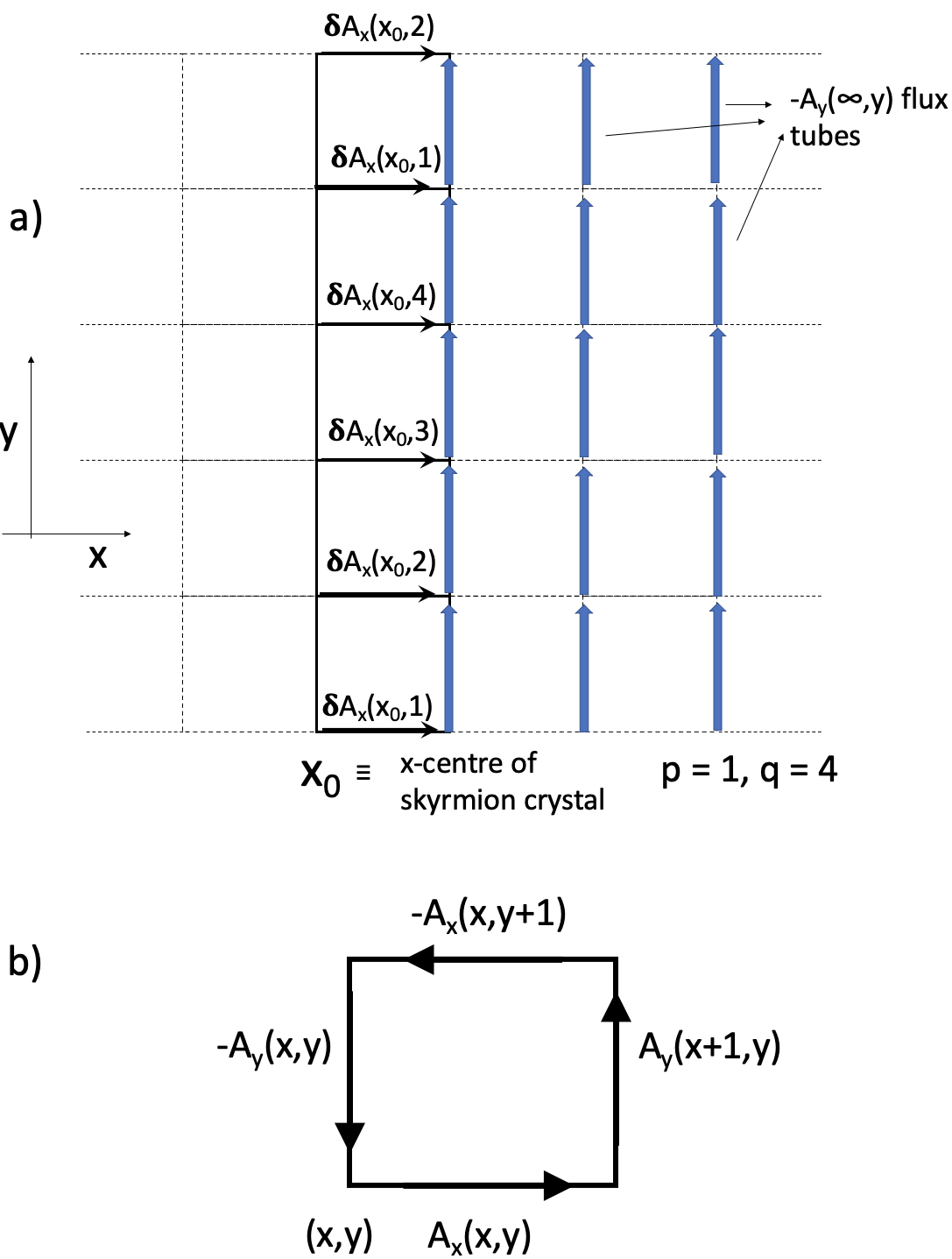}
    \caption{Gauge fixing procedure to ensure zero vector potential in the left and right ends. a)Illustration of the procedure explained in this section for the case $p = 1, q = 4$ b) Sign convention and notation for components of vector potential}
    \label{gaugefix}
\end{figure}
To mirror the problem of the experimentally relevant situation of a skyrmion crystal sandwiched between two ferromagnets, our heuristic model has to comprise a junction with zero vector potential on either side with a finite and varying vector potential in the central region. The heuristic model Hamiltonian with a finite and modulating magnetic field in the central region induces a vector potential which increases from zero to a non-zero finite value. However one can make a gauge transformation to ensure that the vector potential vanishes in the both the ends. In the Landau gauge 
\begin{equation}
    \Tilde{A}_y(x,y) = \int_{-\infty}^{x} B(x',y)dx', \ \Tilde{A}_x = 0
\end{equation}
while this vector potential has the same periodicity in $y$ as the magnetic field, $A_y(\infty,y) \neq 0$ and is also dependent on $y$. One can fix this, while keeping the same $y-$period $a/2$ of the magnetic field, provided the total flux within an infinite strip along $x$ of width $a/2$ along $y$ is an integer $p$ (in units of the flux quantum). Such a procedure is only required for the heuristic model and not the actual skyrmion crystal problem, since in the latter 
we can choose local frames $\bm{e}_{1,2}(\bm{r})$ such that the associated vector potential vanishes far away from interfaces inside both ferromagnetic regions. We also note that in a skyrmion crystal the condition for integer units of flux quantum within an infinite strip of half period is satisfied. \\
For the procedure, we introduce the following notations for the lattice discretization of the problem. We define the magnetic field associated with a plaquette as 
\begin{equation}
    B(x,y) \equiv A_x(x,y) + A_y(x+1,y) - A_x(x,y+1) - A_y(x,y)
\end{equation}
where $A_{(x/y)}$ is the vector field along the $\hat{x}/\hat{y}$ direction on the link originating from the lattice point $(x,y)$ (refer to Fig. \ref{gaugefix}b for sign convention). Now, we put flux tubes each carrying flux $-1$ at $x = x_0$ and $y = y_i + ma/2$, with $0 \leq y_1 < y_2 < ... < y_p \leq a/2-1$, $x_0$ the mid-point of the central region and $m$ an arbitrary integer. This singular flux configuration is described by the vector potential:
\begin{equation}
\begin{aligned}
    \delta A_x (x,y) &= 0, \ x \neq x_0 \\
    \delta A_y(x,y) &= 0, \ x \leq x_0 \\
    \delta A_y(x,y) &= -A_y(\infty,y), \ x \geq x_0 + 1
\end{aligned}
\end{equation}
The condition on fluxes reads:
\begin{equation}
    \delta A_x(x_0,y) - \delta A_x(x_0,y+1) - A_y(\infty,y) = -\sum_{i=1}{p}\sum_{m} \delta_{y,y_i + mq}
\end{equation}
Starting from an arbitrary $\delta A_x(x_0,y)$, these equations determine successively $\delta A_x (x_0,y \pm 1)$,$\delta A_x(x_0,y \pm 2)$ and so on. See Fig. \ref{gaugefix}a for a pictorial description of the flux addition procedure.

\section{Multi-channel scattering - role of interference}
\begin{figure}[t]
    \centering
    \includegraphics[scale = 0.3]{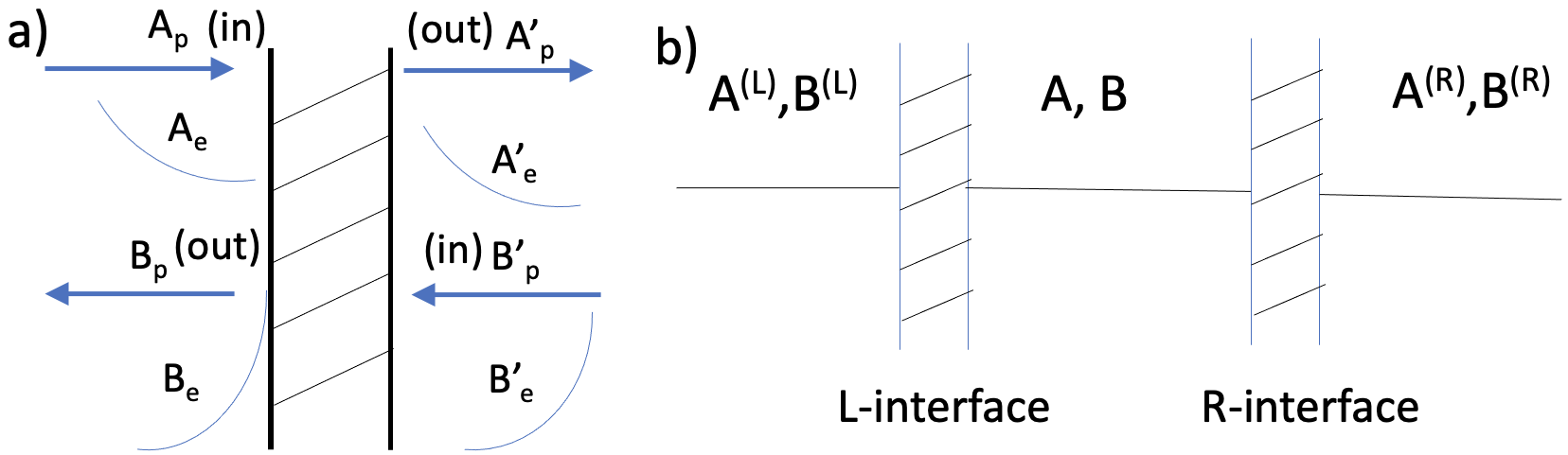}
    \caption{Incoming and outgoing amplitudes for a) A single interface problem and b) A double interface problem.}
    \label{intface}
\end{figure}
To understand the role of interference between channels, for simplicity, we will consider a 1D system with coordinate $x$ and $N$ internal states (transverse positions, for example). Consider $N$-component wave functions $\Psi (x) \in \mathcal{C}^N$, subjected to the Hamiltonian
\begin{equation}
    H = \frac{1}{2}(\overrightarrow{\partial_x} + iA(x))M^{-1}(x)(\overrightarrow{\partial_x} + iA(x)) + V(x)
\end{equation}
where $M(x)$ is a positive definite, real and symmetric $N \times N$ matrix, a space dependent effective mass. $V(x) = V^{\dagger} (x)$ plays the role of a local potential together with "hopping terms" in the transverse direction. Then $A(x) = A^{\dagger}(x)$ encodes an orbital generalized magnetic field.\\
Let $\Psi$ satisfy $H \Psi = E \Psi$ with $E \in \mathbb{R}$. Then, we have a current $J(x) = \frac{1}{2i} (\Psi^{\dagger} (x) M^{-1}(x) (\overrightarrow{\partial_x} + iA)\Psi(x) - \Psi^{\dagger}(x)(\overleftarrow{\partial_x} - iA) M^{-1}(x) \Psi(x))$ 
which is independent of $x$. \\
The Schrodinger equation $H \psi = E \psi$ is linear and second order in $\frac{d}{dx}$, therefore it has a $2N$-dimensional space of solutions. For any point $x$, a solution is uniquely determined by specifying $\Psi(x)$ and $\frac{d \Psi(x)}{dx}$ in $\mathbb{C}^N$. The current $J(x)$ may be seen as a Hermitian form 
\begin{equation}
    J = \frac{1}{2}\begin{pmatrix}
    \psi^{\dagger} &  \psi^{\dagger}(\overleftarrow{\partial_x} - iA)
    \end{pmatrix}\begin{pmatrix}
        0 & -iM^{-1} \\
        iM^{-1} & 0
    \end{pmatrix}
    \begin{pmatrix}
        \psi \\
        (\overrightarrow{\partial_x} + iA)\psi
    \end{pmatrix}
\end{equation}
The linear map $(\psi$,  $\partial_x \psi)^T \rightarrow (\psi $, $ (\partial_x + iA)\psi)^T$ is one-to-one, so as a hermitian form, J has the same signature as the middle matrix in the above equation. In a basis where $M^{-1}$ is diagonal, we see that the eigenvalues of the matrix are $\pm m_1, ...,\pm m_N$, where $m_i > 0$ for $1 \leq i \leq N$. So we get the important result that the signature of $J$ is of the type ($N,N$), at every point, independently of possible spatial variations of $M(x)$ and $A(x)$.\\
For a translationally invariant medium, we can look for plane-wave solutions $\psi(x) = e^{ikx}\psi$, with $k \in \mathbb{C}$. $\psi \in \mathbb{C}^N$ satisfies the eigenvalue equation
\begin{equation}
    \frac{1}{2}(k \mathds{1}^N + A)M^{-1}(k \mathds{1}^N + A)\psi + V\psi = E\psi
\end{equation}
Let us consider the current carried by such eigenstates. Since the current is conserved, it vanishes unless the wavevector is real. However, we can also get a finite current from evanescent modes if we take linear superpositions $\psi(x) = e^{ik_1x}\psi_1 + e^{ik_2x}\psi_2$ when $k_1 = k_2^*$. Since $J$ has signature $(N,N)$, we have $2p$ propagating modes ($0 \leq p \leq N$), with $p$ carrying a positive current and the other $p$ carrying a negative current. The remaining $2(N-p)$ evanescent modes are grouped in pairs of modes with complex conjugate momenta.\\
The propagating modes with positive current have momenta $k_1, \cdots, k_p$, and those with negative current have momenta 
$k'_1, \cdots, k'_p$. In a time-reversal invariant effective medium (as it is the case when the vector potential vanishes),
it is possible to label these momenta so that $k_j + k'_j = 0$. But in the presence of a non-zero vector potential, as inside a
Skyrmion crystal, there is no simple relation between the sets of $k_j$ and of $k'_j$ values. For evanescent modes, the corresponding momenta
form pairs $(k_j,k'_j)$ with $k'_j=k_j^*$ for $p+1 \leq j \leq N$. We shall always assume that $\Im(k_j) > 0$ for such modes.\\
A general scattering solution at energy $E$ can then be written in terms of $2N$ complex amplitudes $A_j,B_j$ where $1 \leq j \leq N$ as 
\begin{equation}
 \psi(x) = \sum_{j =1}^{N} A_j e^{ik_jx}\psi_j + \sum_{j=1}^{N}B_je^{k'_jx}\psi'_j   
\end{equation}
It is possible to normalize the eigenstates $\psi_j$,$\psi'_j$ such that 
\begin{align}
\begin{split}
    (\psi_j,J\psi_j) &= 1 = -(\psi'_j,J\psi'_j); 1 \leq j \leq p \\
    (\psi'_j,J\psi_j) &= 1 = (\psi_j,J\psi'_j); p+1 \leq j \leq N
\end{split}
\end{align}
and all other bilinears vanish. Then, using this normalization one gets 
\begin{equation}
    (\psi,J\psi) = \sum_{j=1}^{p}(|A_j|^2 - |B_j|^2) + \sum_{j = p+1}^{N}(A_j^*B_j + B_j^*A_j)
    \label{current_normalized}
\end{equation}\\

Let us first consider the problem of a single interface as shown in Fig. \ref{intface}(a). While the sign of the current determines the incoming and outgoing waves for the propagating channels, for evanescent channels we choose the waves which decay towards the interface as incoming and the ones which decay away from the interface as outgoing. The scattering matrix is then defined as 
\begin{align}
\begin{split}
    \begin{pmatrix}
        A'_p \\
        A'_e
    \end{pmatrix} &= \begin{pmatrix}
        t_{pp} & t_{pe} \\
        t_{ep} & t_{pp}
    \end{pmatrix} \begin{pmatrix}
        A_p \\
        A_e
    \end{pmatrix} + \begin{pmatrix}
        r'_{pp} & r'_{pe}\\
        r'_{ep} & r'_{ee}
    \end{pmatrix} \begin{pmatrix}
        B'_p \\
        B'_e
    \end{pmatrix}\\
    \begin{pmatrix}
        B_p \\
        B_e
    \end{pmatrix} &= \begin{pmatrix}
        r_{pp} & r_{pe} \\
        r_{ep} & r_{pp}
    \end{pmatrix} \begin{pmatrix}
        A_p \\
        A_e
    \end{pmatrix} + \begin{pmatrix}
        t'_{pp} & t'_{pe}\\
        t'_{ep} & t'_{ee}
    \end{pmatrix} \begin{pmatrix}
        B'_p \\
        B'_e
    \end{pmatrix}
\end{split}
\end{align}
Imposing $(\psi,J\psi) = (\psi',J'\psi')$ for any choice of incoming amplitudes and using Eq. \ref{current_normalized} gives the unitarity relations:
\begin{equation}
\begin{aligned}
    t^{\dagger}_{pp}t_{pp} + r^{\dagger}_{pp} r_{pp} &= \mathds{1}\\
    t^{\dagger}_{pe}t_{pp} + r^{\dagger}_{pe}r_{pp} &= r_{ep}\\
    t^{\dagger}_{pe}t_{pe} + r^{\dagger}_{pe}r_{pe} &= r_{ee} + r^{\dagger}_{ee}\\
\end{aligned}
\label{unitarity1}
\end{equation}
\begin{equation}
\begin{aligned}
    t^{\dagger}_{pp}r'_{pp} + r^{\dagger}_{pp} t'_{pp} &= 0\\
    t^{\dagger}_{pe}r'_{pp} + r^{\dagger}_{pe}t'_{pp} &= t'_{ep}\\
    t^{\dagger}_{pp}r'_{pe} + r^{\dagger}_{pp}t'_{pe} &= -t^{\dagger}_{ep} \\
    t^{\dagger}_{pe}r'_{pe} + r^{\dagger}_{pe}t'_{pe} &= t'_{ee} - t^{\dagger}_{ee}
\end{aligned}
\label{unitarity2}
\end{equation}
For the relevant problem of two interfaces (Fig. \ref{intface}(b)), one can write down a composition rule. To keep track of the distance $L$ between the two interfaces we write $Ae^{ikx}$ as $Ae^{ikL}e^{ik(x-L)}$. We may write 
\begin{equation}
    \begin{pmatrix}
        e^{ik^{(r)}L} A^{(r)} \\
        e^{ik'L} B
    \end{pmatrix} = \begin{pmatrix}
    t^{(r)} & r'^{(r)}\\
    r^{(r)} & t'^{(r)}
    \end{pmatrix} \begin{pmatrix}
        e^{ikL} A \\
        e^{ik'^{(r)}L} B^{(r)}
    \end{pmatrix}
\end{equation}
In particular $B = e^{-ik'L}r^{(r)}e^{ikL}A = r^{(r)}(L)A$, if $B^{(r)} = 0$. Our choice $\Im(k_j)>0$ and $\Im (k'_j) = -\Im (k_j) < 0$ for evanescent channels ensures that $r^{(r)}_{pe}(L)$, $r^{(r)}_{ep}(L)$, and $r^{(r)}_{ee}(L)$ decay exponentially with $L$. 
This is also the case for $t^{(r)}_{pe}(L)$ whereas $t^{(r)}_{pp}(L)$ oscillates with $L$. \\
The general composition law reads 
\begin{equation}
    t = t^{(r)}(L)(\mathds{1} - r'^{(l)}r^{(r)}(L))^{-1}t^{(l)}
\end{equation}
In the limit where $L \Im(k_j) \gg 1$ for all $p+1 \leq j \leq N$, we get for large $L$ 
\begin{equation}
    t \approx t^{(r)}(L) \Pi_p(\mathds{1} - r'^{(l)}_{pp}r^{(r)}_{pp}(L))^{-1}\Pi_p t^{(l)}
    \label{large_L_trans}
\end{equation}
where we have introduced the rank $p$ projector $\Pi_p$ on the subset of propagating channels inside the intermediate region.
From the above equation we can directly see that if $p = 0$, i.e all channels are evanescent, in the large $L$ limit, there is no transmission. Moreover, if there is only one propagating channel, $p = 1$, and $|r^{(l)}_{pp}|$ and $|r^{(r)}_{pp}|$ are close to 1, sharp resonances with maximal transmission are possible. However, when $p \geq 2$, interference between the various propagating channels in the intermediate region decreases the maximal transmission at resonances.\\
To illustrate this point further, we note that the unitary relations \ref{unitarity1}, \ref{unitarity2} imply that the scattering sub-matrix
$S_{pp}$ associated to an interface and defined by:
\begin{equation}
S_{pp}=
\begin{pmatrix}
        r_{pp} & t'_{pp} \\
        t_{pp} & r'_{pp}
\end{pmatrix}
\end{equation}
is unitary. 

From Eq. \ref{large_L_trans}, we see that resonances may occur when $t^{(l)}_{pp}$ and $t^{(r)}_{pp}$ are small. When $t$ is small, we can write a unitary scattering matrix as
\begin{equation}
    S \cong \begin{pmatrix}
        r_0 (\mathds{1}  - \frac{1}{2}t^{\dagger}t) & -r_0 t^{\dagger} r'_0 \\
        t & (\mathds{1} - \frac{1}{2}tt^{\dagger})r'_0
    \end{pmatrix}
\end{equation}
where $r_0$ and $r'_0$ are unitary matrices and all entries of $t$ are small of order $\epsilon$. Then $S^{\dagger}S = \mathds{1}+O(\epsilon^3)$. Using this parameterization for $S^{(l)}_{pp}$ and $S^{(r)}_{pp}$ and dropping the $pp$ subscript for notational convenience, Eq. \ref{large_L_trans} becomes:
\begin{multline}
    t \cong t^{(r)}e^{ikL}(\mathds{1} - (\mathds{1} - t^{(l)}t^{\dagger (l)}/2)r'^{(l)}_0e^{-ik'L}r^{(r)}_0 \\
    \times
    (\mathds{1} - \frac{1}{2}t^{\dagger(r)}t^{(r)})e^{ikL})^{-1}t^{(L)}
\end{multline}
So, the condition for resonance now selects the energies at which the unitary matrix $r'^{(l)}_0 e^{-ik'(E)L}r_0^{(r)}e^{ik(E)L}$ has an eigenvalue equal to 1. When the number of propagating channels is at least two, we expect that the behavior of the transmission
near these resonances is going to be significantly more complex than for a single propagating channel. 

\section{Tight-Binding model and forms of the matrix elements}
In the main text we saw that the Schr\"odinger equation could be expressed as a tight binding equation relating the wavefunctions of a slice to those to its left and right. On discretizing the energy functional and then taking the derivative we get a tight-binding problem with nearest and next nearest neighbour hoppings. In this section we give illustrative examples of how the matrix elements for the matrices relating the different slices look. The nearest and next nearest neighbour from the right contributions will enter as matrix elements in the $A_R$ matrix, similar contributions from the left will enter in the $A_L$ matrix and onsite contributions and nearest neighbour contributions from above and below will enter in the $D$ matrix. Let us look at some of the forms of these matrix elements. \\
First consider the exchange term. One can discretize this term by simply writing the contribution from the $i^{\rm{th}}$ site as 
\begin{equation}
    E_{Ji} = J(\bm{n}_i - \bm{n}_{i+1})^2
\end{equation}
exapnding the above equation we get constants plus an $\bm{n}_i \cdot \bm{n}_{i+1}$ term. One can expand this term by expressing the $\bm{n}_i$ in terms of $\chi_1$ and $\chi_2$ by using Eqs. (\ref{1rstorder}), (\ref{2ndorder}) in appendix A. On doing so, and keeping upto $\rm{O}(\chi_{1/2}^2)$ terms one finds that
\begin{align}
    \frac{\partial E_{ji}}{\partial \chi_{1i}} &= \bm{n}_{0i} \cdot \bm{n}_{0j} \chi_{1i} - \bm{e}_{1i} \cdot (\chi_{1j}\bm{e}_{1j} + \chi_{2j}\bm{e}_{2j}) \\
    \frac{\partial E_{ji}}{\partial \chi_{2i}} &= \bm{n}_{0i} \cdot \bm{n}_{0j} \chi_{2i} - \bm{e}_{2i} \cdot (\chi_{1j}\bm{e}_{1j} + \chi_{2j}\bm{e}_{2j}) 
\end{align}
from the above two equations one can directly read out the nearest neighbour and on-site contributions from the coefficients of $\chi_{1/2 j}$ and $\chi_{1/2 i}$ respectively. The exchange term does not induce next-nearest neighbor hopping. Now, after discretizing the topological charge terms as in the main text, we obtained a tight-binding model with hopping terms up to second nearest neighbours. Let us consider one term from the first line in Eq. \ref{varsoltight} of the main text. The derivative of this term can be expressed as 
\begin{align}
\begin{split}
    \frac{\partial \delta \Omega_{01}^2}{\partial \chi_{10}} &= 2 \delta \Omega_{01} \frac{\partial \delta \Omega_{01}}{\partial \chi_{10}}\\
    &= 2 \delta \Omega_{01} (-f_{\alpha_{01}}\bm{z}_{01} \cdot \bm{e}_{10})
\end{split}
\end{align}
where $f_{\alpha_{01}} = \sin(\alpha_{01})/(1+\cos(\alpha_{01}))$, and $\bm{n}_0 \times \bm{n}_1 = \sin(\alpha_{01})\bm{z}_{01}$. One can then expand $\delta \Omega_{01}$ using Eq. \ref{varsolid2} in the main text, and then read off the coefficients same as above. A similar procedure can be used for all the other terms in Eq. \ref{varsoltight}.
\section{Scaling functions for discretization}
Th real space discretization procedure outlined in the main text requires each coupling constant to be scaled by a factor, so that the results are independent of $N$ is the large $N$ limit.\\
From the standard finite-difference type discretization scheme for the exchange terms, we know that the denominator will be $(a/N)^2$ because of the double derivative, where $a/N$ is the grid-size. Therefore, to get the correct continuum limit $J$ should be multiplied  by $(N/a)^2$. Similarly from the expression of the topological charge density, we can see that the denominator will be $(a/N)^4$, hence $g$ should be multiplied by $(N/a)^4$. \\
While the argument above for the exchange term is pretty well known, the argument for the scaling of the $g$ term might be a bit too simplistic. In which case one can also come up with a more sophisticated argument with the same result. Let us denote the topological charge of the $\bm{n}_0$ field generated by the theta functions in section III, over a plaquette, to be $Q_{\square}$. If we change $\bm{n}_0$ to $\bm{n}_0 + \bm{n}_1$, where $\bm{n}_1$ is some small deviation such that $\bm{n}_0 \cdot \bm{n}_1 = 0$ everywhere, $Q_{\square}$ is changed into $Q_{\square} + \triangle Q_{\square}$, where 
\begin{equation}
   \triangle Q_{\square} = \frac{1}{4 \pi} \oint \bm{n}_0 \cdot \bigg(\bm{n}_1 \times \frac{\partial \bm{n}_0}{\partial u}\bigg)du 
\end{equation}
Here, the integral is taken along the boundary of the above square plaquette and $u$ is an arbitrary parameter on this boundary. It is convenient to write $\bm{n}_1 = \bm{v}(\bm{r}) \times \bm{n}_0 (\bm{r})$, where $\bm{v}(\bm{r})$ is an infinitesimal rotation vector. Using the fact that $\bm{n}_0 \cdot \partial \bm{n}_0 / \partial u = 0$, we get
\begin{equation}
    \triangle Q_{\square} =  \frac {1}{4 \pi} \oint \bm{v} \cdot \frac{\partial \bm{n}_0}{\partial u}du
\end{equation}
Using Green's equation one can express the above as an integral over the whole plaquette as:
\begin{equation}
    \triangle Q_{\square} =  \frac{1}{4 \pi} \int \int_{\square} dxdy \bigg(\frac{\partial \bm{v}}{\partial x} \cdot\frac{\partial \bm{v}}{\partial y} - \frac{\partial \bm{v}}{\partial y} \cdot\frac{\partial \bm{n}_0}{\partial x}\bigg)
\end{equation}
As in the next section we check that if $v$ is constant in space (global rotation in spin space), $\triangle Q_0 = 0$. Also, we see that $\triangle Q_{\square}$ is expected to be proportional to the plaquette area $(a/N)^2$, when $N$ is large and $\bm{n}_0$ and $\bm{v}$ are smooth fields. Therefore, we may write
\begin{equation}
\triangle Q_{\square} = \delta \rho (a/N)^2, 
\end{equation}
with $\delta \rho$ being the variation of the local topological charge density. Therefore the 1st term in Eq. \ref{finalE} should scale as (in the large $N$ limit, which is the relevant limit for numerics)
\begin{align}
    \begin{split}
        g \int (\delta \rho)^2 dxdy &\approx g \sum_{\rm{plaq}}[\triangle Q_{\square} (\frac{N}{a}^2)]^2(\frac{a}{N})^2 \\
        &\approx g' \sum_{\rm{plaq}} (\triangle Q_{\square})^2 (\frac{a}{N})^2
    \end{split}
\end{align}
Hence, we see that the scaled version should be $g' = g(N/a)^4$
\section{Test for topological charge discretization scheme}
To test whether our geodesic scheme for discretizing the topological charge density is correct, we perform the following non-trivial check. As in the last section we take an infinitesimal rotation vector $\bm{v}$, constant in space, and rotate the ground state spin vector $\bm{n}_0({\bm{r}})$. On doing so, we can define new variables $\chi'_1$ and $\chi'_2$ which are related to the old variables by
\begin{align}
    \begin{split}
        \chi'_1 (\bm{r}) &= \bm{v} \cdot \bm{e}_2(\bm{r})\\
        \chi'_2(\bm{r}) &= -\bm{v} \cdot \bm{e}_1(\bm{r})
    \end{split}
\end{align}
Using the above expressions we form a column vector of the $\chi' (\bm{r}) = \chi'_1(\bm{r}) + i \chi'_2(\bm{r})$ and $\bar{\chi}'(\bm{r}) =  \chi'_1(\bm{r})(\bm{r}) - i \chi'_2(\bm{r})$ from all the sites. We then right multiply the Hamiltonian constructed from only the topological charge density term ($J = 0$) and multiply it with this vector. If the discretization scheme is correct, then this product should be zero, since an infinitesimal global rotation should not induce any variation of the topological charge density.  We have checked this in our calculations and indeed it does return a column of values which are for all intents and purposes zero ($O(1e-16)$).
\section{Boundary conditions for spectra and relation between $\chi$ and $\bar{\chi}$}
To obtain the spectra of the skyrmion crystal, we considered periodic boundary conditions along both $x$ and $y$- axes of an $a \times a$ unit cell. For the tight-binding model after taking the derivative of the discretized energy functional, this implies that the for the right(left)-most site, the right(left) nearest neighbor contribution will pick up a $e^{iq_xa}(e^{-iq_{x}a})$ phase-factor and similarly for the top(bottom)-most site, the top(bottom) nearest neighbor contribution will pick up a $e^{iq_y a}(e^{-iq_ya})$ phase factor. The phase factors encode how the momentum dependence enters the Hamiltonian matrix. The Hamiltonian is constructed in the site basis, so if there are $N$ rows and columns each in the unit cell, the Hamiltonian has a size $2N^2 \times 2N^2$, where the factor of $2$ comes because of the presence of both $\chi$ and $\bar{\chi}$. Each diagonal $2N \times 2N$ block of the Hamiltonian comprises the on-site, right and left nearest neighbor contributions that come from that particular row. The off-diagonal blocks comprise the up and down nearest neighbor terms as well as the second nearest neighbor contributions. \\
An important point to note while doing these calculations for the spectra and the scattering problem is that $\chi$ and $\bar{\chi}$ aren't always complex conjugates of one another. To see this remember that from appendix A, $\chi = \chi_1 + i\chi_2$ and $\bar{\chi} = \chi_1 - i\chi_2$, however, both $\chi_1$ and $\chi_2$ pickup complex phase factors $e^{\pm iq_{x/y}a}$ due to the boundary conditions as described above. For $q_x,q_y = 0$, the relation $\chi = \bar{\chi}^*$ holds since $\chi_1$ and $\chi_2$ are real. However, generally, this is not the case, since $\chi_1,\chi_2 \in \mathbb{C}$, and so $\chi \neq \bar{\chi}^*$.
\section{Lie-algebra structure in non-linear sigma model}
In this appendix we give details on some of the calculations to show the Lie-algebra structure of the vector fields mentioned in section X of the main text. We showed that $\bm{x} \rightarrow X_{\bm{\epsilon}(\bm{x})}$ is a vector field associated to the infinitesimal left rotation $\exp(-i \bm{\epsilon} \cdot \bm{\sigma}/2)$. We know that vector fields form a Lie-algebra under the Lie-bracket. The Lie-bracket is defined as
\begin{equation}
    L_{[\bm{X},\bm{Y}]}(f) = (L_{\bm{X}}L_{\bm{Y}} - L_{\bm{Y}}L_{\bm{X}})f 
\end{equation}
for any arbitrary function $f$, where $L_{\bm{X}}f \equiv \sum_i X^i\partial_if$ denotes the Lie derivative of f along vector field $\bm{X}$. We can write the Lie derivative as
\begin{align}
\begin{split}
    &(L_{\bm{X}}L_{\bm{Y}} - L_{\bm{Y}}L_{\bm{X}})f = X^i \partial_i (Y^j \partial_jf) - Y^j \partial_j(X^i \partial_i f) \\
    &=(X^i \partial_iY^j - Y^i\partial_iX^j)\partial_j f \equiv  [\bm{X},\bm{Y}]^j \partial_j f
    \end{split}
\end{align}
therefore we get $[\bm{X},\bm{Y}] = \bm{X} \cdot \bm{\nabla}\bm{Y} - \bm{Y} \cdot \bm{\nabla}\bm{X} \equiv Y'(X) - X'(Y)$ where $\bm{X}'$ denotes the Jacobian matrix $(X')^j_i \equiv \partial_iX^j$. Now, let us compute $[\bm{X}_{\bm{\epsilon}},\bm{X}_{\bm{\eta}}]$. Using Eq. \ref{defXepsilon} from the main text we can write
\begin{align}
\begin{split}
    \bm{X}'_{\bm{\eta}}(\bm{X}_{\bm{\epsilon}}) &= -\dfrac{\bm{x}}{\sqrt{1-\bm{x}^2}} \cdot (\sqrt{1-\bm{x}^2}\frac{\bm{\epsilon}}{2} - \frac{\bm{\epsilon}}{2} \times \bm{x})\frac{\bm{\eta}}{2}\\
    &- \dfrac{\bm{\eta}}{2} \times (\sqrt{1-\bm{x}^2}\frac{\bm{\epsilon}}{2} - \frac{\bm{\epsilon}}{2} \times \bm{x})\\
    &= -\frac{1}{4}\bigg((\bm{x}\cdot \bm{\epsilon})\bm{\eta} + \sqrt{1-\bm{x}^2}\,\bm{\eta} \times \bm{\epsilon} - \bm{\eta} \times (\bm{\epsilon}\times \bm{x}) \bigg)
\end{split}
\end{align}
Similarly one can also write 
\begin{equation}
    \bm{X}'_{\bm{\epsilon}}(\bm{X}_{\bm{\eta}}) = -\frac{1}{4}\bigg((\bm{x}\cdot \bm{\eta})\bm{\epsilon} + \sqrt{1-\bm{x}^2}\,\bm{\epsilon} \times \bm{\eta} - \bm{\epsilon} \times (\bm{\eta}\times \bm{x}) \bigg)
\end{equation}
Using the above two equations one gets 
\begin{align}
    \begin{split}       [\bm{X}_{\bm{\epsilon}},\bm{X}_{\bm{\eta}}] &= \frac{1}{4}\bigg(\big((\bm{x}\cdot \bm{\eta})\bm{\epsilon}-(\bm{x}\cdot \bm{\epsilon})\bm{\eta}\big) + 2\sqrt{1-\bm{x}^2}\bm{\epsilon} \times \bm{\eta} \\
    &- \big(\bm{\epsilon} \times (\bm{\eta}\times \bm{x}) -\bm{\eta} \times (\bm{\epsilon}\times \bm{x}) \big) \bigg)\\
    &= \sqrt{1-\bm{x}^2}\,\dfrac{\bm{\epsilon}\times \bm{\eta}}{2} - \frac{1}{2}(\bm{\epsilon}\times \bm{\eta})\times \bm{x}\\
    &= \bm{X}_{\bm{\eta} \times \bm{\epsilon}}
    \end{split}
\end{align}
and hence, we recover the Lie-algebra structure of $\rm{SO}(3)$. Similarly for the right action we get,
\begin{equation}
    [\bm{Y}_{\bm{\epsilon}},\bm{Y}_{\bm{\eta}}] = -\bm{Y}_{\bm{\epsilon} \times \bm{\eta}} = \bm{Y}_{\bm{\eta} \times \bm{\epsilon}}
\end{equation}\\
Let us examine the correspondence between Poisson brackets $\{g,h\}$ and Lie brackets $[X_g,X_h]$ of their associated 
Hamiltonian vector fields $X_g$ and $X_h$. Hamilton's equations, relating $X_g$ to $g$,  
are equivalent to requiring $L_{X_g}f = \{f,g\}$ for any function $f$ over phase-space. Then we have that \begin{align}
\begin{split}
    L_{X_g}L_{X_h} (f) &= \{\{f,h\},g\} \\
    L_{X_h}L_{X_g} (f) &= \{\{f,g\},h\}
\end{split}
\end{align}
hence one can express the Lie-bracket as
\begin{align}
\begin{split}
    L_{[X_g,X_h]}(f) &= \{\{f,h\},g\} + \{\{g,f\},h\} = -\{\{h,g\},f\}  \\
    &= = \{f,\{h,g\}\} = L_{X_{\{h,g\}}}(f)\\
    [X_g,X_h] &= X_{\{h,g\}}
\end{split}
\end{align}
where in the 2nd equality of the first line in the above equation we have used the Jacobi identity.
\bibliography{Bib.bib}

\begin{thebibliography}{62}%
\makeatletter
\providecommand \@ifxundefined [1]{%
 \@ifx{#1\undefined}
}%
\providecommand \@ifnum [1]{%
 \ifnum #1\expandafter \@firstoftwo
 \else \expandafter \@secondoftwo
 \fi
}%
\providecommand \@ifx [1]{%
 \ifx #1\expandafter \@firstoftwo
 \else \expandafter \@secondoftwo
 \fi
}%
\providecommand \natexlab [1]{#1}%
\providecommand \enquote  [1]{``#1''}%
\providecommand \bibnamefont  [1]{#1}%
\providecommand \bibfnamefont [1]{#1}%
\providecommand \citenamefont [1]{#1}%
\providecommand \href@noop [0]{\@secondoftwo}%
\providecommand \href [0]{\begingroup \@sanitize@url \@href}%
\providecommand \@href[1]{\@@startlink{#1}\@@href}%
\providecommand \@@href[1]{\endgroup#1\@@endlink}%
\providecommand \@sanitize@url [0]{\catcode `\\12\catcode `\$12\catcode
  `\&12\catcode `\#12\catcode `\^12\catcode `\_12\catcode `\%12\relax}%
\providecommand \@@startlink[1]{}%
\providecommand \@@endlink[0]{}%
\providecommand \url  [0]{\begingroup\@sanitize@url \@url }%
\providecommand \@url [1]{\endgroup\@href {#1}{\urlprefix }}%
\providecommand \urlprefix  [0]{URL }%
\providecommand \Eprint [0]{\href }%
\providecommand \doibase [0]{http://dx.doi.org/}%
\providecommand \selectlanguage [0]{\@gobble}%
\providecommand \bibinfo  [0]{\@secondoftwo}%
\providecommand \bibfield  [0]{\@secondoftwo}%
\providecommand \translation [1]{[#1]}%
\providecommand \BibitemOpen [0]{}%
\providecommand \bibitemStop [0]{}%
\providecommand \bibitemNoStop [0]{.\EOS\space}%
\providecommand \EOS [0]{\spacefactor3000\relax}%
\providecommand \BibitemShut  [1]{\csname bibitem#1\endcsname}%
\let\auto@bib@innerbib\@empty
%</preamble>
\bibitem [{\citenamefont {Anderson}(2018)}]{anderson2018basic}%
  \BibitemOpen
  \bibfield  {author} {\bibinfo {author} {\bibfnamefont {P.~W.}\ \bibnamefont
  {Anderson}},\ }\href@noop {} {\emph {\bibinfo {title} {Basic notions of
  condensed matter physics}}}\ (\bibinfo  {publisher} {CRC Press},\ \bibinfo
  {year} {2018})\BibitemShut {NoStop}%
\bibitem [{\citenamefont {Moessner}\ and\ \citenamefont
  {Moore}(2021)}]{moessner2021topological}%
  \BibitemOpen
  \bibfield  {author} {\bibinfo {author} {\bibfnamefont {R.}~\bibnamefont
  {Moessner}}\ and\ \bibinfo {author} {\bibfnamefont {J.~E.}\ \bibnamefont
  {Moore}},\ }\href@noop {} {\emph {\bibinfo {title} {Topological phases of
  matter}}}\ (\bibinfo  {publisher} {Cambridge University Press},\ \bibinfo
  {year} {2021})\BibitemShut {NoStop}%
\bibitem [{\citenamefont {Sondhi}\ \emph {et~al.}(1993)\citenamefont {Sondhi},
  \citenamefont {Karlhede}, \citenamefont {Kivelson},\ and\ \citenamefont
  {Rezayi}}]{sondhiskyr}%
  \BibitemOpen
  \bibfield  {author} {\bibinfo {author} {\bibfnamefont {S.~L.}\ \bibnamefont
  {Sondhi}}, \bibinfo {author} {\bibfnamefont {A.}~\bibnamefont {Karlhede}},
  \bibinfo {author} {\bibfnamefont {S.~A.}\ \bibnamefont {Kivelson}}, \ and\
  \bibinfo {author} {\bibfnamefont {E.~H.}\ \bibnamefont {Rezayi}},\ }\href
  {\doibase 10.1103/PhysRevB.47.16419} {\bibfield  {journal} {\bibinfo
  {journal} {Phys. Rev. B}\ }\textbf {\bibinfo {volume} {47}},\ \bibinfo
  {pages} {16419} (\bibinfo {year} {1993})}\BibitemShut {NoStop}%
\bibitem [{\citenamefont {Brey}\ \emph {et~al.}(1995)\citenamefont {Brey},
  \citenamefont {Fertig}, \citenamefont {C\^ot\'e},\ and\ \citenamefont
  {MacDonald}}]{Breyskyr}%
  \BibitemOpen
  \bibfield  {author} {\bibinfo {author} {\bibfnamefont {L.}~\bibnamefont
  {Brey}}, \bibinfo {author} {\bibfnamefont {H.~A.}\ \bibnamefont {Fertig}},
  \bibinfo {author} {\bibfnamefont {R.}~\bibnamefont {C\^ot\'e}}, \ and\
  \bibinfo {author} {\bibfnamefont {A.~H.}\ \bibnamefont {MacDonald}},\ }\href
  {\doibase 10.1103/PhysRevLett.75.2562} {\bibfield  {journal} {\bibinfo
  {journal} {Phys. Rev. Lett.}\ }\textbf {\bibinfo {volume} {75}},\ \bibinfo
  {pages} {2562} (\bibinfo {year} {1995})}\BibitemShut {NoStop}%
\bibitem [{\citenamefont {Rajaraman}(1982)}]{rajaraman1982solitons}%
  \BibitemOpen
  \bibfield  {author} {\bibinfo {author} {\bibfnamefont {R.}~\bibnamefont
  {Rajaraman}},\ }\href@noop {} {\bibfield  {journal} {\bibinfo  {journal}
  {Solitons and instantons}\ } (\bibinfo {year} {1982})}\BibitemShut {NoStop}%
\bibitem [{\citenamefont {Mühlbauer}\ \emph {et~al.}(2009)\citenamefont
  {Mühlbauer}, \citenamefont {Binz}, \citenamefont {Jonietz}, \citenamefont
  {Pfleiderer}, \citenamefont {Rosch}, \citenamefont {Neubauer}, \citenamefont
  {Georgii},\ and\ \citenamefont {Böni}}]{muhlbauer2009skyrmion}%
  \BibitemOpen
  \bibfield  {author} {\bibinfo {author} {\bibfnamefont {S.}~\bibnamefont
  {Mühlbauer}}, \bibinfo {author} {\bibfnamefont {B.}~\bibnamefont {Binz}},
  \bibinfo {author} {\bibfnamefont {F.}~\bibnamefont {Jonietz}}, \bibinfo
  {author} {\bibfnamefont {C.}~\bibnamefont {Pfleiderer}}, \bibinfo {author}
  {\bibfnamefont {A.}~\bibnamefont {Rosch}}, \bibinfo {author} {\bibfnamefont
  {A.}~\bibnamefont {Neubauer}}, \bibinfo {author} {\bibfnamefont
  {R.}~\bibnamefont {Georgii}}, \ and\ \bibinfo {author} {\bibfnamefont
  {P.}~\bibnamefont {Böni}},\ }\href
  {https://www.science.org/doi/abs/10.1126/science.1166767} {\bibfield
  {journal} {\bibinfo  {journal} {Science}\ }\textbf {\bibinfo {volume}
  {323}},\ \bibinfo {pages} {915} (\bibinfo {year} {2009})}\BibitemShut
  {NoStop}%
\bibitem [{\citenamefont {Yu}\ \emph {et~al.}(2010)\citenamefont {Yu},
  \citenamefont {Onose}, \citenamefont {Kanazawa}, \citenamefont {Park},
  \citenamefont {Han}, \citenamefont {Matsui}, \citenamefont {Nagaosa},\ and\
  \citenamefont {Tokura}}]{yu2010real}%
  \BibitemOpen
  \bibfield  {author} {\bibinfo {author} {\bibfnamefont {X.}~\bibnamefont
  {Yu}}, \bibinfo {author} {\bibfnamefont {Y.}~\bibnamefont {Onose}}, \bibinfo
  {author} {\bibfnamefont {N.}~\bibnamefont {Kanazawa}}, \bibinfo {author}
  {\bibfnamefont {J.~H.}\ \bibnamefont {Park}}, \bibinfo {author}
  {\bibfnamefont {J.}~\bibnamefont {Han}}, \bibinfo {author} {\bibfnamefont
  {Y.}~\bibnamefont {Matsui}}, \bibinfo {author} {\bibfnamefont
  {N.}~\bibnamefont {Nagaosa}}, \ and\ \bibinfo {author} {\bibfnamefont
  {Y.}~\bibnamefont {Tokura}},\ }\href
  {https://www.nature.com/articles/nature09124} {\bibfield  {journal} {\bibinfo
   {journal} {Nature}\ }\textbf {\bibinfo {volume} {465}},\ \bibinfo {pages}
  {901} (\bibinfo {year} {2010})}\BibitemShut {NoStop}%
\bibitem [{\citenamefont {Gervais}\ \emph {et~al.}(2005)\citenamefont
  {Gervais}, \citenamefont {Stormer}, \citenamefont {Tsui}, \citenamefont
  {Kuhns}, \citenamefont {Moulton}, \citenamefont {Reyes}, \citenamefont
  {Pfeiffer}, \citenamefont {Baldwin},\ and\ \citenamefont
  {West}}]{gervais2005evidence}%
  \BibitemOpen
  \bibfield  {author} {\bibinfo {author} {\bibfnamefont {G.}~\bibnamefont
  {Gervais}}, \bibinfo {author} {\bibfnamefont {H.}~\bibnamefont {Stormer}},
  \bibinfo {author} {\bibfnamefont {D.}~\bibnamefont {Tsui}}, \bibinfo {author}
  {\bibfnamefont {P.}~\bibnamefont {Kuhns}}, \bibinfo {author} {\bibfnamefont
  {W.}~\bibnamefont {Moulton}}, \bibinfo {author} {\bibfnamefont
  {A.}~\bibnamefont {Reyes}}, \bibinfo {author} {\bibfnamefont
  {L.}~\bibnamefont {Pfeiffer}}, \bibinfo {author} {\bibfnamefont
  {K.}~\bibnamefont {Baldwin}}, \ and\ \bibinfo {author} {\bibfnamefont
  {K.}~\bibnamefont {West}},\ }\href
  {https://journals.aps.org/prl/abstract/10.1103/PhysRevLett.94.196803}
  {\bibfield  {journal} {\bibinfo  {journal} {Physical review letters}\
  }\textbf {\bibinfo {volume} {94}},\ \bibinfo {pages} {196803} (\bibinfo
  {year} {2005})}\BibitemShut {NoStop}%
\bibitem [{\citenamefont {Desrat}\ \emph {et~al.}(2002)\citenamefont {Desrat},
  \citenamefont {Maude}, \citenamefont {Potemski}, \citenamefont {Portal},
  \citenamefont {Wasilewski},\ and\ \citenamefont
  {Hill}}]{desrat2002resistively}%
  \BibitemOpen
  \bibfield  {author} {\bibinfo {author} {\bibfnamefont {W.}~\bibnamefont
  {Desrat}}, \bibinfo {author} {\bibfnamefont {D.}~\bibnamefont {Maude}},
  \bibinfo {author} {\bibfnamefont {M.}~\bibnamefont {Potemski}}, \bibinfo
  {author} {\bibfnamefont {J.}~\bibnamefont {Portal}}, \bibinfo {author}
  {\bibfnamefont {Z.}~\bibnamefont {Wasilewski}}, \ and\ \bibinfo {author}
  {\bibfnamefont {G.}~\bibnamefont {Hill}},\ }\href
  {https://journals.aps.org/prl/abstract/10.1103/PhysRevLett.88.256807}
  {\bibfield  {journal} {\bibinfo  {journal} {Physical review letters}\
  }\textbf {\bibinfo {volume} {88}},\ \bibinfo {pages} {256807} (\bibinfo
  {year} {2002})}\BibitemShut {NoStop}%
\bibitem [{\citenamefont {Bayot}\ \emph {et~al.}(1996)\citenamefont {Bayot},
  \citenamefont {Grivei}, \citenamefont {Melinte}, \citenamefont {Santos},\
  and\ \citenamefont {Shayegan}}]{Bayotprl}%
  \BibitemOpen
  \bibfield  {author} {\bibinfo {author} {\bibfnamefont {V.}~\bibnamefont
  {Bayot}}, \bibinfo {author} {\bibfnamefont {E.}~\bibnamefont {Grivei}},
  \bibinfo {author} {\bibfnamefont {S.}~\bibnamefont {Melinte}}, \bibinfo
  {author} {\bibfnamefont {M.~B.}\ \bibnamefont {Santos}}, \ and\ \bibinfo
  {author} {\bibfnamefont {M.}~\bibnamefont {Shayegan}},\ }\href {\doibase
  10.1103/PhysRevLett.76.4584} {\bibfield  {journal} {\bibinfo  {journal}
  {Phys. Rev. Lett.}\ }\textbf {\bibinfo {volume} {76}},\ \bibinfo {pages}
  {4584} (\bibinfo {year} {1996})}\BibitemShut {NoStop}%
\bibitem [{\citenamefont {Gallais}\ \emph {et~al.}(2008)\citenamefont
  {Gallais}, \citenamefont {Yan}, \citenamefont {Pinczuk}, \citenamefont
  {Pfeiffer},\ and\ \citenamefont {West}}]{Gallaisraman}%
  \BibitemOpen
  \bibfield  {author} {\bibinfo {author} {\bibfnamefont {Y.}~\bibnamefont
  {Gallais}}, \bibinfo {author} {\bibfnamefont {J.}~\bibnamefont {Yan}},
  \bibinfo {author} {\bibfnamefont {A.}~\bibnamefont {Pinczuk}}, \bibinfo
  {author} {\bibfnamefont {L.~N.}\ \bibnamefont {Pfeiffer}}, \ and\ \bibinfo
  {author} {\bibfnamefont {K.~W.}\ \bibnamefont {West}},\ }\href {\doibase
  10.1103/PhysRevLett.100.086806} {\bibfield  {journal} {\bibinfo  {journal}
  {Phys. Rev. Lett.}\ }\textbf {\bibinfo {volume} {100}},\ \bibinfo {pages}
  {086806} (\bibinfo {year} {2008})}\BibitemShut {NoStop}%
\bibitem [{\citenamefont {Zhu}\ \emph {et~al.}(2010)\citenamefont {Zhu},
  \citenamefont {Sambandamurthy}, \citenamefont {Chen}, \citenamefont {Jiang},
  \citenamefont {Engel}, \citenamefont {Tsui}, \citenamefont {Pfeiffer},\ and\
  \citenamefont {West}}]{Hanzhuprl}%
  \BibitemOpen
  \bibfield  {author} {\bibinfo {author} {\bibfnamefont {H.}~\bibnamefont
  {Zhu}}, \bibinfo {author} {\bibfnamefont {G.}~\bibnamefont {Sambandamurthy}},
  \bibinfo {author} {\bibfnamefont {Y.~P.}\ \bibnamefont {Chen}}, \bibinfo
  {author} {\bibfnamefont {P.}~\bibnamefont {Jiang}}, \bibinfo {author}
  {\bibfnamefont {L.~W.}\ \bibnamefont {Engel}}, \bibinfo {author}
  {\bibfnamefont {D.~C.}\ \bibnamefont {Tsui}}, \bibinfo {author}
  {\bibfnamefont {L.~N.}\ \bibnamefont {Pfeiffer}}, \ and\ \bibinfo {author}
  {\bibfnamefont {K.~W.}\ \bibnamefont {West}},\ }\href {\doibase
  10.1103/PhysRevLett.104.226801} {\bibfield  {journal} {\bibinfo  {journal}
  {Phys. Rev. Lett.}\ }\textbf {\bibinfo {volume} {104}},\ \bibinfo {pages}
  {226801} (\bibinfo {year} {2010})}\BibitemShut {NoStop}%
\bibitem [{\citenamefont {Wei}\ \emph {et~al.}(2018)\citenamefont {Wei},
  \citenamefont {Van Der~Sar}, \citenamefont {Lee}, \citenamefont {Watanabe},
  \citenamefont {Taniguchi}, \citenamefont {Halperin},\ and\ \citenamefont
  {Yacoby}}]{wei2018electrical}%
  \BibitemOpen
  \bibfield  {author} {\bibinfo {author} {\bibfnamefont {D.~S.}\ \bibnamefont
  {Wei}}, \bibinfo {author} {\bibfnamefont {T.}~\bibnamefont {Van Der~Sar}},
  \bibinfo {author} {\bibfnamefont {S.~H.}\ \bibnamefont {Lee}}, \bibinfo
  {author} {\bibfnamefont {K.}~\bibnamefont {Watanabe}}, \bibinfo {author}
  {\bibfnamefont {T.}~\bibnamefont {Taniguchi}}, \bibinfo {author}
  {\bibfnamefont {B.~I.}\ \bibnamefont {Halperin}}, \ and\ \bibinfo {author}
  {\bibfnamefont {A.}~\bibnamefont {Yacoby}},\ }\href
  {https://www.science.org/doi/abs/10.1126/science.aar4061} {\bibfield
  {journal} {\bibinfo  {journal} {Science}\ }\textbf {\bibinfo {volume}
  {362}},\ \bibinfo {pages} {229} (\bibinfo {year} {2018})}\BibitemShut
  {NoStop}%
\bibitem [{\citenamefont {Zhou}\ \emph {et~al.}(2021)\citenamefont {Zhou},
  \citenamefont {Carmiggelt}, \citenamefont {G{\"a}chter}, \citenamefont
  {Esterlis}, \citenamefont {Sels}, \citenamefont {St{\"o}hr}, \citenamefont
  {Du}, \citenamefont {Fernandez}, \citenamefont {Rodriguez-Nieva},
  \citenamefont {B{\"u}ttner} \emph {et~al.}}]{zhou2021magnon}%
  \BibitemOpen
  \bibfield  {author} {\bibinfo {author} {\bibfnamefont {T.~X.}\ \bibnamefont
  {Zhou}}, \bibinfo {author} {\bibfnamefont {J.~J.}\ \bibnamefont
  {Carmiggelt}}, \bibinfo {author} {\bibfnamefont {L.~M.}\ \bibnamefont
  {G{\"a}chter}}, \bibinfo {author} {\bibfnamefont {I.}~\bibnamefont
  {Esterlis}}, \bibinfo {author} {\bibfnamefont {D.}~\bibnamefont {Sels}},
  \bibinfo {author} {\bibfnamefont {R.~J.}\ \bibnamefont {St{\"o}hr}}, \bibinfo
  {author} {\bibfnamefont {C.}~\bibnamefont {Du}}, \bibinfo {author}
  {\bibfnamefont {D.}~\bibnamefont {Fernandez}}, \bibinfo {author}
  {\bibfnamefont {J.~F.}\ \bibnamefont {Rodriguez-Nieva}}, \bibinfo {author}
  {\bibfnamefont {F.}~\bibnamefont {B{\"u}ttner}},  \emph {et~al.},\ }\href
  {https://www.pnas.org/doi/full/10.1073/pnas.2019473118} {\bibfield  {journal}
  {\bibinfo  {journal} {Proceedings of the National Academy of Sciences}\
  }\textbf {\bibinfo {volume} {118}},\ \bibinfo {pages} {e2019473118} (\bibinfo
  {year} {2021})}\BibitemShut {NoStop}%
\bibitem [{\citenamefont {Zhou}\ \emph {et~al.}(2022)\citenamefont {Zhou},
  \citenamefont {Huang}, \citenamefont {Wei}, \citenamefont {Taniguchi},
  \citenamefont {Watanabe}, \citenamefont {Zaletel}, \citenamefont {Papi{\'c}},
  \citenamefont {MacDonald},\ and\ \citenamefont {Young}}]{zhou2022strong}%
  \BibitemOpen
  \bibfield  {author} {\bibinfo {author} {\bibfnamefont {H.}~\bibnamefont
  {Zhou}}, \bibinfo {author} {\bibfnamefont {C.}~\bibnamefont {Huang}},
  \bibinfo {author} {\bibfnamefont {N.}~\bibnamefont {Wei}}, \bibinfo {author}
  {\bibfnamefont {T.}~\bibnamefont {Taniguchi}}, \bibinfo {author}
  {\bibfnamefont {K.}~\bibnamefont {Watanabe}}, \bibinfo {author}
  {\bibfnamefont {M.~P.}\ \bibnamefont {Zaletel}}, \bibinfo {author}
  {\bibfnamefont {Z.}~\bibnamefont {Papi{\'c}}}, \bibinfo {author}
  {\bibfnamefont {A.~H.}\ \bibnamefont {MacDonald}}, \ and\ \bibinfo {author}
  {\bibfnamefont {A.~F.}\ \bibnamefont {Young}},\ }\href
  {https://journals.aps.org/prx/abstract/10.1103/PhysRevX.12.021060} {\bibfield
   {journal} {\bibinfo  {journal} {Physical Review X}\ }\textbf {\bibinfo
  {volume} {12}},\ \bibinfo {pages} {021060} (\bibinfo {year}
  {2022})}\BibitemShut {NoStop}%
\bibitem [{\citenamefont {Pierce}\ \emph {et~al.}(2022)\citenamefont {Pierce},
  \citenamefont {Xie}, \citenamefont {Lee}, \citenamefont {Forrester},
  \citenamefont {Wei}, \citenamefont {Watanabe}, \citenamefont {Taniguchi},
  \citenamefont {Halperin},\ and\ \citenamefont
  {Yacoby}}]{pierce2022thermodynamics}%
  \BibitemOpen
  \bibfield  {author} {\bibinfo {author} {\bibfnamefont {A.~T.}\ \bibnamefont
  {Pierce}}, \bibinfo {author} {\bibfnamefont {Y.}~\bibnamefont {Xie}},
  \bibinfo {author} {\bibfnamefont {S.~H.}\ \bibnamefont {Lee}}, \bibinfo
  {author} {\bibfnamefont {P.~R.}\ \bibnamefont {Forrester}}, \bibinfo {author}
  {\bibfnamefont {D.~S.}\ \bibnamefont {Wei}}, \bibinfo {author} {\bibfnamefont
  {K.}~\bibnamefont {Watanabe}}, \bibinfo {author} {\bibfnamefont
  {T.}~\bibnamefont {Taniguchi}}, \bibinfo {author} {\bibfnamefont {B.~I.}\
  \bibnamefont {Halperin}}, \ and\ \bibinfo {author} {\bibfnamefont
  {A.}~\bibnamefont {Yacoby}},\ }\href
  {https://www.nature.com/articles/s41567-021-01421-x} {\bibfield  {journal}
  {\bibinfo  {journal} {Nature Physics}\ }\textbf {\bibinfo {volume} {18}},\
  \bibinfo {pages} {37} (\bibinfo {year} {2022})}\BibitemShut {NoStop}%
\bibitem [{\citenamefont {Assouline}\ \emph {et~al.}(2021)\citenamefont
  {Assouline}, \citenamefont {Jo}, \citenamefont {Brasseur}, \citenamefont
  {Watanabe}, \citenamefont {Taniguchi}, \citenamefont {Jolicoeur},
  \citenamefont {Glattli}, \citenamefont {Kumada}, \citenamefont {Roche},
  \citenamefont {Parmentier} \emph {et~al.}}]{assouline2021excitonic}%
  \BibitemOpen
  \bibfield  {author} {\bibinfo {author} {\bibfnamefont {A.}~\bibnamefont
  {Assouline}}, \bibinfo {author} {\bibfnamefont {M.}~\bibnamefont {Jo}},
  \bibinfo {author} {\bibfnamefont {P.}~\bibnamefont {Brasseur}}, \bibinfo
  {author} {\bibfnamefont {K.}~\bibnamefont {Watanabe}}, \bibinfo {author}
  {\bibfnamefont {T.}~\bibnamefont {Taniguchi}}, \bibinfo {author}
  {\bibfnamefont {T.}~\bibnamefont {Jolicoeur}}, \bibinfo {author}
  {\bibfnamefont {D.}~\bibnamefont {Glattli}}, \bibinfo {author} {\bibfnamefont
  {N.}~\bibnamefont {Kumada}}, \bibinfo {author} {\bibfnamefont
  {P.}~\bibnamefont {Roche}}, \bibinfo {author} {\bibfnamefont
  {F.}~\bibnamefont {Parmentier}},  \emph {et~al.},\ }\href
  {https://www.nature.com/articles/s41567-021-01411-z} {\bibfield  {journal}
  {\bibinfo  {journal} {Nature Physics}\ }\textbf {\bibinfo {volume} {17}},\
  \bibinfo {pages} {1369} (\bibinfo {year} {2021})}\BibitemShut {NoStop}%
\bibitem [{\citenamefont {Zhang}\ \emph {et~al.}(2005)\citenamefont {Zhang},
  \citenamefont {Tan}, \citenamefont {Stormer},\ and\ \citenamefont
  {Kim}}]{zhang2005experimental}%
  \BibitemOpen
  \bibfield  {author} {\bibinfo {author} {\bibfnamefont {Y.}~\bibnamefont
  {Zhang}}, \bibinfo {author} {\bibfnamefont {Y.-W.}\ \bibnamefont {Tan}},
  \bibinfo {author} {\bibfnamefont {H.~L.}\ \bibnamefont {Stormer}}, \ and\
  \bibinfo {author} {\bibfnamefont {P.}~\bibnamefont {Kim}},\ }\href
  {https://www.nature.com/articles/nature04235} {\bibfield  {journal} {\bibinfo
   {journal} {nature}\ }\textbf {\bibinfo {volume} {438}},\ \bibinfo {pages}
  {201} (\bibinfo {year} {2005})}\BibitemShut {NoStop}%
\bibitem [{\citenamefont {Novoselov}\ \emph {et~al.}(2007)\citenamefont
  {Novoselov}, \citenamefont {Jiang}, \citenamefont {Zhang}, \citenamefont
  {Morozov}, \citenamefont {Stormer}, \citenamefont {Zeitler}, \citenamefont
  {Maan}, \citenamefont {Boebinger}, \citenamefont {Kim},\ and\ \citenamefont
  {Geim}}]{novoselov2007room}%
  \BibitemOpen
  \bibfield  {author} {\bibinfo {author} {\bibfnamefont {K.~S.}\ \bibnamefont
  {Novoselov}}, \bibinfo {author} {\bibfnamefont {Z.}~\bibnamefont {Jiang}},
  \bibinfo {author} {\bibfnamefont {Y.}~\bibnamefont {Zhang}}, \bibinfo
  {author} {\bibfnamefont {S.}~\bibnamefont {Morozov}}, \bibinfo {author}
  {\bibfnamefont {H.~L.}\ \bibnamefont {Stormer}}, \bibinfo {author}
  {\bibfnamefont {U.}~\bibnamefont {Zeitler}}, \bibinfo {author} {\bibfnamefont
  {J.}~\bibnamefont {Maan}}, \bibinfo {author} {\bibfnamefont {G.}~\bibnamefont
  {Boebinger}}, \bibinfo {author} {\bibfnamefont {P.}~\bibnamefont {Kim}}, \
  and\ \bibinfo {author} {\bibfnamefont {A.~K.}\ \bibnamefont {Geim}},\ }\href
  {https://www.science.org/doi/full/10.1126/science.1137201?casa_token=bc6GClkFZv4AAAAA%3AwiiTZXKWC6QtJbckrT00xY7uJhHUQpH7IHTUINuhKf_CkGBJZGJPi_SybxotWS4mzfDQbP04xYOHCKU}
  {\bibfield  {journal} {\bibinfo  {journal} {science}\ }\textbf {\bibinfo
  {volume} {315}},\ \bibinfo {pages} {1379} (\bibinfo {year}
  {2007})}\BibitemShut {NoStop}%
\bibitem [{\citenamefont {Girvin}(2002)}]{girvin2002quantum}%
  \BibitemOpen
  \bibfield  {author} {\bibinfo {author} {\bibfnamefont {S.~M.}\ \bibnamefont
  {Girvin}},\ }\bibfield  {booktitle} {\emph {\bibinfo {booktitle} {Aspects
  topologiques de la physique en basse dimension. Topological aspects of low
  dimensional systems: Session LXIX. 7--31 July 1998}},\ }\href@noop {} {\ ,\
  \bibinfo {pages} {53} (\bibinfo {year} {2002})}\BibitemShut {NoStop}%
\bibitem [{\citenamefont {Hasan}\ and\ \citenamefont
  {Kane}(2010)}]{KaneHasanTI}%
  \BibitemOpen
  \bibfield  {author} {\bibinfo {author} {\bibfnamefont {M.~Z.}\ \bibnamefont
  {Hasan}}\ and\ \bibinfo {author} {\bibfnamefont {C.~L.}\ \bibnamefont
  {Kane}},\ }\href {\doibase 10.1103/RevModPhys.82.3045} {\bibfield  {journal}
  {\bibinfo  {journal} {Rev. Mod. Phys.}\ }\textbf {\bibinfo {volume} {82}},\
  \bibinfo {pages} {3045} (\bibinfo {year} {2010})}\BibitemShut {NoStop}%
\bibitem [{\citenamefont {Zhou}\ \emph {et~al.}(2020)\citenamefont {Zhou},
  \citenamefont {Polshyn}, \citenamefont {Taniguchi}, \citenamefont
  {Watanabe},\ and\ \citenamefont {Young}}]{zhou2020solids}%
  \BibitemOpen
  \bibfield  {author} {\bibinfo {author} {\bibfnamefont {H.}~\bibnamefont
  {Zhou}}, \bibinfo {author} {\bibfnamefont {H.}~\bibnamefont {Polshyn}},
  \bibinfo {author} {\bibfnamefont {T.}~\bibnamefont {Taniguchi}}, \bibinfo
  {author} {\bibfnamefont {K.}~\bibnamefont {Watanabe}}, \ and\ \bibinfo
  {author} {\bibfnamefont {A.}~\bibnamefont {Young}},\ }\href
  {https://www.nature.com/articles/s41567-019-0729-8#Bib1} {\bibfield
  {journal} {\bibinfo  {journal} {Nature Physics}\ }\textbf {\bibinfo {volume}
  {16}},\ \bibinfo {pages} {154} (\bibinfo {year} {2020})}\BibitemShut
  {NoStop}%
\bibitem [{\citenamefont {Debarre}(2005)}]{debarre2005complex}%
  \BibitemOpen
  \bibfield  {author} {\bibinfo {author} {\bibfnamefont {O.}~\bibnamefont
  {Debarre}},\ }\href@noop {} {\emph {\bibinfo {title} {Complex tori and
  abelian varieties}}},\ \bibinfo {number} {6}\ (\bibinfo  {publisher}
  {American Mathematical Soc.},\ \bibinfo {year} {2005})\BibitemShut {NoStop}%
\bibitem [{\citenamefont {Dou\ifmmode~\mbox{\c{c}}\else \c{c}\fi{}ot}\ \emph
  {et~al.}(2008)\citenamefont {Dou\ifmmode~\mbox{\c{c}}\else \c{c}\fi{}ot},
  \citenamefont {Goerbig}, \citenamefont {Lederer},\ and\ \citenamefont
  {Moessner}}]{Doucotent}%
  \BibitemOpen
  \bibfield  {author} {\bibinfo {author} {\bibfnamefont {B.}~\bibnamefont
  {Dou\ifmmode~\mbox{\c{c}}\else \c{c}\fi{}ot}}, \bibinfo {author}
  {\bibfnamefont {M.~O.}\ \bibnamefont {Goerbig}}, \bibinfo {author}
  {\bibfnamefont {P.}~\bibnamefont {Lederer}}, \ and\ \bibinfo {author}
  {\bibfnamefont {R.}~\bibnamefont {Moessner}},\ }\href {\doibase
  10.1103/PhysRevB.78.195327} {\bibfield  {journal} {\bibinfo  {journal} {Phys.
  Rev. B}\ }\textbf {\bibinfo {volume} {78}},\ \bibinfo {pages} {195327}
  (\bibinfo {year} {2008})}\BibitemShut {NoStop}%
\bibitem [{\citenamefont {Pendry}(1994)}]{Pendryphot}%
  \BibitemOpen
  \bibfield  {author} {\bibinfo {author} {\bibfnamefont {J.}~\bibnamefont
  {Pendry}},\ }\href {\doibase 10.1080/09500349414550281} {\bibfield  {journal}
  {\bibinfo  {journal} {Journal of Modern Optics}\ }\textbf {\bibinfo {volume}
  {41}},\ \bibinfo {pages} {209} (\bibinfo {year} {1994})}\BibitemShut
  {NoStop}%
\bibitem [{\citenamefont {Iwasaki}\ \emph {et~al.}(2014)\citenamefont
  {Iwasaki}, \citenamefont {Beekman},\ and\ \citenamefont
  {Nagaosa}}]{Iwasakimag}%
  \BibitemOpen
  \bibfield  {author} {\bibinfo {author} {\bibfnamefont {J.}~\bibnamefont
  {Iwasaki}}, \bibinfo {author} {\bibfnamefont {A.~J.}\ \bibnamefont
  {Beekman}}, \ and\ \bibinfo {author} {\bibfnamefont {N.}~\bibnamefont
  {Nagaosa}},\ }\href {\doibase 10.1103/PhysRevB.89.064412} {\bibfield
  {journal} {\bibinfo  {journal} {Phys. Rev. B}\ }\textbf {\bibinfo {volume}
  {89}},\ \bibinfo {pages} {064412} (\bibinfo {year} {2014})}\BibitemShut
  {NoStop}%
\bibitem [{\citenamefont {Sch\"utte}\ and\ \citenamefont
  {Garst}(2014)}]{Schuttemag}%
  \BibitemOpen
  \bibfield  {author} {\bibinfo {author} {\bibfnamefont {C.}~\bibnamefont
  {Sch\"utte}}\ and\ \bibinfo {author} {\bibfnamefont {M.}~\bibnamefont
  {Garst}},\ }\href {\doibase 10.1103/PhysRevB.90.094423} {\bibfield  {journal}
  {\bibinfo  {journal} {Phys. Rev. B}\ }\textbf {\bibinfo {volume} {90}},\
  \bibinfo {pages} {094423} (\bibinfo {year} {2014})}\BibitemShut {NoStop}%
\bibitem [{\citenamefont {Atteia}\ \emph {et~al.}(2022)\citenamefont {Atteia},
  \citenamefont {Parmentier}, \citenamefont {Roulleau},\ and\ \citenamefont
  {Goerbig}}]{atteia2022beating}%
  \BibitemOpen
  \bibfield  {author} {\bibinfo {author} {\bibfnamefont {J.}~\bibnamefont
  {Atteia}}, \bibinfo {author} {\bibfnamefont {F.}~\bibnamefont {Parmentier}},
  \bibinfo {author} {\bibfnamefont {P.}~\bibnamefont {Roulleau}}, \ and\
  \bibinfo {author} {\bibfnamefont {M.~O.}\ \bibnamefont {Goerbig}},\ }\href
  {https://arxiv.org/abs/2204.12795} {\bibfield  {journal} {\bibinfo  {journal}
  {arXiv:2204.12795}\ } (\bibinfo {year} {2022})}\BibitemShut {NoStop}%
\bibitem [{\citenamefont {Wei}\ \emph {et~al.}(2021)\citenamefont {Wei},
  \citenamefont {Huang},\ and\ \citenamefont {MacDonald}}]{Weimagn}%
  \BibitemOpen
  \bibfield  {author} {\bibinfo {author} {\bibfnamefont {N.}~\bibnamefont
  {Wei}}, \bibinfo {author} {\bibfnamefont {C.}~\bibnamefont {Huang}}, \ and\
  \bibinfo {author} {\bibfnamefont {A.~H.}\ \bibnamefont {MacDonald}},\ }\href
  {\doibase 10.1103/PhysRevLett.126.117203} {\bibfield  {journal} {\bibinfo
  {journal} {Phys. Rev. Lett.}\ }\textbf {\bibinfo {volume} {126}},\ \bibinfo
  {pages} {117203} (\bibinfo {year} {2021})}\BibitemShut {NoStop}%
\bibitem [{\citenamefont {Moon}\ \emph {et~al.}(1995)\citenamefont {Moon},
  \citenamefont {Mori}, \citenamefont {Yang}, \citenamefont {Girvin},
  \citenamefont {MacDonald}, \citenamefont {Zheng}, \citenamefont {Yoshioka},\
  and\ \citenamefont {Zhang}}]{MoonPRB}%
  \BibitemOpen
  \bibfield  {author} {\bibinfo {author} {\bibfnamefont {K.}~\bibnamefont
  {Moon}}, \bibinfo {author} {\bibfnamefont {H.}~\bibnamefont {Mori}}, \bibinfo
  {author} {\bibfnamefont {K.}~\bibnamefont {Yang}}, \bibinfo {author}
  {\bibfnamefont {S.~M.}\ \bibnamefont {Girvin}}, \bibinfo {author}
  {\bibfnamefont {A.~H.}\ \bibnamefont {MacDonald}}, \bibinfo {author}
  {\bibfnamefont {L.}~\bibnamefont {Zheng}}, \bibinfo {author} {\bibfnamefont
  {D.}~\bibnamefont {Yoshioka}}, \ and\ \bibinfo {author} {\bibfnamefont
  {S.-C.}\ \bibnamefont {Zhang}},\ }\href {\doibase 10.1103/PhysRevB.51.5138}
  {\bibfield  {journal} {\bibinfo  {journal} {Phys. Rev. B}\ }\textbf {\bibinfo
  {volume} {51}},\ \bibinfo {pages} {5138} (\bibinfo {year}
  {1995})}\BibitemShut {NoStop}%
\bibitem [{\citenamefont {Kovrizhin}\ \emph
  {et~al.}(2013{\natexlab{a}})\citenamefont {Kovrizhin}, \citenamefont
  {Dou\ifmmode~\mbox{\c{c}}\else \c{c}\fi{}ot},\ and\ \citenamefont
  {Moessner}}]{Dimaskyrmion}%
  \BibitemOpen
  \bibfield  {author} {\bibinfo {author} {\bibfnamefont {D.~L.}\ \bibnamefont
  {Kovrizhin}}, \bibinfo {author} {\bibfnamefont {B.}~\bibnamefont
  {Dou\ifmmode~\mbox{\c{c}}\else \c{c}\fi{}ot}}, \ and\ \bibinfo {author}
  {\bibfnamefont {R.}~\bibnamefont {Moessner}},\ }\href {\doibase
  10.1103/PhysRevLett.110.186802} {\bibfield  {journal} {\bibinfo  {journal}
  {Phys. Rev. Lett.}\ }\textbf {\bibinfo {volume} {110}},\ \bibinfo {pages}
  {186802} (\bibinfo {year} {2013}{\natexlab{a}})}\BibitemShut {NoStop}%
\bibitem [{\citenamefont {Haldane}\ and\ \citenamefont
  {Rezayi}(1985)}]{Haldanetheta}%
  \BibitemOpen
  \bibfield  {author} {\bibinfo {author} {\bibfnamefont {F.~D.~M.}\
  \bibnamefont {Haldane}}\ and\ \bibinfo {author} {\bibfnamefont {E.~H.}\
  \bibnamefont {Rezayi}},\ }\href {\doibase 10.1103/PhysRevB.31.2529}
  {\bibfield  {journal} {\bibinfo  {journal} {Phys. Rev. B}\ }\textbf {\bibinfo
  {volume} {31}},\ \bibinfo {pages} {2529} (\bibinfo {year}
  {1985})}\BibitemShut {NoStop}%
\bibitem [{\citenamefont {Pichard}\ and\ \citenamefont
  {Sarma}(1981)}]{pichard1981finite}%
  \BibitemOpen
  \bibfield  {author} {\bibinfo {author} {\bibfnamefont {J.-L.}\ \bibnamefont
  {Pichard}}\ and\ \bibinfo {author} {\bibfnamefont {G.}~\bibnamefont
  {Sarma}},\ }\href
  {https://iopscience.iop.org/article/10.1088/0022-3719/14/21/004} {\bibfield
  {journal} {\bibinfo  {journal} {Journal of Physics C: Solid State Physics}\
  }\textbf {\bibinfo {volume} {14}},\ \bibinfo {pages} {L617} (\bibinfo {year}
  {1981})}\BibitemShut {NoStop}%
\bibitem [{\citenamefont {MacKinnon}\ and\ \citenamefont
  {Kramer}(1981)}]{Mackramer}%
  \BibitemOpen
  \bibfield  {author} {\bibinfo {author} {\bibfnamefont {A.}~\bibnamefont
  {MacKinnon}}\ and\ \bibinfo {author} {\bibfnamefont {B.}~\bibnamefont
  {Kramer}},\ }\href {\doibase 10.1103/PhysRevLett.47.1546} {\bibfield
  {journal} {\bibinfo  {journal} {Phys. Rev. Lett.}\ }\textbf {\bibinfo
  {volume} {47}},\ \bibinfo {pages} {1546} (\bibinfo {year}
  {1981})}\BibitemShut {NoStop}%
\bibitem [{\citenamefont {Fisher}\ and\ \citenamefont {Lee}(1981)}]{Fisherlee}%
  \BibitemOpen
  \bibfield  {author} {\bibinfo {author} {\bibfnamefont {D.~S.}\ \bibnamefont
  {Fisher}}\ and\ \bibinfo {author} {\bibfnamefont {P.~A.}\ \bibnamefont
  {Lee}},\ }\href {\doibase 10.1103/PhysRevB.23.6851} {\bibfield  {journal}
  {\bibinfo  {journal} {Phys. Rev. B}\ }\textbf {\bibinfo {volume} {23}},\
  \bibinfo {pages} {6851} (\bibinfo {year} {1981})}\BibitemShut {NoStop}%
\bibitem [{\citenamefont {Dou{\c{c}}ot}\ \emph {et~al.}(2018)\citenamefont
  {Dou{\c{c}}ot}, \citenamefont {Kovrizhin},\ and\ \citenamefont
  {Moessner}}]{douccot2018zero}%
  \BibitemOpen
  \bibfield  {author} {\bibinfo {author} {\bibfnamefont {B.}~\bibnamefont
  {Dou{\c{c}}ot}}, \bibinfo {author} {\bibfnamefont {D.~L.}\ \bibnamefont
  {Kovrizhin}}, \ and\ \bibinfo {author} {\bibfnamefont {R.}~\bibnamefont
  {Moessner}},\ }\href
  {https://www.sciencedirect.com/science/article/pii/S0003491618302707?casa_token=vu1im9EZ_84AAAAA:ta2R4d0_cEFtMajVB8QyjC2bfLYIVpCftHNLiJ6ZtpHxYbOkry7vei56m_3lObMZXI-83-UFww}
  {\bibfield  {journal} {\bibinfo  {journal} {Annals of Physics}\ }\textbf
  {\bibinfo {volume} {399}},\ \bibinfo {pages} {239} (\bibinfo {year}
  {2018})}\BibitemShut {NoStop}%
\bibitem [{\citenamefont {Kharitonov}(2012)}]{kharithall}%
  \BibitemOpen
  \bibfield  {author} {\bibinfo {author} {\bibfnamefont {M.}~\bibnamefont
  {Kharitonov}},\ }\href {\doibase 10.1103/PhysRevB.85.155439} {\bibfield
  {journal} {\bibinfo  {journal} {Phys. Rev. B}\ }\textbf {\bibinfo {volume}
  {85}},\ \bibinfo {pages} {155439} (\bibinfo {year} {2012})}\BibitemShut
  {NoStop}%
\bibitem [{\citenamefont {C\^ot\'e}\ \emph
  {et~al.}(2007{\natexlab{a}})\citenamefont {C\^ot\'e}, \citenamefont
  {Boisvert}, \citenamefont {Bourassa}, \citenamefont {Boissonneault},\ and\
  \citenamefont {Fertig}}]{Cotecoll}%
  \BibitemOpen
  \bibfield  {author} {\bibinfo {author} {\bibfnamefont {R.}~\bibnamefont
  {C\^ot\'e}}, \bibinfo {author} {\bibfnamefont {D.~B.}\ \bibnamefont
  {Boisvert}}, \bibinfo {author} {\bibfnamefont {J.}~\bibnamefont {Bourassa}},
  \bibinfo {author} {\bibfnamefont {M.}~\bibnamefont {Boissonneault}}, \ and\
  \bibinfo {author} {\bibfnamefont {H.~A.}\ \bibnamefont {Fertig}},\ }\href
  {\doibase 10.1103/PhysRevB.76.125320} {\bibfield  {journal} {\bibinfo
  {journal} {Phys. Rev. B}\ }\textbf {\bibinfo {volume} {76}},\ \bibinfo
  {pages} {125320} (\bibinfo {year} {2007}{\natexlab{a}})}\BibitemShut
  {NoStop}%
\bibitem [{\citenamefont {Kovrizhin}\ \emph
  {et~al.}(2013{\natexlab{b}})\citenamefont {Kovrizhin}, \citenamefont
  {Dou\ifmmode~\mbox{\c{c}}\else \c{c}\fi{}ot},\ and\ \citenamefont
  {Moessner}}]{Dimaskyr}%
  \BibitemOpen
  \bibfield  {author} {\bibinfo {author} {\bibfnamefont {D.~L.}\ \bibnamefont
  {Kovrizhin}}, \bibinfo {author} {\bibfnamefont {B.}~\bibnamefont
  {Dou\ifmmode~\mbox{\c{c}}\else \c{c}\fi{}ot}}, \ and\ \bibinfo {author}
  {\bibfnamefont {R.}~\bibnamefont {Moessner}},\ }\href {\doibase
  10.1103/PhysRevLett.110.186802} {\bibfield  {journal} {\bibinfo  {journal}
  {Phys. Rev. Lett.}\ }\textbf {\bibinfo {volume} {110}},\ \bibinfo {pages}
  {186802} (\bibinfo {year} {2013}{\natexlab{b}})}\BibitemShut {NoStop}%
\bibitem [{\citenamefont {C\^ot\'e}\ \emph
  {et~al.}(2007{\natexlab{b}})\citenamefont {C\^ot\'e}, \citenamefont
  {Boisvert}, \citenamefont {Bourassa}, \citenamefont {Boissonneault},\ and\
  \citenamefont {Fertig}}]{CoteCP3}%
  \BibitemOpen
  \bibfield  {author} {\bibinfo {author} {\bibfnamefont {R.}~\bibnamefont
  {C\^ot\'e}}, \bibinfo {author} {\bibfnamefont {D.~B.}\ \bibnamefont
  {Boisvert}}, \bibinfo {author} {\bibfnamefont {J.}~\bibnamefont {Bourassa}},
  \bibinfo {author} {\bibfnamefont {M.}~\bibnamefont {Boissonneault}}, \ and\
  \bibinfo {author} {\bibfnamefont {H.~A.}\ \bibnamefont {Fertig}},\ }\href
  {\doibase 10.1103/PhysRevB.76.125320} {\bibfield  {journal} {\bibinfo
  {journal} {Phys. Rev. B}\ }\textbf {\bibinfo {volume} {76}},\ \bibinfo
  {pages} {125320} (\bibinfo {year} {2007}{\natexlab{b}})}\BibitemShut
  {NoStop}%
\bibitem [{\citenamefont {Kim}\ \emph {et~al.}(2020)\citenamefont {Kim},
  \citenamefont {Xu}, \citenamefont {Berdyugin}, \citenamefont {Principi},
  \citenamefont {Slizovskiy}, \citenamefont {Xin}, \citenamefont
  {Kumaravadivel}, \citenamefont {Kuang}, \citenamefont {Hamer}, \citenamefont
  {Krishna~Kumar} \emph {et~al.}}]{kim2020control}%
  \BibitemOpen
  \bibfield  {author} {\bibinfo {author} {\bibfnamefont {M.}~\bibnamefont
  {Kim}}, \bibinfo {author} {\bibfnamefont {S.}~\bibnamefont {Xu}}, \bibinfo
  {author} {\bibfnamefont {A.}~\bibnamefont {Berdyugin}}, \bibinfo {author}
  {\bibfnamefont {A.}~\bibnamefont {Principi}}, \bibinfo {author}
  {\bibfnamefont {S.}~\bibnamefont {Slizovskiy}}, \bibinfo {author}
  {\bibfnamefont {N.}~\bibnamefont {Xin}}, \bibinfo {author} {\bibfnamefont
  {P.}~\bibnamefont {Kumaravadivel}}, \bibinfo {author} {\bibfnamefont
  {W.}~\bibnamefont {Kuang}}, \bibinfo {author} {\bibfnamefont
  {M.}~\bibnamefont {Hamer}}, \bibinfo {author} {\bibfnamefont
  {R.}~\bibnamefont {Krishna~Kumar}},  \emph {et~al.},\ }\href
  {https://www.nature.com/articles/s41467-020-15829-1} {\bibfield  {journal}
  {\bibinfo  {journal} {Nature communications}\ }\textbf {\bibinfo {volume}
  {11}},\ \bibinfo {pages} {2339} (\bibinfo {year} {2020})}\BibitemShut
  {NoStop}%
\bibitem [{\citenamefont {Fertig}\ \emph {et~al.}(1997)\citenamefont {Fertig},
  \citenamefont {Brey}, \citenamefont {C\^ot\'e}, \citenamefont {MacDonald},
  \citenamefont {Karlhede},\ and\ \citenamefont {Sondhi}}]{Fertigskyrmions}%
  \BibitemOpen
  \bibfield  {author} {\bibinfo {author} {\bibfnamefont {H.~A.}\ \bibnamefont
  {Fertig}}, \bibinfo {author} {\bibfnamefont {L.}~\bibnamefont {Brey}},
  \bibinfo {author} {\bibfnamefont {R.}~\bibnamefont {C\^ot\'e}}, \bibinfo
  {author} {\bibfnamefont {A.~H.}\ \bibnamefont {MacDonald}}, \bibinfo {author}
  {\bibfnamefont {A.}~\bibnamefont {Karlhede}}, \ and\ \bibinfo {author}
  {\bibfnamefont {S.~L.}\ \bibnamefont {Sondhi}},\ }\href {\doibase
  10.1103/PhysRevB.55.10671} {\bibfield  {journal} {\bibinfo  {journal} {Phys.
  Rev. B}\ }\textbf {\bibinfo {volume} {55}},\ \bibinfo {pages} {10671}
  (\bibinfo {year} {1997})}\BibitemShut {NoStop}%
\bibitem [{\citenamefont {Cao}\ \emph {et~al.}(2018{\natexlab{a}})\citenamefont
  {Cao}, \citenamefont {Fatemi}, \citenamefont {Demir}, \citenamefont {Fang},
  \citenamefont {Tomarken}, \citenamefont {Luo}, \citenamefont
  {Sanchez-Yamagishi}, \citenamefont {Watanabe}, \citenamefont {Taniguchi},
  \citenamefont {Kaxiras} \emph {et~al.}}]{cao2018correlated}%
  \BibitemOpen
  \bibfield  {author} {\bibinfo {author} {\bibfnamefont {Y.}~\bibnamefont
  {Cao}}, \bibinfo {author} {\bibfnamefont {V.}~\bibnamefont {Fatemi}},
  \bibinfo {author} {\bibfnamefont {A.}~\bibnamefont {Demir}}, \bibinfo
  {author} {\bibfnamefont {S.}~\bibnamefont {Fang}}, \bibinfo {author}
  {\bibfnamefont {S.~L.}\ \bibnamefont {Tomarken}}, \bibinfo {author}
  {\bibfnamefont {J.~Y.}\ \bibnamefont {Luo}}, \bibinfo {author} {\bibfnamefont
  {J.~D.}\ \bibnamefont {Sanchez-Yamagishi}}, \bibinfo {author} {\bibfnamefont
  {K.}~\bibnamefont {Watanabe}}, \bibinfo {author} {\bibfnamefont
  {T.}~\bibnamefont {Taniguchi}}, \bibinfo {author} {\bibfnamefont
  {E.}~\bibnamefont {Kaxiras}},  \emph {et~al.},\ }\href
  {https://www.nature.com/articles/nature26154} {\bibfield  {journal} {\bibinfo
   {journal} {Nature}\ }\textbf {\bibinfo {volume} {556}},\ \bibinfo {pages}
  {80} (\bibinfo {year} {2018}{\natexlab{a}})}\BibitemShut {NoStop}%
\bibitem [{\citenamefont {Cao}\ \emph {et~al.}(2018{\natexlab{b}})\citenamefont
  {Cao}, \citenamefont {Fatemi}, \citenamefont {Fang}, \citenamefont
  {Watanabe}, \citenamefont {Taniguchi}, \citenamefont {Kaxiras},\ and\
  \citenamefont {Jarillo-Herrero}}]{cao2018unconventional}%
  \BibitemOpen
  \bibfield  {author} {\bibinfo {author} {\bibfnamefont {Y.}~\bibnamefont
  {Cao}}, \bibinfo {author} {\bibfnamefont {V.}~\bibnamefont {Fatemi}},
  \bibinfo {author} {\bibfnamefont {S.}~\bibnamefont {Fang}}, \bibinfo {author}
  {\bibfnamefont {K.}~\bibnamefont {Watanabe}}, \bibinfo {author}
  {\bibfnamefont {T.}~\bibnamefont {Taniguchi}}, \bibinfo {author}
  {\bibfnamefont {E.}~\bibnamefont {Kaxiras}}, \ and\ \bibinfo {author}
  {\bibfnamefont {P.}~\bibnamefont {Jarillo-Herrero}},\ }\href
  {https://www.nature.com/articles/nature26160%3C} {\bibfield  {journal}
  {\bibinfo  {journal} {Nature}\ }\textbf {\bibinfo {volume} {556}},\ \bibinfo
  {pages} {43} (\bibinfo {year} {2018}{\natexlab{b}})}\BibitemShut {NoStop}%
\bibitem [{\citenamefont {Khalaf}\ \emph {et~al.}(2021)\citenamefont {Khalaf},
  \citenamefont {Chatterjee}, \citenamefont {Bultinck}, \citenamefont
  {Zaletel},\ and\ \citenamefont {Vishwanath}}]{khalaf2021charged}%
  \BibitemOpen
  \bibfield  {author} {\bibinfo {author} {\bibfnamefont {E.}~\bibnamefont
  {Khalaf}}, \bibinfo {author} {\bibfnamefont {S.}~\bibnamefont {Chatterjee}},
  \bibinfo {author} {\bibfnamefont {N.}~\bibnamefont {Bultinck}}, \bibinfo
  {author} {\bibfnamefont {M.~P.}\ \bibnamefont {Zaletel}}, \ and\ \bibinfo
  {author} {\bibfnamefont {A.}~\bibnamefont {Vishwanath}},\ }\href
  {https://www.science.org/doi/full/10.1126/sciadv.abf5299} {\bibfield
  {journal} {\bibinfo  {journal} {Science advances}\ }\textbf {\bibinfo
  {volume} {7}},\ \bibinfo {pages} {eabf5299} (\bibinfo {year}
  {2021})}\BibitemShut {NoStop}%
\bibitem [{\citenamefont {Khalaf}\ and\ \citenamefont
  {Vishwanath}(2022)}]{khalaf2022baby}%
  \BibitemOpen
  \bibfield  {author} {\bibinfo {author} {\bibfnamefont {E.}~\bibnamefont
  {Khalaf}}\ and\ \bibinfo {author} {\bibfnamefont {A.}~\bibnamefont
  {Vishwanath}},\ }\href {https://www.nature.com/articles/s41467-022-33673-3}
  {\bibfield  {journal} {\bibinfo  {journal} {Nature Communications}\ }\textbf
  {\bibinfo {volume} {13}},\ \bibinfo {pages} {6245} (\bibinfo {year}
  {2022})}\BibitemShut {NoStop}%
\bibitem [{\citenamefont {Kwan}\ \emph {et~al.}(2022)\citenamefont {Kwan},
  \citenamefont {Wagner}, \citenamefont {Bultinck}, \citenamefont {Simon},\
  and\ \citenamefont {Parameswaran}}]{kwan2022skyrmions}%
  \BibitemOpen
  \bibfield  {author} {\bibinfo {author} {\bibfnamefont {Y.~H.}\ \bibnamefont
  {Kwan}}, \bibinfo {author} {\bibfnamefont {G.}~\bibnamefont {Wagner}},
  \bibinfo {author} {\bibfnamefont {N.}~\bibnamefont {Bultinck}}, \bibinfo
  {author} {\bibfnamefont {S.~H.}\ \bibnamefont {Simon}}, \ and\ \bibinfo
  {author} {\bibfnamefont {S.}~\bibnamefont {Parameswaran}},\ }\href
  {https://journals.aps.org/prx/abstract/10.1103/PhysRevX.12.031020} {\bibfield
   {journal} {\bibinfo  {journal} {Physical Review X}\ }\textbf {\bibinfo
  {volume} {12}},\ \bibinfo {pages} {031020} (\bibinfo {year}
  {2022})}\BibitemShut {NoStop}%
\bibitem [{\citenamefont {Chatterjee}\ \emph {et~al.}(2020)\citenamefont
  {Chatterjee}, \citenamefont {Bultinck},\ and\ \citenamefont
  {Zaletel}}]{chatterjee2020symmetry}%
  \BibitemOpen
  \bibfield  {author} {\bibinfo {author} {\bibfnamefont {S.}~\bibnamefont
  {Chatterjee}}, \bibinfo {author} {\bibfnamefont {N.}~\bibnamefont
  {Bultinck}}, \ and\ \bibinfo {author} {\bibfnamefont {M.~P.}\ \bibnamefont
  {Zaletel}},\ }\href
  {https://journals.aps.org/prb/abstract/10.1103/PhysRevB.101.165141}
  {\bibfield  {journal} {\bibinfo  {journal} {Physical Review B}\ }\textbf
  {\bibinfo {volume} {101}},\ \bibinfo {pages} {165141} (\bibinfo {year}
  {2020})}\BibitemShut {NoStop}%
\bibitem [{\citenamefont {Grover}\ \emph {et~al.}(2022)\citenamefont {Grover},
  \citenamefont {Bocarsly}, \citenamefont {Uri}, \citenamefont {Stepanov},
  \citenamefont {Di~Battista}, \citenamefont {Roy}, \citenamefont {Xiao},
  \citenamefont {Meltzer}, \citenamefont {Myasoedov}, \citenamefont {Pareek}
  \emph {et~al.}}]{grover2022chern}%
  \BibitemOpen
  \bibfield  {author} {\bibinfo {author} {\bibfnamefont {S.}~\bibnamefont
  {Grover}}, \bibinfo {author} {\bibfnamefont {M.}~\bibnamefont {Bocarsly}},
  \bibinfo {author} {\bibfnamefont {A.}~\bibnamefont {Uri}}, \bibinfo {author}
  {\bibfnamefont {P.}~\bibnamefont {Stepanov}}, \bibinfo {author}
  {\bibfnamefont {G.}~\bibnamefont {Di~Battista}}, \bibinfo {author}
  {\bibfnamefont {I.}~\bibnamefont {Roy}}, \bibinfo {author} {\bibfnamefont
  {J.}~\bibnamefont {Xiao}}, \bibinfo {author} {\bibfnamefont {A.~Y.}\
  \bibnamefont {Meltzer}}, \bibinfo {author} {\bibfnamefont {Y.}~\bibnamefont
  {Myasoedov}}, \bibinfo {author} {\bibfnamefont {K.}~\bibnamefont {Pareek}},
  \emph {et~al.},\ }\href {https://www.nature.com/articles/s41567-022-01635-7}
  {\bibfield  {journal} {\bibinfo  {journal} {Nature physics}\ }\textbf
  {\bibinfo {volume} {18}},\ \bibinfo {pages} {885} (\bibinfo {year}
  {2022})}\BibitemShut {NoStop}%
\bibitem [{\citenamefont {Xu}\ \emph {et~al.}(2021)\citenamefont {Xu},
  \citenamefont {Ray}, \citenamefont {Shao}, \citenamefont {Jiang},
  \citenamefont {Weber}, \citenamefont {Goldberger}, \citenamefont {Watanabe},
  \citenamefont {Taniguchi}, \citenamefont {Muller}, \citenamefont {Mak} \emph
  {et~al.}}]{xu2021emergence}%
  \BibitemOpen
  \bibfield  {author} {\bibinfo {author} {\bibfnamefont {Y.}~\bibnamefont
  {Xu}}, \bibinfo {author} {\bibfnamefont {A.}~\bibnamefont {Ray}}, \bibinfo
  {author} {\bibfnamefont {Y.-T.}\ \bibnamefont {Shao}}, \bibinfo {author}
  {\bibfnamefont {S.}~\bibnamefont {Jiang}}, \bibinfo {author} {\bibfnamefont
  {D.}~\bibnamefont {Weber}}, \bibinfo {author} {\bibfnamefont {J.~E.}\
  \bibnamefont {Goldberger}}, \bibinfo {author} {\bibfnamefont
  {K.}~\bibnamefont {Watanabe}}, \bibinfo {author} {\bibfnamefont
  {T.}~\bibnamefont {Taniguchi}}, \bibinfo {author} {\bibfnamefont {D.~A.}\
  \bibnamefont {Muller}}, \bibinfo {author} {\bibfnamefont {K.~F.}\
  \bibnamefont {Mak}},  \emph {et~al.},\ }\href
  {https://arxiv.org/abs/2103.09850} {\bibfield  {journal} {\bibinfo  {journal}
  {arXiv:2103.09850}\ } (\bibinfo {year} {2021})}\BibitemShut {NoStop}%
\bibitem [{\citenamefont {Akram}\ \emph {et~al.}(2021)\citenamefont {Akram},
  \citenamefont {LaBollita}, \citenamefont {Dey}, \citenamefont {Kapeghian},
  \citenamefont {Erten},\ and\ \citenamefont {Botana}}]{akram2021moire}%
  \BibitemOpen
  \bibfield  {author} {\bibinfo {author} {\bibfnamefont {M.}~\bibnamefont
  {Akram}}, \bibinfo {author} {\bibfnamefont {H.}~\bibnamefont {LaBollita}},
  \bibinfo {author} {\bibfnamefont {D.}~\bibnamefont {Dey}}, \bibinfo {author}
  {\bibfnamefont {J.}~\bibnamefont {Kapeghian}}, \bibinfo {author}
  {\bibfnamefont {O.}~\bibnamefont {Erten}}, \ and\ \bibinfo {author}
  {\bibfnamefont {A.~S.}\ \bibnamefont {Botana}},\ }\href
  {https://pubs.acs.org/doi/abs/10.1021/acs.nanolett.1c02096} {\bibfield
  {journal} {\bibinfo  {journal} {Nano Letters}\ }\textbf {\bibinfo {volume}
  {21}},\ \bibinfo {pages} {6633} (\bibinfo {year} {2021})}\BibitemShut
  {NoStop}%
\bibitem [{\citenamefont {Brey}\ \emph {et~al.}(1996)\citenamefont {Brey},
  \citenamefont {Fertig}, \citenamefont {C\^ot\'e},\ and\ \citenamefont
  {MacDonald}}]{Breymeron}%
  \BibitemOpen
  \bibfield  {author} {\bibinfo {author} {\bibfnamefont {L.}~\bibnamefont
  {Brey}}, \bibinfo {author} {\bibfnamefont {H.~A.}\ \bibnamefont {Fertig}},
  \bibinfo {author} {\bibfnamefont {R.}~\bibnamefont {C\^ot\'e}}, \ and\
  \bibinfo {author} {\bibfnamefont {A.~H.}\ \bibnamefont {MacDonald}},\ }\href
  {\doibase 10.1103/PhysRevB.54.16888} {\bibfield  {journal} {\bibinfo
  {journal} {Phys. Rev. B}\ }\textbf {\bibinfo {volume} {54}},\ \bibinfo
  {pages} {16888} (\bibinfo {year} {1996})}\BibitemShut {NoStop}%
\bibitem [{\citenamefont {Kamilla}\ \emph {et~al.}(1996)\citenamefont
  {Kamilla}, \citenamefont {Wu},\ and\ \citenamefont
  {Jain}}]{kamilla1996skyrmions}%
  \BibitemOpen
  \bibfield  {author} {\bibinfo {author} {\bibfnamefont {R.}~\bibnamefont
  {Kamilla}}, \bibinfo {author} {\bibfnamefont {X.}~\bibnamefont {Wu}}, \ and\
  \bibinfo {author} {\bibfnamefont {J.}~\bibnamefont {Jain}},\ }\href
  {https://www.sciencedirect.com/science/article/pii/0038109896001263}
  {\bibfield  {journal} {\bibinfo  {journal} {Solid state communications}\
  }\textbf {\bibinfo {volume} {99}},\ \bibinfo {pages} {289} (\bibinfo {year}
  {1996})}\BibitemShut {NoStop}%
\bibitem [{\citenamefont {Balram}\ \emph {et~al.}(2015)\citenamefont {Balram},
  \citenamefont {Wurstbauer}, \citenamefont {W{\'o}js}, \citenamefont
  {Pinczuk},\ and\ \citenamefont {Jain}}]{balram2015fractionally}%
  \BibitemOpen
  \bibfield  {author} {\bibinfo {author} {\bibfnamefont {A.~C.}\ \bibnamefont
  {Balram}}, \bibinfo {author} {\bibfnamefont {U.}~\bibnamefont {Wurstbauer}},
  \bibinfo {author} {\bibfnamefont {A.}~\bibnamefont {W{\'o}js}}, \bibinfo
  {author} {\bibfnamefont {A.}~\bibnamefont {Pinczuk}}, \ and\ \bibinfo
  {author} {\bibfnamefont {J.}~\bibnamefont {Jain}},\ }\href
  {https://www.nature.com/articles/ncomms9981} {\bibfield  {journal} {\bibinfo
  {journal} {Nature communications}\ }\textbf {\bibinfo {volume} {6}},\
  \bibinfo {pages} {8981} (\bibinfo {year} {2015})}\BibitemShut {NoStop}%
\bibitem [{\citenamefont {Doretto}\ \emph {et~al.}(2005)\citenamefont
  {Doretto}, \citenamefont {Goerbig}, \citenamefont {Lederer}, \citenamefont
  {Caldeira},\ and\ \citenamefont {Smith}}]{doretto2005spin}%
  \BibitemOpen
  \bibfield  {author} {\bibinfo {author} {\bibfnamefont {R.~L.}\ \bibnamefont
  {Doretto}}, \bibinfo {author} {\bibfnamefont {M.~O.}\ \bibnamefont
  {Goerbig}}, \bibinfo {author} {\bibfnamefont {P.}~\bibnamefont {Lederer}},
  \bibinfo {author} {\bibfnamefont {A.~O.}\ \bibnamefont {Caldeira}}, \ and\
  \bibinfo {author} {\bibfnamefont {C.~M.}\ \bibnamefont {Smith}},\ }\href
  {https://journals.aps.org/prb/abstract/10.1103/PhysRevB.72.035341} {\bibfield
   {journal} {\bibinfo  {journal} {Physical Review B}\ }\textbf {\bibinfo
  {volume} {72}},\ \bibinfo {pages} {035341} (\bibinfo {year}
  {2005})}\BibitemShut {NoStop}%
\bibitem [{\citenamefont {W{\'o}js}\ and\ \citenamefont
  {Quinn}(2002)}]{wojs2002spin}%
  \BibitemOpen
  \bibfield  {author} {\bibinfo {author} {\bibfnamefont {A.}~\bibnamefont
  {W{\'o}js}}\ and\ \bibinfo {author} {\bibfnamefont {J.~J.}\ \bibnamefont
  {Quinn}},\ }\href
  {https://journals.aps.org/prb/abstract/10.1103/PhysRevB.66.045323} {\bibfield
   {journal} {\bibinfo  {journal} {Physical Review B}\ }\textbf {\bibinfo
  {volume} {66}},\ \bibinfo {pages} {045323} (\bibinfo {year}
  {2002})}\BibitemShut {NoStop}%
\bibitem [{\citenamefont {Lian}\ and\ \citenamefont {Goerbig}(2017)}]{Lianent}%
  \BibitemOpen
  \bibfield  {author} {\bibinfo {author} {\bibfnamefont {Y.}~\bibnamefont
  {Lian}}\ and\ \bibinfo {author} {\bibfnamefont {M.~O.}\ \bibnamefont
  {Goerbig}},\ }\href {\doibase 10.1103/PhysRevB.95.245428} {\bibfield
  {journal} {\bibinfo  {journal} {Phys. Rev. B}\ }\textbf {\bibinfo {volume}
  {95}},\ \bibinfo {pages} {245428} (\bibinfo {year} {2017})}\BibitemShut
  {NoStop}%
\bibitem [{\citenamefont {Weber}\ \emph {et~al.}(2022)\citenamefont {Weber},
  \citenamefont {Fobes}, \citenamefont {Waizner}, \citenamefont {Steffens},
  \citenamefont {Tucker}, \citenamefont {Bohm}, \citenamefont {Beddrich},
  \citenamefont {Franz}, \citenamefont {Gabold}, \citenamefont {Bewley} \emph
  {et~al.}}]{weber2022emergent}%
  \BibitemOpen
  \bibfield  {author} {\bibinfo {author} {\bibfnamefont {T.}~\bibnamefont
  {Weber}}, \bibinfo {author} {\bibfnamefont {D.~M.}\ \bibnamefont {Fobes}},
  \bibinfo {author} {\bibfnamefont {J.}~\bibnamefont {Waizner}}, \bibinfo
  {author} {\bibfnamefont {P.}~\bibnamefont {Steffens}}, \bibinfo {author}
  {\bibfnamefont {G.}~\bibnamefont {Tucker}}, \bibinfo {author} {\bibfnamefont
  {M.}~\bibnamefont {Bohm}}, \bibinfo {author} {\bibfnamefont {L.}~\bibnamefont
  {Beddrich}}, \bibinfo {author} {\bibfnamefont {C.}~\bibnamefont {Franz}},
  \bibinfo {author} {\bibfnamefont {H.}~\bibnamefont {Gabold}}, \bibinfo
  {author} {\bibfnamefont {R.}~\bibnamefont {Bewley}},  \emph {et~al.},\ }\href
  {https://www.science.org/doi/10.1126/science.abe4441} {\bibfield  {journal}
  {\bibinfo  {journal} {Science}\ }\textbf {\bibinfo {volume} {375}} (\bibinfo
  {year} {2022})}\BibitemShut {NoStop}%
\bibitem [{\citenamefont {Yan}\ \emph {et~al.}(2011)\citenamefont {Yan},
  \citenamefont {Wang},\ and\ \citenamefont {Wang}}]{Yantorque}%
  \BibitemOpen
  \bibfield  {author} {\bibinfo {author} {\bibfnamefont {P.}~\bibnamefont
  {Yan}}, \bibinfo {author} {\bibfnamefont {X.~S.}\ \bibnamefont {Wang}}, \
  and\ \bibinfo {author} {\bibfnamefont {X.~R.}\ \bibnamefont {Wang}},\ }\href
  {\doibase 10.1103/PhysRevLett.107.177207} {\bibfield  {journal} {\bibinfo
  {journal} {Phys. Rev. Lett.}\ }\textbf {\bibinfo {volume} {107}},\ \bibinfo
  {pages} {177207} (\bibinfo {year} {2011})}\BibitemShut {NoStop}%
\bibitem [{\citenamefont {Kim}\ \emph {et~al.}(2014)\citenamefont {Kim},
  \citenamefont {Tserkovnyak},\ and\ \citenamefont
  {Tchernyshyov}}]{Kimpropulsion}%
  \BibitemOpen
  \bibfield  {author} {\bibinfo {author} {\bibfnamefont {S.~K.}\ \bibnamefont
  {Kim}}, \bibinfo {author} {\bibfnamefont {Y.}~\bibnamefont {Tserkovnyak}}, \
  and\ \bibinfo {author} {\bibfnamefont {O.}~\bibnamefont {Tchernyshyov}},\
  }\href {\doibase 10.1103/PhysRevB.90.104406} {\bibfield  {journal} {\bibinfo
  {journal} {Phys. Rev. B}\ }\textbf {\bibinfo {volume} {90}},\ \bibinfo
  {pages} {104406} (\bibinfo {year} {2014})}\BibitemShut {NoStop}%
\bibitem [{\citenamefont {{\v{S}}mejkal}\ \emph {et~al.}(2018)\citenamefont
  {{\v{S}}mejkal}, \citenamefont {Mokrousov}, \citenamefont {Yan},\ and\
  \citenamefont {MacDonald}}]{vsmejkal2018topological}%
  \BibitemOpen
  \bibfield  {author} {\bibinfo {author} {\bibfnamefont {L.}~\bibnamefont
  {{\v{S}}mejkal}}, \bibinfo {author} {\bibfnamefont {Y.}~\bibnamefont
  {Mokrousov}}, \bibinfo {author} {\bibfnamefont {B.}~\bibnamefont {Yan}}, \
  and\ \bibinfo {author} {\bibfnamefont {A.~H.}\ \bibnamefont {MacDonald}},\
  }\href {https://www.nature.com/articles/s41567-018-0064-5} {\bibfield
  {journal} {\bibinfo  {journal} {Nature physics}\ }\textbf {\bibinfo {volume}
  {14}},\ \bibinfo {pages} {242} (\bibinfo {year} {2018})}\BibitemShut
  {NoStop}%
\bibitem [{\citenamefont {Lee}\ \emph {et~al.}(2022)\citenamefont {Lee},
  \citenamefont {Nakata}, \citenamefont {Tchernyshyov},\ and\ \citenamefont
  {Kim}}]{lee2022magnon}%
  \BibitemOpen
  \bibfield  {author} {\bibinfo {author} {\bibfnamefont {S.}~\bibnamefont
  {Lee}}, \bibinfo {author} {\bibfnamefont {K.}~\bibnamefont {Nakata}},
  \bibinfo {author} {\bibfnamefont {O.}~\bibnamefont {Tchernyshyov}}, \ and\
  \bibinfo {author} {\bibfnamefont {S.~K.}\ \bibnamefont {Kim}},\ }\href
  {https://arxiv.org/abs/2211.00030} {\bibfield  {journal} {\bibinfo  {journal}
  {arXiv:2211.00030}\ } (\bibinfo {year} {2022})}\BibitemShut {NoStop}%
\end{thebibliography}%
\end{document}